\documentclass[12pt]{article}

\usepackage[titletoc]{appendix}
\usepackage[margin={1.0in}]{geometry}
\usepackage{rotating}
\usepackage{lscape}
\usepackage{graphicx}
\usepackage{color}
\usepackage{subfig}
\usepackage{comment}
\renewcommand{\comment}[1]{}

\usepackage{threeparttable}
\usepackage{booktabs}
\usepackage{longtable}
\usepackage{threeparttablex}
\usepackage{tabularx}
\usepackage{pdflscape} 
\usepackage{natbib}

\usepackage{authblk}
\usepackage{mathtools}
\usepackage{dsfont}
\usepackage{lmodern} 
\usepackage{anyfontsize} 
\usepackage[sc]{mathpazo}
\usepackage{courier}
\usepackage[utf8]{inputenc}
\usepackage[T1]{fontenc}
\usepackage{setspace}
\usepackage{amssymb}
\usepackage{amsmath}
\usepackage{mathrsfs}
\usepackage{amsthm}
\usepackage{bbm}
\usepackage{microtype}
\usepackage{float} 
\usepackage{multirow}
\usepackage{enumerate}
\usepackage{hyperref} 

\usepackage{tikz}
\usepackage[nameinlink]{cleveref} 
\usepackage{afterpage}

\usepackage{array}
\usepackage{adjustbox}

\usepackage{xfrac}

\ExplSyntaxOn
\cs_new:Npn \expandableinput #1
  { \use:c { @@input } { \file_full_name:n {#1} } }
\ExplSyntaxOff

\hypersetup{
	colorlinks=true, 
	linktoc=all,        
	linkcolor=black,  
	urlcolor  = blue,
	citecolor = black,
    pdftitle={Transportation Technology and Gentrification: Evidence from the entry of Ridesharing Services},
    pdfauthor={Sumit Agarwal, Shashwat Alok, Sergio Correia, Deepa Mani, Bernardo Morais},
    pdfstartview={Fit}
}

\usepackage[nottoc,notlot,notlof]{tocbibind}

\hypersetup{pdfstartview={Fit}} 
\RequirePackage[font=small,format=plain,labelfont=bf,textfont=it]{caption}

\def\eatcell#1\unskip{}
\newcolumntype{E}{>{\eatcell}c@{}}
\newcolumntype{H}{>{\lrbox0}c<{\endlrbox}@{}}

\begin{document}
	
\title{\textbf{Transportation Technology and Gentrification: \\
 Evidence from Ridesharing Services}}

\author{\vspace{0.5cm}Sumit Agarwal, Shashwat Alok, Sergio Correia, \\
Deepa Mani, and Bernardo Morais\thanks{Corresponding Author. Agarwal: National University of Singapore, \href{mailto:ushakri@yahoo.com}{ushakri@yahoo.com}; Alok: Indian School of Business, \href{mailto:shashwat\_alok@isb.edu}{shashwat\_alok@isb.edu}; Correia: Board of Governors of the Federal Reserve System, \href{mailto:sergio.a.correia@frb.gov}{sergio.a.correia@frb.gov}; Mani:  Indian School of Business, \href{mailto:deepa\_mani@isb.edu}{deepa\_mani@isb.edu};  Morais, Board of Governors of the Federal Reserve System, \href{mailto:bernardo.c.morais@frb.gov}{bernardo.c.morais@frb.gov}.

We thank Etienne Gagnon, Caitlin Gorback, Andrew Haughwout, Jose Luis Peydro, Wilbert van der Klaauw, Stijn Van Nieuwerburgh, Christopher Waller, and seminar participants at the Federal Reserve Board and New York Fed's System Equitable Growth Conference for comments and suggestions.

The opinions expressed in this paper do not necessarily reflect those of the Board of Governors of the Federal Reserve System.
}}
	
\date{\vspace{0.5cm} \today}

\pagenumbering{gobble}
\maketitle

\begin{abstract}
\fontsize{12.0pt}{15.0pt}\selectfont
   We analyze the staggered entry of rideshare services across U.S. metropolitan areas, estimating its effect on the spatial redistribution and real outcomes of residents. Ridesharing services gentrify urban areas—especially those with ex-ante lower housing values—causing housing prices to rise 9 percent, with the in-migration of rich-younger individuals more than offsetting the out-migration of incumbent residents and reduced in-migration of poorer individuals. Impact on incumbent residents is conditional on ex-ante homeownership. For homeowners, there is no displacement and a decline in delinquency rates. For non-homeowners, displacement and delinquency rates rise 11 percent and 42 percent, respectively. Our study emphasizes how the private provision of high-end transportation technologies can increase urbanization and exacerbate inequality.
\end{abstract}
	
	\vspace{0.5cm}
	
	\vspace{0.5cm}
    \vspace{0.5cm}
	
    \newpage
    \setcounter{page}{1}

\doublespacing
\clearpage
\pagenumbering{arabic}
 
    \section{Introduction}

    Urban areas in the United States have undergone remarkable gentrification over the past two decades, transforming from low-income, low-education to high-income, high-education zones \citep{couture2023neighborhood}. This urban gentrification, a phenomenon in which higher-income individuals migrate into denser areas, has significant implications for social welfare as the arriving high-income individuals drive up prices of services (e.g. housing), potentially causing the displacement of lower-income residents. Such spatial sorting exacerbates wealth inequality, inter-generational mobility, and long-term life outcomes such as education, access to credit, income, and crime \citep{boustan2010effect, boustan2010postwar, baum2019long, couture2019income, chyn2021neighborhoods, kulkarni2022homeownership,  cook2024urban}. Consequently, the causes and effects of urban gentrification are a subject of considerable academic and policy interest.
	
    In this paper, we expand the literature relating transportation innovations with gentrification.\footnote{See \citet{boustan2013urbanization} and \citet{couture2023neighborhood} for a survey.} While previous research emphasizes the effect of low-cost mass public transit on the spatial redistribution of individuals \citep{baum2007did, glaeser2008poor, gupta2022take, lee2023urban, severen2023commuting} we focus on the impact that private ridesharing services—a technology offering convenient personal transportation—have on urban gentrification. We do this for three main reasons. First, while the adoption of rideshare services has been large and widespread across metropolitan statistical areas (MSAs), its use has been largely driven by younger and higher-income individuals given technological requirements and relatively high cost \citep{hall2018uber}. Second, the use of rideshare services is concentrated in urban areas, as opposed to relatively more sprawled suburban areas, as the denser cores increase the frequency of use, helping reduce operational costs (e.g., matching) \citep{cramer2016disruptive, gorback2020your, li2022demand, agarwal2023impact}.\footnote{Rideshare services also eliminate costs associated with car ownership (e.g, parking) which are larger in urban areas \citep{glaeser2008poor}, and provide last-mile connectivity for public transit, which predominantly serves urban areas \citep{hall2018uber}.} Finally, the advent of autonomous taxis should further expand the supply of rideshare services at a lower cost, potentially reinforcing the effects caused by the entry of driver-based rideshare services.\footnote{For example, \textit{Waymo} already operates an autonomous taxi service in Phoenix and San Francisco. While the cost of autonomous taxi services is currently similar to those of equivalent driver-based rideshare services, it is estimated that by 2030, its cost is likely to be 50 percent lower per trip \citep{mckinsey_weymo}.}  
 
    Summarizing our results, we find that the entry of rideshare services increases urban gentrification. Upon its entry, there is a gradual increase in the number of high-income individuals in urban zip codes, with a smaller opposite effect for low-income individuals. This in-migration is driven by young individuals without dependents and is muted in suburban areas. Mortgage originations rise—especially in urban areas with lower housing values—as the increase in high-income mortgage applicants more than offsets the decline in subsidized mortgages, leading to a sharp increase in housing prices.\footnote{Importantly, and corroborating our identification strategy, we find that the documented impact of the entry of rideshare services, was reversed when rideshare services temporarily exited Austin in 2016.} The impact of the price appreciation on incumbent residents depends on their ex-ante homeownership status. For homeowners, there is no impact on relocation and a significant improvement in their financial health. Conversely, non-homeowners experience a sharp increase in out-migration and credit delinquency.  Collectively, these results point to a rideshare-driven urban gentrification within metropolitan statistical areas (MSAs) and to an increase in inequality between incumbent homeowners and lower-income renters.

    To examine the impact of rideshare services on urban gentrification, we use the staggered entry of \textit{Uber} and \textit{Lyft} (henceforth rideshare services) into MSAs across the United States and use the difference-in-differences estimator proposed by \citet{borusyak2021revisiting}. Specifically, we compare observations in MSAs with rideshare services with not-yet-treated observations.\footnote{This method circumvents the recently documented limitations of two-way fixed effect (TWFE) models when treatment effects are heterogeneous and time to treatment varies \citep{borusyak2021revisiting, callaway2021difference, de2020two, goodman2021difference, jakiela2021simple, sun2021estimating}.}	This difference-in-differences specification is augmented with a multitude of detailed controls, granular location fixed effects at the zip-code and MSA level, and broader location*time (e.g., state*year) fixed effects. For additional robustness, we also control for time-varying factors at a local level using MSA-time fixed effects, classifying urban zip-codes as the treated group. These fixed effects help allay worries that our findings are influenced by unobservable time-varying confounders at the MSA level. Finally, a concern regarding our identification strategy is that entry of rideshare services may be correlated with time-varying factors influencing the local economy. However, we show that the entry of rideshare services is mainly determined by the population in the MSA and is unrelated to local economic trends.\footnote{These findings are in line with literature documenting that entry decisions of rideshare services are primarily based on population rank-order and orthogonal to economic activity \citep{hall2018uber, barrios2022launching, fos2019gig, buchak2019financing}.} 
    
    We start our analysis by uncovering the impact of rideshare entry on the composition of households across urban and suburban zip-codes. Since gentrification is associated with the in-migration of high-income individuals \citep{baum2019long}, we examine the effects of ridesharing services on the number of high- and low-income households in treated MSAs. Consistent with gentrification, we observe a gradual annual increase averaging 3.5 percent in the number of high-income households in urban zip-codes in the five years following the entry of rideshare services. Conversely, we find a slight reduction in the number of low-income households of around 0.8 percent in urban zip-codes. We do not observe these patterns in our control group of suburban zip-codes. To shed further light on changes in the demographic composition of treated MSAs, and given that younger individuals are likelier to use rideshare services \citep{hall2018uber} we examine the age composition of individuals in a zip-code. As expected, we find a decrease in the average age of individuals of around 0.8 percent in treated urban zip-codes, with no impact on suburban areas. Finally, we analyze the fraction of households claiming a child tax credit, a proxy for being a household with children, and find that rideshare entry is associated with a decline of roughly 6 percent in the fraction of households claiming a child tax credit, with no effect on suburban areas. Overall, these aggregate results are consistent with an in-migration of high-income/younger individuals into urban zip-codes, at the expense of a smaller out-migration of lower-income households. 

    We then examine the impact of rideshare services at a more granular level. Starting with the impact on housing markets, and given that a significant proportion of residential property acquisitions are financed through mortgages \citep{garriga_2020}, we focus on the number and the characteristics of mortgage originations. We find a significant increase in the number of mortgages in treated MSAs, with originations increasing by 8.1 percent following the entry of rideshare services. In suburban zip-codes the effect is significantly more muted.\footnote{Similarly, and related to the increase in demand for housing, we also find a decline in the number of vacant homes in urban zip codes, with vacancy rates dropping 11 percent in the years after treatment. Again, the effect is weak in suburban areas.} Consistent with the in-migration of high-income individuals into urban zip-codes, we find that the average income of mortgage applicants increases by 6.5 percent in the years after the entry of ride-share services. Relatedly, we also find a sharp reduction of roughly 30 percent in the share of (subsidized) FHA-insured mortgages. This is consistent with the reduction in opportunities for lower-income groups to live in a gentrified area \citep{couture2023neighborhood} and indicates a decline in the inflow of low-income individuals into urban MSAs, further intensifying spatial economic segregation. Again, we detect no distinct pre-trends and no effect on suburban zip-codes for all outcomes of interest. 
    
    Subsequently, we analyze whether new mortgages cluster in more affordable areas within urban zip-codes in line with the mechanism proposed by \citet{guerrieri2013endogenous}. Indeed, we find that within urban areas, mortgages originated in zip codes with 2.2 percent lower house prices on average, suggesting that an improvement in amenities—like the one provided by the entry of ride-hailing services—can engender an in-migration of higher-income individuals into lower-priced neighborhoods adjacent to richer areas. Finally, we evaluate the impact that increased housing demand in urban neighborhoods has on home prices, especially in neighborhoods with ex-ante lower prices. The entry of rideshare services leads to an average 9 percent increase in house prices in urban areas in the four years after treatment. Similarly, we also find that rental prices increase on average 6 percent in the years following the entry of rideshare services.As a robustness check, we use the temporary exit of rideshare services from Austin in May 2016 to examine if it led to a reversal in housing demand in urban areas. Focusing on rental prices, which tend to be more sensitive in the short term to fluctuations in demand, we find an economically and statistically significant impact on rental prices in urban areas of around 2 percent.
    
    Finally, and given that gentrification tends to increase the price of amenities \citep{berkes2023, couture2019income}, we conclude our paper by analyzing the impact that this urban gentrification has on incumbent residents, more concretely on their residence and their financial health. Furthermore, given that gentrification led to an increase in house prices, we split our analysis into existing homeowners and non-homeowners. The underlying idea is that rideshare-led in-migration is likely to cause financial stress for low-income renters in urban areas, while homeowners benefit from house-price appreciation. Overall, we find a significant increase in the likelihood of individuals moving out of urban zip-codes in treated MSAs, with a gradual increase averaging 11 percent of the number of people moving out of urban zip-codes in treated MSAs.  Furthermore, we find that the subsample of non-homeowners drives these effects. Insofar as home ownership and income are correlated, these results support the hypothesis that rideshare entry is associated with urban gentrification and the displacement of lower-income borrowers. We find a gradual increase in the likelihood of late repayments of incumbent residents in treated urban zip-codes averaging 11 percent. However, this increase is driven by non-homeowners, which more than offsets the decline in late repayments of homeowners. More concretely, non-homeowners increased their late repayments by 42 percent, likely due to increased housing costs, while homeowners had a decline in their late repayments of around 36 percent, likely due to the increase in their property value and net worth. 
    
    Overall, ours is the first study emphasizing the impact that the provision of high-end personal transportation by ride-hailing services has on the spatial reorganization of households and on residential mortgage and real estate markets. Specifically, ride-hailing services increased the relative desirability of poorer zip-codes in treated urban MSAs, thereby encouraging higher-income individuals to move into such areas. An important implication is that ride-hailing services may facilitate the gentrification of low-income neighborhoods by increasing access to convenient transportation.  These findings are consistent with the theory and empirical evidence presented in \citet{guerrieri2013endogenous},  showing that in the presence of neighborhood externalities, housing demand shocks induce migration of higher-income individuals into poorer neighborhoods with lower housing values that are adjacent to wealthier neighborhoods. The resultant increase in demand for housing in these areas causes home prices to rise and displacement of low-income renters, which may exacerbate inequality.


	Our research contributes to multiple strands of the literature. First, we contribute to the research by examining the causes and consequences of urbanization and the spatial sorting of households. Prior work identifies changes in public transportation technology, natural amenities, and income distribution as significant determinants of spatial sorting \citep{glaeser2008poor, couture2019income, lee2018natural,  severen2023commuting, su2022rising}. This research has predominantly examined the impact of subsidized public transportation and commute time on the residential location of individuals with low and high incomes. Specifically, \citet{baum2007did} find that highway infrastructure in the U.S.  contributed significantly to suburbanization, leading to a marked decline in higher-income population in core urban areas. Empirical evidence on the effects of urban public transit infrastructure is mixed.  \citet{glaeser2008poor} document that better access to public transportation in urban areas makes them more attractive to low-income individuals and explains the urbanization of poverty. In contrast, \citet{lee2023urban} find that the downtown transit line in Singapore resulted in spatial relocation of low-income jobs to less attractive locations and exacerbated inequality while \citet{severen2023commuting} document inelastic labor supply and housing supply response to LA Metro Rail construction and concludes that it had negligible impact on individuals' willingness to relocate spatially. While  \citet{su2022rising} concludes that the rising value of time is a significant driver of urban gentrification, particularly among high-skilled workers. In contrast, ours is the first study to document the proliferation of private for-profit transportation services brought about by ridesharing technology increased urban gentrification. 
 
    Second, our work speaks to the literature on the welfare and distributional consequences of urban spatial and gentrification. This literature collectively documents that rising income inequality, especially among the wealthy, improved amenities, construction of new apartments, improved urban transit infrastructure, and the rise in the knowledge economy have driven the gentrification of urban cores, leading to higher housing prices and the displacement of poorer residents. This has exacerbated welfare inequality, with the rich benefiting from improved amenities and reduced commute times while the poor face higher costs and consequent displacement \citep{appel2016pockets, guerrieri2013endogenous, berkes2023, baum2019long, boustan2019does, couture2019income, chyn2021neighborhoods, asquith2021local, lee2023urban}. Our study expands this literature and documents that rideshare entry led to an in-migration of higher-income households in urban areas, leading to a sharp increase in housing prices with asymmetric welfare consequences for homeowners and non-homeowners. Homeowners experience a significant improvement in their financial health proxied by lower credit delinquency. In contrast, non-homeowners experience a sharp increase in out-migration and credit delinquency. Further, rideshare entry exacerbated economic segregation by both displacing low-income residents and curbing the inflow of low-income individuals in urban MSAs.

    Third, our work contributes to extant work on the factors influencing mortgage markets and house price dynamics \citep{cellini2010value, hurst2016regional, badarinza2018home, favilukis2021out, duca2021drives, korevaar2023reaching}. Across the papers, a consistent theme emerges: public infrastructure and investment in local amenities, such as transit systems and school facilities, significantly impact housing prices. In terms of economic significance, the Second Avenue Subway extension increased property prices by 8\% (\citet{gupta2022take}), while school facility investments are associated with a 6\% rise in local housing prices (\citet{cellini2010value}). Further, \citet{favilukis2021out} show that an inflow of out-of-town buyers further inflates housing prices, exacerbating affordability issues for local residents. House prices increase by 6.3 percent in the short run and 4.8 percent in the long run, while rents go up by 10.9 percent in the short run and by 8.5 percent in the long run. More directly related to our work, \citet{gorback2020your} documents that \textit{UberX} entry is associated with increased restaurant establishments, house prices, and rental rates in previously inaccessible areas.  In contrast, this paper aims to investigate the impact of the prospective convenience and affordability of transportation brought about by ride-sharing companies on urban gentrification and, consequently, residential real estate and mortgage markets. We find that rideshare entry leads to an average 9 percent increase in house prices and a 6 percent increase in rental prices in urban areas. Consistent with house price dynamics, mortgage originations increased by 8 percent. Importantly, these effects on house prices and mortgage origination are especially concentrated among treated urban regions with lower housing values, indicating an increase in gentrification in ex-ante poorer neighborhoods.

    Finally, we contribute to the novel theoretical literature forecasting the consequences of incoming autonomous vehicles on urban/non-urban spatial population organization, housing markets, and welfare \citep{zakharenko2016,larson2016, gelauff2019spatial}. There are two distinct and opposing effects that are contingent upon the private ownership of automated vehicles and rideshare-operated autonomous vehicles. On the one hand, the automation of private vehicles will enhance the comfort of long-distance automobile travel, enabling drivers to participate in other activities during the commute. This decrease in the perceived cost of travel time may result in suburbanization as lengthier commutes become more acceptable. Consequently, there could be a population shift from urban centers to suburban and non-urban areas. On the other hand, shared-ride automated vehicles (SAVs) are likely to make public transit more convenient and efficient, especially in densely populated urban areas. This could increase the attractiveness of urban centers, leading to population clustering and gentrification. Our findings speak to the public transit role of autonomous vehicles and indicate that SAVs are likely to exacerbate the trend of urban gentrification.  This can worsen housing affordability and result in greater economic segregation and inequality between the rich and the poor. Our study has important implications for policymakers: the overall welfare benefits and distributional consequences of vehicle automation are likely to vary significantly depending on how these technologies are implemented and the resulting shifts in population distribution.

    \section{ Data and Summary Statistics }
	
    In this paper, we use data from a series of sources described below. Furthermore, given that rideshare services started operating in 2010, we begin our analysis in 2009 to reduce potential outliers from the global financial crisis. Similarly, given the large negative effects on demand for rideshare services due to the COVID-19 pandemic, we end our analysis in 2019.
 
    Data on the urbanness of zip-codes is obtained from the "Urbanization Perceptions Small Area Index" (UPSAI) compiled in 2017 by the U.S. Department of Housing and Urban Development. The UPSAI is a metric used to quantify and assess the perceived level of urbanization in small geographic areas. This index incorporates various indicators (e.g.,  density, land use, infrastructure) to gauge how \textit{urban}, \textit{suburban}, or \textit{rural} an area is perceived to be. For comparability purposes, and given our focus in this paper on the use of rideshare services, we restrict our sample to urban and suburban zip-codes within MSAs. In this context, suburban areas have lower population density, are primarily residential, have limited public services—such as public transportation and healthcare facilities—have higher car dependency, and have more green space and recreational facilities. Data is compiled at the census tract level, and we aggregate it at a zip-code level, weighing tracts by population. The top panel of \Cref{tab:summary_stats} reports the summary statistics of the urban perception index. \footnote{As we discuss in more detail in subsections \ref{subsec:rideshare_supply} and \ref{subsec:rideshare_demand}, both the supply and demand for rideshare services are relatively higher in urban areas relative to suburban ones.} We define a zip-code as urban (suburban) if it is in the top (bottom) quartile of the index of perceived urbanicity.\footnote{\Cref{tab:urban_suburban} in the Appendix contrasts the main characteristics of urban and suburban zip-codes. On average, urban zip-codes have a population density nine times higher, its households have a 17 p.p. lower share of car ownership, a 20 percent lower commute time, 45 percent higher parking costs, significant higher amenities and transportation scores, and account for 63.6 percent of the population across MSAs.} 
 
    Information on the entry of rideshare companies in MSAs was obtained directly from both \textit{Uber} and \textit{Lyft}. Overall, we have the entry of rideshare apps (\textit{Uber} and \textit{Lyft}) for 312 MSAs from 2010 to 2019. In \Cref{fig:appentry}, we display the geographical and temporal distribution of the entry of rideshare apps across MSAs. As we discuss in more detail in subsection \ref{subsec:rideshare entry} the entry was staggered across time with rideshare-service providers sorting their entry by population size.

    Given the unavailability of detailed data on the number and revenue of rideshare services across time and MSAs, we use more aggregated data from the Census Bureau's Non-employer Statistics (NES). This is an annual series providing subnational economic data by industry.  NES presents data for businesses that have no paid employees (e.g., sole proprietorships, partnerships, and other forms of small businesses) and are subject to federal income tax. Data is available at the year and MSA level and covers NAICS sectors at the 4-digit level. For our purposes, we focus on sector 4853 \textit{Taxi and Limousine Service}, which includes both rideshare service drivers as well as taxi and limousine drivers. We use the yearly data from 2009 to 2019. Summary statistics on the number and revenue of drivers are reported in the \textit{Rideshare Entry and Drivers} panel of \Cref{tab:summary_stats}. As we discuss in subsection \ref{subsec:rideshare_supply}, the entry of rideshare services into an MSA led to a significant increase in the number and revenue of businesses in NAICS sector 4853 (which includes drivers rideshare services) operating in the MSA.
	
    Data on the number of Uber trips within an MSA are compiled from regulatory data collected from five different regions, namely Massachusetts, Chicago, New York, Los Angeles, and San Francisco. For a more exhaustive coverage, we also include data on internet searches obtained from Google Trends on the terms \textit{Uber}, \textit{Lyft}, or \textit{Rideshare app} in a zip-code in the month prior to the entrance of the rideshare services in the MSA. Summary statistics on both of these measures are displayed in the third panel of \Cref{tab:summary_stats}. As explained in subsection \ref{subsec:rideshare_demand}, within MSAs, rideshare services are disproportionately used in urban areas compared to suburban ones.  

    Data on housing prices is obtained from \textit{CoreLogic}. The \textit{CoreLogic} home price data consists of yearly housing price indices at the zip-code level using a weighted repeat-sales methodology. We use data from 2009 to 2019. Data on rental prices is obtained from \textit{Zillow}. It is a smoothed yearly measure of the typical observed market rate rent in a zip-code. It is a repeat-rent index that is weighted to the rental housing stock to ensure representativeness across the entire zip-code, not just those homes currently listed for-rent. The index is dollar-denominated by computing the mean of listed rents that fall into the 40th to 60th percentile range for all homes and apartments in a given region. We use data from 2009 to 2019. 
    
    Finally, data on vacancies is obtained from the Housing and Urban Development Authority (HUD).  Using information from the United States Postal Service, HUD compiles data on addresses identified as having been "Vacant" in a given census tract. Vacant addresses include addresses that delivery staff have identified as being vacant for 90 days or longer. For this project, we will use data from 2012 (the earliest available data) to 2019. Data is compiled at the Census Tract Level, and we aggregated at a zip-code level, weighing tracts by population.  The summary statistics of these housing statistics are displayed in the \textit{Housing Statistics} panel of \Cref{tab:summary_stats}.
	
    Information on the number of households per income group and zip-code is obtained from IRS's Statistics of Income. This is a yearly dataset collected by the IRS at the zipcode*income-group level based on individual income tax returns filed from 2009 to 2019. The data includes items such as the number of returns (which approximates the number of households), adjusted gross income and salaries, as well as mortgage interest deductions and child tax credits. The summary statistics of the main variables of interest from this dataset are displayed in the \textit{IRS Statistics} panel of \Cref{tab:summary_stats}.

    Mortgage information is obtained from the confidential Home Mortgage Disclosure Act (HMDA) dataset, containing regulatory information reported by depository institutions and certain for-profit, non-depository institutions from 2009 to 2019. The data is compiled yearly by the Federal Financial Institutions Examination Council and includes records on loan originations and loan applications that did not result in originations. Mortgage information includes property location, borrower demographics (e.g., applicant income, information on co-applicants, race), whether the loan is government guaranteed (e.g., FHA), and loan amount. Further mortgage information on home value and on the credit score of the applicant is obtained from McDash. Summary statistics of the main variables of interest are displayed in the \textit{Mortgage Statistics} panel of \Cref{tab:summary_stats}.
 
    Information on consumer credit usage and performance is obtained from the FRB-New York’s Equifax Consumer Credit Panel. This database contains credit‐reporting data for a nationally representative 5 percent sample of all adults with a social security number and a credit report. The dataset is a quarterly panel, with snapshots of consumers' credit profiles captured at the end of each quarter. It includes variables on certain demographics of individuals (e.g., age, current zip-code of their address, and FICO scores) along with debt information on housing, credit cards, and auto, as well as on loan performance. We use its fourth-quarter data from 2009 to 2019. The summary statistics of the main variables of interest are displayed in the \textit{Equifax Statistics} panel of {tab:table1A}.
	
    Zip-code information on the level of poverty and racial composition Information is obtained from the Census Bureau's Current Population Survey (CPS).  The CPS includes comprehensive yearly statistics from 2009 to 2019 on the percentage of individuals living below the poverty line and the demographic breakdown of minority populations within each zip-code. Summary statistics of the main variables of interest are displayed in the \textit{Other Statistics} panel of {tab:table1A}.

    \textit{Walkscore} and \textit{Transitscore} datasets were obtained from \textit{Redfin}. \textit{Walkscore} provides information on the walkability of locations. It measures how easily residents can access amenities like restaurants, shops, parks, and schools within walking distance. It is a numerical score from 0 to 100, where higher scores indicate better walkability. \textit{Transitscore} measures how well a location is served by public transit on a scale from 0 (``Minimal Transit'') to 100 (``Rider’s Paradise''). It calculates a score for a specific point by summing the relative "usefulness" of nearby routes, depending on the distance to the nearest stop on the route, the frequency of the route, and the type of route. We use the available waves collected in 2015 and 2017.
	
    \section{Empirical Strategy} \label{sec:empirical_strategy}
	
	We use a staggered difference-in-differences approach to estimate the effects of the entry of rideshare services in an MSA. Relying on the estimator proposed by \citet{borusyak2021revisiting}, we compare observations in MSAs with rideshare services with not-yet-treated observations. This method circumvents the recently documented limitations of two-way fixed effect (TWFE) models when treatment effects are heterogeneous and time to treatment varies \citep{borusyak2021revisiting, callaway2021difference, de2020two, goodman2021difference, jakiela2021simple, sun2021estimating}. When units are treated at different points in time, the canonical TWFE regressions make \textit{clean} comparisons between treated and not-yet treated units as well as \textit{forbidden} comparisons that can receive negative weights, which are differences-in-differences that use early-treated units as a control group. In the presence of dynamic treatment effects, such comparisons can lead to biases and erroneous conclusions about the impact of an intervention \citep{baker2022much, goodman2021difference, rambachan2023more}.
 
	In our context, diagnostics indicate that such limitations are likely to be present \citep{ jakiela2021simple} with the effects of the rideshare entry varying over time, as the supply of drivers gradually increases following the entry of rideshare services as operations become more established (\Cref{fig:supply_cabdrivers}). 
	
	To overcome these limitations, the main building block of the staggered difference-in-difference approach involves calculating the average effect of the entry of rideshare services \textit{t} for the cohort of zip-codes in the MSAs starting the program in each period \textit{g}, ATT(g,t). Identification is obtained by comparing the expected change in outcomes between periods \textit{g-1} and \textit{t} of two groups: MSAs in cohort \textit{g} and a control group of not-yet-treated MSAs. The estimated ATT(g,t) cohort-time parameters avoid ``forbidden comparisons'' and can be aggregated using specified weights to obtain the overall average effect of the program, as well as its impact over time. We use the \textit{did\_imputation} command in Stata, described in \citet{borusyak2021revisiting}. Here, an MSA remains in the control group until the entry of rideshare services into the MSA. 

    Formally, our baseline difference-in-differences (DID) specification examines whether MSAs associated with the entry of rideshare services  experience differential changes in our variables of interest using the following specification:

	\begin{equation}
		Y_{z,m,y}=\beta_0+\sum_{-1}^{-2} \theta_k \times Pre[k]_{z,m,y} + \sum_{0}^{4} \theta_k \times Post[k]_{z,m,y} + \delta_{z}+\gamma_{s,y}
	\end{equation} \label{eq:dynamic}

	$Y_{z,m,t}$ is the dependent variable in a zip-code \textit{z} located in MSA \textit{m} in year \textit{y}. $Post[k] 
    (Pre[k])$ is an indicator that  year \textit{y} is \textit{k} years from (to) rideshare entry into the MSA \textit{m}. The coefficient $\theta_{0}$ measures the immediate DID effect of rideshare's entry. The \emph{marginal} coefficients $\theta_{1}$ ($\theta_{2}$) measure the \emph{additional} marginal responses one year (two years) after rideshare entry. Similarly, coefficients $\theta_{-1}$, and $\theta_{-2}$ capture the difference in the dependent variables between the treatment group and the control group in each of the two pre-treatment years. Finally, $\delta_{z}$ are zip-code fixed effects to control for time-invariant characteristics of the zip-code, while $\gamma_{s,y}$ are state*year fixed effects that control for average movements at the state*year level. 
    
    Given that rideshare services are overwhelmingly used in urban areas (as we discuss in subsection \ref{subsec:rideshare_demand}) our benchmark analysis compares the impact of the entry of rideshare services in urban and suburban zip-codes—akin to our treatment and control zip-codes. In particular, \textit{urban} and \textit{suburban} zip-codes are defined as being in the top and bottom quartiles, respectively, of our measure of perceived urbanicity. Furthermore, and as a robustness check for all outcomes of interest, we also include a specification where we pool all zip-codes and include MSA*year fixed effects $\gamma_{m,y}$ to control for average changes in the outcome variable at the MSA*year level. Therefore, this specification tests for variations \textit{within} the MSA across urban and suburban zip-codes. For this specification, we assume that rideshare services only operate in urban zip-codes within the MSA. In other words, suburban zip-codes are the untreated control group.

    As with the canonical difference-in-differences methodology, there are two key identifying assumptions for MSAs in the control group (i.e., not-yet treated MSAs) to provide a valid counterfactual for the evolution of outcomes that treated MSAs would have followed in the absence of the entry of rideshare services. The first assumption requires that the average outcomes of MSAs with rideshare services entering at different points in time follow parallel trends absent rideshare entry. The second identifying assumption is that of no anticipation, which implies that rideshare services have no causal effect prior to their entry. In the next two subsections, we provide evidence that suggests that these assumptions hold in our setting.
	
    \subsection{Entry Decision of Rideshare Services}\label{subsec:rideshare entry}
	
\begin{figure}[!ht]
    \caption{\textbf{Entry of Rideshare Services Across MSAs.} \label{fig:appentry}} 
    \footnotesize{{ Panel (a) displays the entry of rideshare services across time by Core Based Statistical Area (CBSA). Each CBSA is colored according to the year a rideshare service (either \textit{Uber} or \textit{Lyft}) first appeared. CBSAs include both metropolitan statistical areas (MSAs) and micropolitan statistical areas. Panel (b) displays the population size of an MSA along with the month in which rideshare services entered an MSA.} 
    \vspace{2pt} 
    \hrule height 1pt 
    \vspace{1.5pt} 
    \hrule height 0.5pt 
    \begin{center}
	\subfloat[Entry of rideshare services—Uber and Lyft.]{
        \includegraphics[width=0.49\textwidth]{figures/fig\_1a.pdf}}  
        \hfill
     \subfloat[Entry month of rideshare services and population of MSA.]{
        \includegraphics[width=0.49\textwidth]{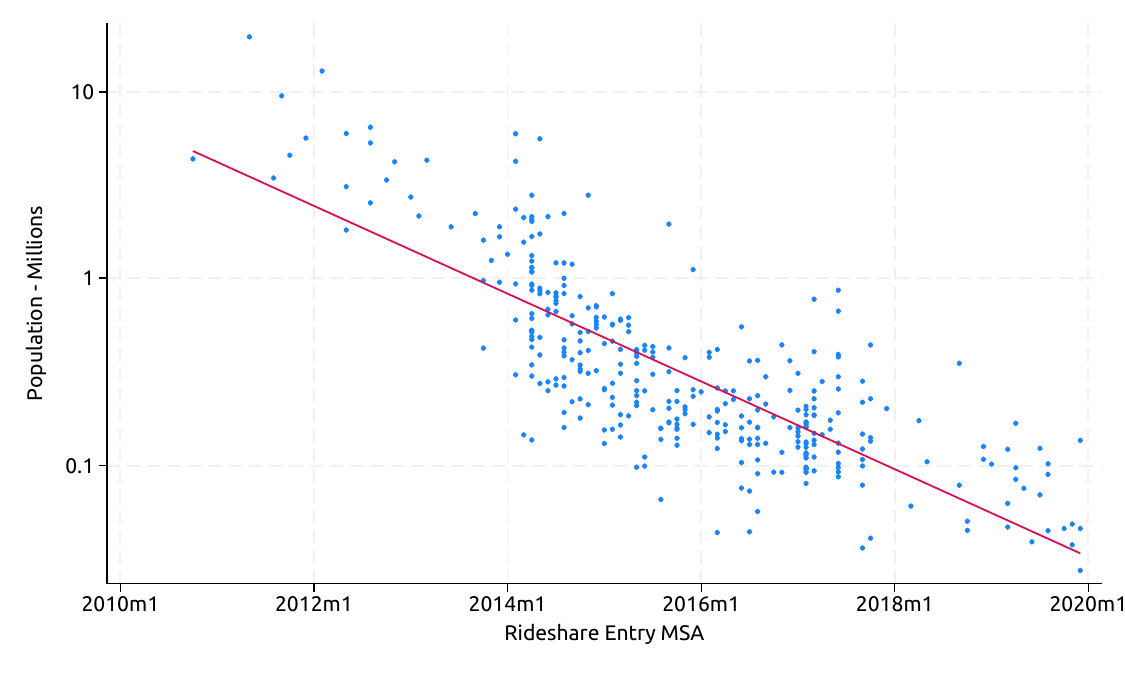}}  
        \end{center}
    \rule{\textwidth}{0.4pt} %
}
\end{figure}

	The left panel of \Cref{fig:appentry} displays the frequency and geography of the entry of rideshare services—\textit{Uber} and \textit{Lyft}—across U.S. MSAs. It shows that there is a large variation in the timing and geography of entry of rideshare services across MSAs. 
 
    Given that our estimation exploits this staggered entry, a concern could be that the entry decision may be correlated with time-varying factors related with our variables of interest (e.g., in housing or rental markets). Previous work, relying on the same identification strategy, has shown that the entry decision of rideshare services was primarily based on the population size of the MSA and was orthogonal to other dynamics like economic or population growth \citep{hall2018uber,barrios2022launching,fos2019gig,buchak2019financing}. This relation between population size and entry is graphically displayed in the right panel of \Cref{fig:appentry}.\footnote{\citet{anil2017regulatory} also find that \textit{Uber} likelier to enter MSAs with higher regulatory barriers for entry of conventional cab services. Insofar as ex-ante regulations governing the entry of taxi operators are orthogonal to local economic conditions, this finding adds another layer of exogenous variation in the entry decision of rideshare services.} The timing of entry of rideshare services in an MSA is strongly negatively correlated with the MSAs population size, with rideshare services entering areas with larger populations earlier.	For robustness, we use a hazard model to verify the relationship between rideshare entry, the population of an MSA, as well as other economic and demographic variables of interest. Results of this test are displayed in \Cref{tab:entry} of the Appendix. The dependent variable $Entry_{m,y}$ is a yearly indicator of whether a rideshare service entered MSA \textit{m} in year \textit{y}. Observations of an MSA are dropped for all years after entry. We find that the entry of rideshare services is mainly determined by the population in the MSA and, to a lesser extent, by whether rideshare services were already operating in the state. We find no evidence that the timing of rideshare services was systematically driven by the economic conditions of the MSA, such as average income or unemployment. This finding suggests that not-yet-treated observations should be a valid counterfactual for treated observations. 
	
    \subsection{Rideshare Entry and Supply of Rideshare Services Across MSAs} \label{subsec:rideshare_supply}
	
	Second, we examine the effect of rideshare services' entry into an MSA on the overall supply and revenue of cab drivers. The idea is to use the total cab-drivers' supply in an MSA as a proxy for the supply of ride-hailing services.\footnote{As mentioned in the data section, we use somewhat more aggregated data from the Census Bureau at the year and MSA level on the 4-digit NAICS sector \textit{Taxi and Limousine Service} including both rideshare service drivers as well as taxi and limousine drivers.} The evolution of \Cref{fig:supply_cabdrivers} plots the entire paths of the average share of drivers and revenues within the MSA in the years surrounding the entry of rideshare services. Lines around point estimates depict the corresponding 95 percent confidence intervals. We start by noting that the number and revenues from cab drivers only increase after the entry of rideshare services (\Cref{fig:supply_cabdrivers} and \Cref{tab:driversupply}), suggesting that the entry of these services in an MSA had an important effect on the provision of transportation services. However, following the entry of rideshare services in the MSA, there was a substantial increase in the volume and share of the number and revenue of the sector “Taxi and Limousine Service"  of single-worker businesses. From both panels, we can see that prior to the entry of rideshare services, the sector represented roughly 0.4 percent of the number and 0.3 percent of the revenue, respectively, of businesses with no paid employees in an MSA. However, five years after the entry of rideshare services, the share of businesses and revenue of the "Taxi and Limousine Service" sector more than tripled to 3.1 and 1.1 percent, respectively. Similarly, results in the \Cref{tab:driversupply} reports the point estimates for impulse responses using specification in \ref{eq:dynamic}, with similar qualitative results. These results suggest that the entry of rideshare services had a very large effect on the use of taxi services within an MSA. Furthermore, they do not display any discernible trend prior to the rideshare entry into an MSA.  
	
    \subsection{Use of Rideshare Services Within MSAs}\label{subsec:rideshare_demand}
	
    Previous research has shown that the use of rideshare services is especially higher in urban areas \citep{gorback2020your, li2022demand}. There are several reasons for this. The shorter distances in urban areas lead to lower per-trip prices, resulting in higher demand for rides \citep{cramer2016disruptive}. This higher demand, combined with increased density, raises passenger turnover, reducing operational costs (e.g., matching drivers with passengers)\citep{cohen2016using}. Rideshare services also help avoid costs associated with car ownership (e.g., parking), which are larger in urban areas \citep{glaeser2008poor}.\footnote{In Figure \Cref{fig: caronwershipcosts} in the Internet Appendix, we display the positive relation between our measure of perceived urbanicity of zip codes in New York City and its monthly parking costs.} Finally, rideshare services can provide last-mile connectivity for public transit, which predominantly serves urban areas \citep{hall2018uber}. For all these reasons, we expect the impact of the entry of rideshare services to be relatively stronger in urban areas. 
 
    Next, we verify the positive relation between the use of rideshare services and the urbanness of a zip-code. The main outcome of interest is the number of rideshare trips per capita. However, for regulatory reasons, this measure is only available for zip-codes in Massachusetts, Chicago, San Francisco, and New York City. The explanatory variable is a \textit{Urban$_{z}$} is a continuous variable of the perception of whether zip-code \textit{z} is urban, using data compiled by the \textit{Census' Urbanization Perceptions Small Area Index}. To control for additional drivers of demand, we control for characteristics at the zip-code level (e.g., income). Furthermore, to control for local characteristics (e.g., urban sprawl), we also include MSA fixed effects. The results of this exercise are displayed in \Cref{tab:ridesharedemand}. Focusing on column 1, we find that one standard deviation in the measure of urban perception is associated with a 30 percent increase in the number of trips per capita. In column 2, we include a series of controls at the zip-code level (e.g., average household income, home value, walk score, and transit score), and in column 3, we include MSA fixed effects. In both specifications, results remain qualitatively similar, showing a robust correlation between zip-code urbanization and rideshare usage. Finally, in column 4, we divide the sample of zip-codes into four quartiles of \textit{Urban$_{z}$}. We then compare rideshare usage for MSAs in the top three quartiles with those in the bottom quartile. Rideshare usage is monotonically increasing in urban perception, with the number of trips per capita in the top urban quartile more than double that of the bottom quartile.
	
	Since data on trips per capita is only available for a small subset of MSAs, for robustness, in columns 5-8 of \Cref{tab:ridesharedemand}, we use Google queries for the terms \textit{Uber}, \textit{Lyft}, or \textit{Rideshare app} in a zip-code as a proxy for rideshare demand.  Specifically, \textit{GoogleSearches$_{z}$} indicates the relative number of \textit{Google} queries for the terms \textit{Uber}, \textit{Lyft}, or \textit{Rideshare app} in zip-code \textit{z} in the month following the entry of rideshare services in the MSA. Consistent with the results reported in columns 1-4, Rideshare-related \textit{Google} searches are significantly more prevalent in urban zip-codes.   
	
    Overall, results in this subsection suggest that ridesharing usage is indeed greater in urban areas.

    \section{Main results}

    Given that rideshare services increase the relative attractiveness of urban areas \citep{cramer2016disruptive}, in this section, we display the results of the relative impact of the entry of rideshare services on urban and suburban zip codes.  

    \subsection{Entry of Rideshare Services and Gentrification}\label{subsec:gentrification} 
	
    Given that the adoption of rideshare services has been mainly driven by high-income and young individuals \citep{hall2018uber}, in this section, we evaluate the impact its entry had on any in-migration of individuals in these groups. 

    In particular, we start by examining the effects on the number of high- and low-income households in urban and suburban zip-codes of treated MSAs. We use the IRS data on \textit{Statistics of Income} to analyze the evolution of the number of households in a given zip-code \textit{z} and year \textit{y}, classifying them as having high (low) income if they have an annual income above (below) 50,000 dollars (high income).\footnote{High-income households represent, on average, 41 percent of total households in our study}. Our benchmark results, splitting the sample into \textit{urban} and \textit{suburban} zip-codes, are displayed in columns 1-4 of Table \Cref{tab: gentrification}. The results on the number of high-income households in urban and suburban zip-codes are displayed in columns 1 and 2. Consistent with gentrification, we observe a gradual increase averaging 3.5 percent in the number of high-income households in $urban$ zip-codes in the years after the entry of rideshare services. We do not observe a similar pattern in our control group of suburban zip-codes. In columns 3-4, we display the evolution of low-income households in both urban and suburban locations. In column 3, we find that there is a slight decline in the number of low-income households of around 0.8 percent in the years following the entry of rideshare services. Conversely, results in column 4 indicate that the number of low-income households increased in suburban zip codes roughly 2 percent following the entry of rideshare services. Reassuringly, we do not find any differential pre-trends for the number of both high- and low-income households. Overall, these results support the hypothesis that the entry of rideshare services is associated with an \textit{in-migration of high-income individuals} into urban zip-codes and an out-migration of lower-income individuals to suburban areas. This out-migration of lower-income individuals is likely partly driven by the increase in housing prices that we later document in subsection \ref{subsubsec: house prices}.

    To shed further light on changes in the demographic composition of treated MSAs, we now examine the average age of individuals living in a zip-code as well as the fraction of households claiming a child tax credit, which we use as a proxy for being a household with children. \footnote{Information on the average age of individuals in a zip-code comes from the \textit{Equifax Consumer Credit Panel}, while information on children's deductions is obtained from the \textit{IRS's Statistics of Income}}. Starting with the average age, column 5 shows a decrease in the average age of individuals of around 0.8 percent in treated urban zip codes, while column 6 indicates no impact on suburban areas. These findings are consistent with previous literature documenting a large increase in the number of younger high-income individuals in urban areas over the last two decades \citep{couture2023neighborhood}. Finally, column 7 indicates that rideshare entry is associated with a decline in the fraction of households claiming child tax credit of 0.8 p.p. (around 6 percent), while results in column 8, suggest that this effect is not present in suburban areas. Therefore, urban gentrification—which tends to increase the price of services \citep{gorback2020your}—appears to be reducing the affordability of urban areas for families with children. 

    As we mentioned in the empirical section \ref{sec:empirical_strategy}, to allay concerns that our findings may be driven by variations in local economic conditions, we test all our outcomes of interest using a specification with MSA*year fixed effects, pooling all observations within an MSA using urban zip-codes as the treatment group. Results of this specification are displayed in \Cref{tab:table_demography_pooled}. All results are quantitatively similar to the previous specification comparing urban and suburban zip-codes separately. In particular, the number of high-income (low-income) households increases by 4.7 percent (decrease of 4.7 percent) in treated urban zip codes relative to suburban zip-codes. The average age of individuals declines by 0.9 percent in urban zip-codes, whereas the fraction of households reporting a child tax credit declines by 0.3 p.p. (1.9 percent).
	
    Overall, results in this section suggest that the entry of rideshare services has contributed to urban gentrification.	
	
    \subsection{Entry of Rideshare Services and the Housing Market}\label{subsec:housing_market}
	
    As results in the previous section suggest that the introduction of rideshare services is leading to the gentrification of urban areas, we now examine its impact on housing markets. More concretely, we analyze the impact on the demand for housing, on the demographics of new homeowners, on characteristics of the neighborhoods new homeowners are moving into, as well as on the prices of homes and rents.	
		
    \subsubsection{Demand for Housing}\label{subsubsec:mortgages}
	
    Given that a significant proportion of residential property acquisitions are financed through mortgages \citep{garriga_2020}, our analysis in this section focuses mainly on them. In particular, we analyze the number of mortgages (and of vacant homes), the characteristics of mortgage applicants (e.g., applicant income) as well as on they characteristics of the urban neighborhoods in which homes are located.
    
    \textbf{Mortgage Originations -}  We begin by examining the evolution of mortgage originations, collapsing the number of mortgage originations at the zip-code*year level. Results of this exercise are displayed in \Cref{tab: mortgageoriginations} and \Cref{fig:mortgageorigination}. We find an economically and statistically significant relative increase in the number of mortgages originated in urban zip-codes of treated MSAs after rideshare entry (column 1). Mortgage originations increase gradually over time, averaging around 8.1 percent in the five years following the entry of rideshare services. Conversely, the effect on suburban zip-codes is muted (column 2). These results are consistent with the relative increase in demand for housing in urban areas following the entry of rideshare services. Importantly, these results are not driven by pre-existing differential trends.\footnote{A concern regarding the documented effect on number of originations is that some unobservable time-varying characteristics or policy changes at the MSA-level correlated with rideshare entry may have led to an increase in the approval rate for new mortgage applications. We conduct a similar test to analyze whether there was any change in approval rates after the rideshare entry and do not find any change.}

    \textbf{Vacant Homes - }As a robustness check on the increase in demand for housing in urban areas, we also examine whether urban zip-codes in treated MSAs also experienced a decline in vacant homes. Results of this test are displayed in \Cref{tab:vacancies2} of the Appendix. Consistent with an increase in housing demand, in column 1, we observe a drop in vacancy rates in urban areas by 0.5 p.p. (11 percent) in the five years after treatment. Reassuringly, the impact estimates of vacancy rates prior to rideshare entry are close to zero and do not display a discernible trend. This effect is muted for suburban areas (column 2). In columns 3-4, we use the the number of vacant residences (in logs) in a given zip-code in a year as the dependent variable and obtain qualitatively similar results.\footnote{ \Cref{fig:vacancies} plots the entire path of the coefficient of difference-in-differences treatment effect estimates for vacancy rates.}  Finally, in \Cref{tab:table_vacancies_pooled} we display the results of the specification pooling urban and suburban zip-codes using MSA*year fixed effects. Results are quantitatively similar to the specifications, which analyze urban and suburban vacancies separately. All in all, these results suggest that the entry of rideshare services in an MSA led to a sharp reduction in vacant homes in urban areas.
 
    \textbf{Mortgage Applicants - }Insofar as gentrification is associated with the migration of high-income individuals into urban zip-codes \citep{baum2019long}, we expect new mortgage applicants in these areas to have higher incomes. Therefore, we test the variation of income of households across urban and suburban zip-codes. Results on the income of households with new mortgages are displayed in columns 3 and 4 of \Cref{tab: zip code characteristics} and on \Cref{fig:incomeapplicants}. We find an increase in the average income of households who took out new mortgages in treated urban zip-codes relative to the control group of MSAs. Consistent with the evidence in \Cref{tab: gentrification}, this effect gradually increases with time. In column 3 we see that new mortgages in urban zip-codes are taken out by households with 6.5 percent higher income in treated MSAs, with no distinct pre-trends. In column 4, we see a relatively modest and short-lived increase in the income of households contracting new mortgages in treated suburban zip-codes compared to the control group of suburban MSA. 
    
    A usual consequence of gentrification is the reduction of opportunities for lower-income groups to live in the area \citep{couture2023neighborhood}. Therefore, we also test for the impact that rideshare services have on the number of subsidized mortgages. More concretely, we examine whether the frequency of FHA-insured mortgage origination, a proxy for poor households, has changed.\footnote{FHA insurance is a government-run program that provides insurance to lenders on behalf of borrowers, thereby reducing the risk associated with lending to higher-risk individuals. It is designed to help low-income individuals with limited credit access obtain financing for the purchase of a home.} Results of this test are displayed in columns 5 and 6 of \Cref{tab: zip code characteristics}. Consistent with the increase in the average income of households with mortgages, we see a very sharp reduction in the number of FHA-insured loans in urban zip-codes. In the five years following the entry of rideshare services, the share of FHA-insured loans declined by an average of 6.4 percentage points (around 30 percent) in urban zip-codes. Conversely, we do not see any effect on suburban areas of treated MSAs. 

    As before, as a robustness test, we pool urban and suburban zip-codes using MSA*year fixed effects using urban zip codes as a control group. Results of this specification are displayed in \Cref{tab:table_mortgage_origination_pooled}. All results are qualitatively similar to the specifications, comparing urban and suburban zip codes separately. The number of mortgage originations increases 8.6 percent in urban zip-codes compared to suburban ones, the average income of mortgage applicant rises 2.2 percent, while the fraction of FHA-insured loans declines 1.9 p.p. (10.2 percent).    
	
    \textbf{Zip-Codes of Mortgage Originations -} In this section, we analyze the characteristics of the neighborhoods of the originated mortgages. In particular, we test whether these new mortgages cluster in more affordable areas, proxied by median housing prices, within urban and suburban zip-codes in line with the mechanism proposed in \citet{guerrieri2013endogenous}. Furthermore, and in the context of gentrification, we also check whether new mortgages are concentrated in areas with higher poverty rates and a larger share of minorities. In this specification, we still split the sample into urban and suburban zip-coded, but we now use MSA fixed effects to absorb time-invariant characteristics of urban (suburban) zip-codes in an MSA, as well as state*year fixed effects. Therefore, all our results compare urban (suburban) zip-codes within an MSA.
    
    Results relating mortgage originations with the median housing value in the zip-code are displayed in \Cref{tab: zip code characteristics} and \Cref{fig:homevalue_zip-codes}. In columns 1 and 2, we display the results for median house prices in urban and suburban zip-codes. They show that within urban areas in an MSA, mortgages originated in zip-codes with 2.2 percent lower median house prices in the five years following the entry of rideshare services. Furthermore, this effect is gradual with demand for lower-priced zip-codes increasing through time. By the fifth after entry, mortgage originations in urban areas are for zip-codes with 5 percent lower home values. Results on the average poverty are displayed in columns 3 and 4. After the entry of rideshare services within urban areas, more mortgages tend to originate in higher poverty zip-codes. The average poverty rate of these zip-codes is 0.3 p.p. (1.5 percent) higher. Again, the effect is gradual and we find no effect on suburban zip-codes. Finally, in columns 5 and 6, we analyze the composition of minorities. In urban zip codes, new mortgages originated in zip codes with a 0.7 p.p. (1.5 percent) higher fraction of minorities. Again, effects on suburban areas are significantly lower and exhibit pre-trends. Results of the pooled specification are displayed in \Cref{tab:tableA7}. They show that the urban zip-codes of mortgage originations have 1.1 percent lower house prices, 1.8 percent higher poverty rates, and 2.7 percent higher share of minorities.
	
    In summary, our results suggest that ride-hailing services, by improving access to easy and convenient transportation, increased the relative attractiveness of poorer neighborhoods in treated MSAs. Therefore, an important consequence of the entry of ride-hailing services is that they are conducive to the gentrification of poorer urban neighborhoods. Next, we analyze whether the resulting increase in demand for housing in urban areas leads to an increase in home prices, especially in more affordable urban neighborhoods. 
	
    \subsubsection{Housing and Rental Prices}\label{subsubsec: house prices}
   
    So far, our results have been consistent with an increase in the demand for housing in urban areas. We now analyze whether this higher demand leads to higher home and rental prices given the low elasticity of housing supply in the short-run \citep{glaser2018}. Results of our benchmark specification for house prices and rental prices are displayed in \Cref{tab: house prices} and in \Cref{fig:houseprice}. Columns 1 and 2 report the results on house prices for the sub-samples of urban and suburban zip codes, respectively. Coefficients in column 1 indicate a gradual average house price averaging 9 percent in the five years after treatment, consistent with higher demand for rideshare in urban areas. Conversely, coefficients in column 2 indicate that the impact on suburban prices is muted. Results on rental prices are displayed in columns 3 and 4. In line with the results on house prices, we find an overall differential increase in rental prices in treated MSAs. Following the entry of rideshare services, rental prices in urban zip codes increase on average 6 percent. We also find a small, but statistically significant, effect on suburban zip codes averaging 1.8 percent.\footnote{This positive impact on suburban areas could be driven by the displacement of low-income people from urban areas to suburban areas. Insofar as such individuals are likely to be liquidity-constrained, they may not be able to bid up house prices, resulting in a muted effect on house prices in suburban areas.  Instead, they are likely to shift into rental housing, boosting rental demand and cost in suburban areas.} Results of the specification pooling urban and suburban zip-codes together are displayed in \Cref{tab:table_houseprices_pool}. Qualitatively, results are similar to the ones splitting urban and suburban zip-codes. We find that housing prices and rental prices rise 2.5 percent and 1.9 percent, respectively, in the treated urban zip-codes compared to suburban zip-codes.

    In summary, results in this section indicate that the entry of rideshare services contributed to the sharp rise in housing prices by increasing housing demand from higher-income individuals in a context of limited supply. 
 
    \textbf{Heterogeneity given ex-ante vacancy rate -} If the housing supply were perfectly elastic, we would not have observed a significant impact on home and rental prices. Therefore, in this section, we test whether the effect of rideshare entry on house prices is stronger in regions with lower ex-ante vacant housing supply. For this, we use USPS data on vacant housing units. Results of this test are displayed in \Cref{tab:vacancies1} in the Appendix. We divide the sample into three terciles based on ex-ante vacancy rates of zip-codes. \textit{Vacancy Rate$_{z,y}$} measures the percentage of vacant addresses of zip-code \textit{z} in year \textit{y}.  We find that the effect of rideshare entry on house prices is monotonically decreasing in ex-ante housing vacancies. Specifically, the house price impact is strongest in the lowest tercile of ex-ante vacancy rates and relatively muted for zip-codes in the highest tercile. This suggests that the impact on housing prices of rideshare services is likely to be larger in MSAs with larger constraints (e.g., physical or geographical) on housing supply.
	
    \textbf{Heterogeneity given car ownership costs -} In this section, we check whether the impact of rideshare services on housing prices is higher in treated regions with higher car ownership costs. In other words, we examine whether the liquidity unlocked by the elimination of the car ownership costs is potentially used on housing, resulting in higher house prices  \citet{sabouri2020exploring}.  To test this hypothesis, we collected parking and car insurance quotes from \textit{SpotHero} and \textit{CarInsurance} at the MSA level. We then repeat our baseline analysis, dividing the sample into MSAs with high and low car ownership costs. An MSA has high (low) car ownership costs if its average parking rates and insurance premiums are above (below) the median. \Cref{tab:carownershipcost} presents the results. Overall, results suggest that the impact on housing and rental prices is roughly double that of MSAs with lower car ownership costs.
	
    \textbf{Exit of Rideshare Services - The Case of Austin.} As a robustness check, we check whether the patterns observed following the entry of rideshare services are somewhat reversed after the (temporary) exit of this service. While there are not many incidents of rideshare services exiting an MSA, in April of 2016, both \textit{Uber} and \textit{Lyft} left Austin temporarily. Rideshare services entered the Austin metropolitan area in May of 2014. In December 2015, the Austin City Council passed an ordinance requiring drivers of rideshare companies to receive fingerprint-based background checks. After this ordinance, a \textit{Ridesharing Works} campaign was organized—with financial support from both \textit{Uber} and \textit{Lyft}—creating a citizen-initiative ballot to repeal the ordinance. The voting occurred in May of 2016, and the ordinance was not repealed (i.e., \textit{Uber} and \textit{Lyft} lost). Immediately after the defeat of this campaign, \textit{Uber} and \textit{Lyft} suspended rideshare operations in Austin. Therefore, we use the exit of rideshare services from Austin to examine if it led to a decline in housing demand in urban areas, focusing on a short window around the exit event. In particular, we focus on the impact on rental prices, as these tend to be more sensitive in the short term to fluctuations in demand relative to prices of housing \citep{anenbergdemandshocks}. More concretely, we use monthly rental-price data controlling for zip-code and state-month fixed effects. Results of this test are displayed in \Cref{fig:exit}. Consistent with our baseline results, the exit of rideshare services exit led to a 2 percent drop in rental prices in urban areas. In suburban areas, the effect was indistinguishable from zero.
 
    \subsection{Impact of Rideshare Services on Incumbent Residents}

    Taken together, the results in the previous sections suggest that the entry of rideshare services led to urban gentrification, with the in-migration of high-income individuals leading to an increase in housing prices, causing the out-migration of lower-income residents. 
    
    In this section, we use granular data from the FRB-New York’s Equifax Consumer Credit Panel to analyze changes in the real outcomes of incumbent residents, namely relocation status and their financial health. Furthermore, given that gentrification led to an increase in house prices, we split our analysis between homeowners and non-homeowners (e.g., renters). We make this comparison given that a generalized appreciation in housing prices is likely to differently affect renters and homeowners \citep{autor2014}. In our context, low-income renters should be negatively impacted by increased rental prices and are likely to be forced to move and face financial stress. For homeowners, the outcomes are less straightforward. While increased home values boost incumbents' wealth, which can be accessed either through outright sale or home-equity loans, it is also likely to result in higher property taxes, which can place a financial strain on homeowners with limited incomes.

    We define individuals as homeowners (non-homeowners) depending on whether they have (not) had a mortgage or a home equity loan in their current zip code since 1999 (the first year for which we have data). To limit the number of events in which individuals are misclassified as non-homeowners, given that they have fully paid their mortgage, we limit our analysis to individuals under 55 years.\footnote{We do this given that roughly 85 percent of home acquisitions by individuals under 55 years old are financed through mortgages (National Association of Realtors).}  The main characteristics of homeowners and non-homeowners are displayed in \Cref{tab:sumstathomeowner}. Consistent with the findings in \citet{wolff2000} we find that on average homeowners are younger, have higher credit scores, move less frequently, and live in zip-codes with higher median income and lower poverty. 
 
    \subsubsection{Relocation}
	
    We start by analyzing whether the entry of rideshare services increases the relocation of incumbent residents. For that, we use our benchmark specification and create a variable \textit{Moving Out of Zip-Code$_{i,y}$} indicating whether individual \textit{i} will be living in a different zip-code by year \textit{y+1} relative to current in year \textit{y}. Results of this exercise are exhibited in \Cref{tab: relocation}. Our first test, displayed in column 1, pools homeowners and non-homeowners in urban zip-codes. We observe a significant increase in the likelihood of individuals moving out following the entry of rideshare services. There is on average a 1.2 p.p. (11 percent) gradual increase in the number of people moving out of urban zip-codes in treated MSAs. To understand whether non-homeowners are relatively more impacted than homeowners, we then split or sample into these two groups. The results of this exercise are displayed in columns 2 and 3. We find that the subsample of non-homeowners drives the aggregate results. In columns 3 and 4, we do not find these patterns in the subsample of homeowners residing in treated urban regions and the full sample of individuals residing in treated suburban regions. Finally, we do not observe a statistically significant difference in the pre-treatment period between the treated and control MSAs. 

    Insofar as home ownership and income are correlated, these results confirm and support the evidence discussed in Section~\ref{subsec:gentrification} above that rideshare entry is associated with the displacement of incumbent renters, which tend to have lower incomes.
    
    \subsubsection{Financial Health}
    
    In this section, we investigate the impact of rideshare entry on the short-term financial health of incumbent residents, given the increase in housing prices. Again, we also split our analysis into homeowners and non-homeowners. We proxy for short-term financial health of residents using future information on credit delinquency. In particular, we create an indicator \textit{Late Repayment$_{i,y+2}$} of whether an individual \textit{i} is late on his/her payments in year \textit{y+2}.\footnote{We create an indicator of future delinquency, in this case, two years ahead, as individuals may be moving locations prior to delinquency.} Results of this exercise are displayed in \Cref{tab: credit quality}. In column 1 we present the results for the full sample of homeowners and non-homeowners in urban zip-codes. Its coefficients indicate that there is a slight gradual increase in the likelihood of late repayments following rideshare entry, averaging 0.3 p.p. (11 percent). However, as indicated by columns 2 and 3, this increase in future delinquency is driven by non-homeowners, which more than offset a decline in late repayments of homeowners. From column 2, we find non-homeowners increase their late repayments by 0.9 p.p. (42 percent), likely due to increased rental costs. Conversely, results for homeowners, displayed in column 3, show that this group has a decline in their late repayments by 1.2 p,p (36 percent), likely due to the increase in their property value and consequent net worth, which in principle allows them to expand their credit constraints. Results for the suburban sample of zip-codes, our control group, show no impact of the entry of rideshare services on delinquencies. 

    Results of the specification pooling urban and suburban zip-codes together are displayed in \Cref{tab:tableA13}. Qualitatively, results are similar to the ones splitting urban and suburban zip-codes. We find that following the entry of rideshare services, the delinquency rates have increased by 0.4 p.p. (15 percent). However, as displayed in columns 2 and 3, this increase in delinquency rates is driven by the sample of non-homeowners.    
	
    Overall, results in this section suggest that rideshare entry has uneven effects across residents of urban zip-codes. While homeowners benefit from improved asset returns in the form of housing price appreciation, non-homeowners suffer from an increase in housing prices, which increases the likelihood that they have to move away from their current locations while increasing the probability that they will suffer financial stress in the near future in the form of late credit repayments.

    \section{Conclusion}

This is the first study to provide a comprehensive analysis of the impact of ride-sharing services on the spatial reorganization of high- and low-income households and, consequently, mortgage and housing markets. Our analysis reveals that the introduction of ride-sharing services significantly accelerates urban gentrification by attracting higher-income individuals to previously lower-value urban neighborhoods. This influx drives up housing prices, resulting in the displacement of lower-income residents, particularly non-homeowners, who face increased financial stress and higher rates of out-migration. There is also a decline in the in-migration of poorer individuals into urban MSAs. The findings underscore the role of transportation innovations, specifically private ride-sharing services, in reshaping urban demographics and housing markets, leading to more pronounced economic segregation.

Further, the study highlights the differential impacts on homeowners versus renters. While homeowners benefit from rising property values and improved financial health, non-homeowners experience increased vulnerability to displacement and credit delinquency. This dynamic contributes to a widening gap in economic and financial inequality between these two groups as gentrification and economic segregation intensify in areas served by ride-sharing services.

These findings have important policy implications as the use of autonomous vehicles (AVs) by ridesharing firms is likely to amplify these trends. Policymakers must consider interventions that balance the benefits of improved transportation access with the need to protect vulnerable populations from the negative consequences of gentrification, ensuring that the advantages of AVs do not disproportionately benefit higher-income individuals at the expense of others.

    \comment{
	
	This is the first study to provide a comprehensive analysis of the impact of ride-sharing services on gentrification and, consequently, housing markets. The introduction of ride-sharing apps led to urban gentrification and the displacement of lower-income families with children. The rising demand for urban housing reduces vacancies and drives up home prices and rental rates. Consistent with the dynamics of home prices, there is an increase in both extensive and intensive margins of mortgages. 
	
	Supply-side factors do not drive the increase in mortgage demand as the approval rates remain unchanged. Rather, our analysis implies that the ease of transportation and lower automobile ownership costs enhanced the relative desirability of urban areas and, consequently, the demand for housing in such areas. Furthermore, consistent with gentrification, the new properties acquired in treated urban zip-codes are in poorer neighborhoods with a higher minority population.
	
	The in-migration of high-income persons and the concomitant rise in housing costs and displacement of low-income individuals cause financial distress and increase the likelihood of late repayments among the low-income non-homeowner population. Our study emphasizes the implications of accessible transportation provided by for-profit ridesharing services on gentrification and one of the most important household consumption decisions - real estate purchase.
}	
\clearpage  
{\singlespacing
    \small
    \bibliographystyle{chicago}
    \bibliography{literature}
}

\clearpage
\doublespacing


\begin{scriptsize}
	\begin{table}[ht!]
		\caption{\textbf{Summary Statistics}}
		\footnotesize{ This panel displays the summary statistics of the main variables of interest. See Table \ref{tab: variable definitions} in the Appendix for detailed variable definitions.} \\
		\setlength{\tabcolsep}{9pt}
		\begin{tabular}{lccccccc}
	\midrule
	\midrule
	& Num. Obs. & Mean  & \multicolumn{1}{c}{p10} & \multicolumn{1}{c}{Median} & \multicolumn{1}{c}{p90} & \multicolumn{2}{c}{Std. Dev.} \\
	\midrule
	\textit{\textbf{Urban Perception Index}} &       &       &       &       &       & \multicolumn{2}{c}{} \\
	Urban$_{z}$ & 10,777 & 0.52  & 0.03  & 0.19  & 0.7   &  \multicolumn{2}{c}{0.26} \\
	\\
	\textit{\textbf{Rideshare Entry and Drivers}} &       &       &       &       &       & \multicolumn{2}{c}{} \\
	Entry$_{m,y}$ & 8,614 & 0.05  & 0     & 0     & 0     & \multicolumn{2}{c}{0.21} \\
	Number of Drivers$_{m,y}$ & 5,072 & 4.37  & 2.08  & 4.06  & 7.18  & \multicolumn{2}{c}{2.03} \\
	Revenue Drivers$_{m,y}$ & 5,072 & 7.14  & 4.74  & 6.84  & 9.95  & \multicolumn{2}{c}{2.11} \\
	Share of Number Drivers$_{m,y}$ & 5,072 & 0.004 & 0.001 & 0.003 & 0.009 & \multicolumn{2}{c}{0.004} \\
	Share of Revenue Drivers$_{m,y}$ & 5,072 & 0.011 & 0.002 & 0.007 & 0.028 & \multicolumn{2}{c}{0.011} \\ 
	\\
	\multicolumn{4}{l}{\textit{\textbf{Zip-codes Statistics at Time of Rideshare Services Entry}}} &       &       & \multicolumn{2}{c}{} \\
	Trips Capita$_{z}$ & 711   & 8.4   & 4.6   & 10.4  & 14.1  & \multicolumn{2}{c}{3.7} \\
	Google Searches$_{z}$ & 2,288 & 81.2  & 48.1  & 88    & 100   & \multicolumn{2}{c}{20.8} \\ \\

        \textit{\textbf{IRS Statistics}} &       &       &       &       &       & \multicolumn{2}{c}{} \\
	Number of Households$_{z,y}$ & 52,198 & 9     & 7.9   & 9.2   & 9.9   & \multicolumn{2}{c}{0.8} \\
	Number of Households - High Income$_{z,y}$ & 52,198 & 8     & 6.8   & 8.2   & 9.1   & \multicolumn{2}{c}{0.9} \\
	Number of Households - Low Income$_{z,y}$ & 52,198 & 8.5   & 7.3   & 8.6   & 9.5   & \multicolumn{2}{c}{0.9} \\
	Child Credit$_{z,y}$ & 43,769 & 0.15  & 0.08  & 0.13  & 0.27  & \multicolumn{2}{c}{0.08} \\ \\

        \textit{\textbf{Mortgage Statistics}} &       &       &       &       &       & \multicolumn{2}{c}{} \\
	\#Mortgage Originations$_{z,y}$ & 32,921 & 5     & 3.6   & 5.2   & 6.4   & \multicolumn{2}{c}{1.2} \\
	Applicant Income$_{i,y}$ & 7,804,222 & 4.5   & 3.7   & 4.5   & 5.4   & \multicolumn{2}{c}{0.7} \\
	FHA-Insured$_{i,y}$ & 7,804,222 & 0.19  & 0     & 0     & 1     & \multicolumn{2}{c}{0.39} \\
	Mortgage Value$_{i,y}$ & 7,804,222 & 5.4   & 4.5   & 5.4   & 6.2   & \multicolumn{2}{c}{0.7} \\ \\

        \textit{\textbf{Housing Statistics}} &       &       &       &       &       & \multicolumn{2}{r}{} \\
	House Price Index$_{z,y}$ & 31,210 & 5     & 4.7   & 5     & 5.4   & \multicolumn{2}{c}{0.3} \\
	Rental Prices$_{z,y}$ & 33,254 & 7.4   & 6.9   & 7.3   & 7.9   & \multicolumn{2}{c}{0.4} \\
	Vacancy Rate$_{z,y}$ & 30,813 & 2.9   & 0.1   & 1.9   & 7.2   & \multicolumn{2}{c}{3} \\
	Number of Vacancies$_{z,y}$ & 30,813 & 8.9   & 7.8   & 9.1   & 9.9   & \multicolumn{2}{c}{0.9} \\ \\

        \textit{\textbf{Equifax Statistics}} &       &       &       &       &       & \multicolumn{2}{c}{} \\
	Age$_{z,y}$ & 36,384 & 4.3   & 4     & 4.3   & 4.5   & \multicolumn{2}{c}{0.2} \\
	Move out of Zip-code$_{i,y}$ & 2,473,268 & 0.09  & 0.05  & 0.08  & 0.05  & \multicolumn{2}{c}{0.31} \\
	Late Repayment & 2,473,268 & 0.03  & 0     & 0     & 0     & \multicolumn{2}{c}{0.16} \\ \\

        \textit{\textbf{Other Statistics}} &       &       &       &       &       & \multicolumn{2}{c}{} \\
	Average House Price in Zip-code$_{i,y}$ & 7,804,222 & 12.4  & 11.7  & 12.4  & 13.1  & \multicolumn{2}{c}{0.6} \\
	Average Poverty Rate in Zip-code$_{i,y}$ & 7,804,222 & 0.09  & 0.03  & 0.09  & 0.27  & \multicolumn{2}{c}{0.09} \\
	Average Minority Share in Zip-code$_{i,y}$ & 7,804,222 & 0.27  & 0.06  & 0.19  & 0.59  & \multicolumn{2}{c}{0.2} \\
	\bottomrule
\end{tabular}%
		\label{tab:summary_stats}%
	\end{table}%
\end{scriptsize}
\clearpage

\begin{scriptsize}
	\begin{landscape}
		\begin{table}[ht!]
			\caption[Zip-Codes using Rideshare Services]{\textbf{Urbanicity of Zip-code and Rideshare Usage}}
			\footnotesize 
			In this table, we display the results of regressions relating the perceived urbanicity of a zip-code with the number of trips and \textit{Google-searches} for rideshare services. Sample is restricted to MSAs in which rideshare services operate. \textit{Trips Capita$_{z}$} is the number of trips per capita using rideshare services in zip-code z. Information only available for zip-codes in Massachusetts, Chicago, San Francisco, and New York City.\textit{ Google Searches$_{z}$} is an index of the relative number of searches on \textit{Google} querying for the terms \textit{Uber}, \textit{Lyft} or \textit{Rideshare services} in zip-code \textit{z}, in the month following the entrance of the rideshare services in an MSA. \textit{Urban$_{z}$} is a measure of the urbanicity of zip-code \textit{z} compiled by the Census’\textit{ Urbanization Perceptions Small Area Index}. \textit{Urban$_{z,Qn}$} is an indicator of whether the urban measure of zip-code \textit{z} is in quartile \textit{n}. \textit{Zip-code Controls} include income per household, home value, share of individuals commuting using personal car, commute time, walkscore and transit score. *, **, and *** denote statistical significance at the 10\%, 5\%, and 1\% levels, respectively. See Table \ref{tab: variable definitions} in the Appendix for detailed variable definitions. \\
			\setlength{\tabcolsep}{17.5pt}
			\begin{tabular}{lcccccccc}
	\midrule
	\midrule
	\multicolumn{1}{l}{} & \multicolumn{4}{c}{Trips Capita$_{z}$} & \multicolumn{4}{c}{Google Searches$_{z}$} \\
	\multicolumn{1}{l}{} & \multicolumn{1}{c}{(1)} & \multicolumn{1}{c}{(2)} & \multicolumn{1}{c}{(3)} & \multicolumn{1}{c}{(4)} & \multicolumn{1}{c}{(5)} & \multicolumn{1}{c}{(6)} & \multicolumn{1}{c}{(7)} & \multicolumn{1}{c}{(8)} \\
	\midrule
	\multicolumn{1}{l}{} &       &       &       &       &       &       &       &  \\
	Urban$_{z}$ & \multicolumn{1}{c}{9.62***} & \multicolumn{1}{c}{12.98***} & \multicolumn{1}{c}{9.80***} &       & \multicolumn{1}{c}{8.45***} & \multicolumn{1}{c}{11.70***} & \multicolumn{1}{c}{8.31***} &  \\
	\multicolumn{1}{l}{} & \multicolumn{1}{c}{(1.54)} & \multicolumn{1}{c}{(4.27)} & \multicolumn{1}{c}{(3.31)} &       & \multicolumn{1}{c}{(1.34)} & \multicolumn{1}{c}{(2.60)} & \multicolumn{1}{c}{(1.57)} &  \\
	Urban$_{z,Q2}$ &       &       &       & \multicolumn{1}{c}{1.51} &       &       &       & \multicolumn{1}{c}{2.16***} \\
	\multicolumn{1}{l}{} &       &       &       & \multicolumn{1}{c}{(1.36)} &       &       &       & \multicolumn{1}{c}{(0.79)} \\
	Urban$_{z,Q3}$ &       &       &       & \multicolumn{1}{c}{5.06**} &       &       &       & \multicolumn{1}{c}{2.64***} \\
	\multicolumn{1}{l}{} &       &       &       & \multicolumn{1}{c}{(2.21)} &       &       &       & \multicolumn{1}{c}{(0.96)} \\
	Urban$_{z,Q4}$ &       &       &       & \multicolumn{1}{c}{7.65***} &       &       &       & \multicolumn{1}{c}{5.79***} \\
	\multicolumn{1}{l}{} &       &       &       & \multicolumn{1}{c}{(2.35)} &       &       &       & \multicolumn{1}{c}{(1.18)} \\
	\multicolumn{1}{l}{} &       &       &       &       &       &       &       &  \\
	Observations & 711   & 579   & 579   & 579   & 2,288  & 2,197  & 2,197  & 2,197 \\
	R-squared & 0.07  & 0.18  & 0.53  & 0.53  & 0.01  & 0.18  & 0.76  & 0.76 \\
	Zip-code Controls & \multicolumn{1}{c}{No} & \multicolumn{1}{c}{Yes} & \multicolumn{1}{c}{Yes} & \multicolumn{1}{c}{Yes} & \multicolumn{1}{c}{No} & \multicolumn{1}{c}{Yes} & \multicolumn{1}{c}{Yes} & \multicolumn{1}{c}{Yes} \\
	MSA FE & \multicolumn{1}{c}{No} & \multicolumn{1}{c}{No} & \multicolumn{1}{c}{Yes} & \multicolumn{1}{c}{Yes} & \multicolumn{1}{c}{No} & \multicolumn{1}{c}{No} & \multicolumn{1}{c}{Yes} & \multicolumn{1}{c}{Yes} \\
	Average(Dependent Variable) & 8.4   & 8.4   & 8.4   & 8.4   & 81.2  & 81.2  & 81.2  & 81.2 \\
 	\bottomrule
\end{tabular}%
			\label{tab:ridesharedemand}%
		\end{table}%
	\end{landscape}
\end{scriptsize}
\clearpage

\newpage
\begin{scriptsize}
	\begin{landscape}
		\begin{table}[ht!]
			\caption{\textbf{Entry of Rideshare Services and Income Composition of Households}}
			\footnotesize 
			This table displays the relation between the entry of rideshare services and the number of high- and low-income households in a zip-code*year in urban and suburban zip-codes. \textit{Urban (Suburban)} zip-codes are in the top (bottom) quartile of a measure of perceived urbanicity compiled by the Census’ \textit{Urbanization Perceptions Small Area Index}. \textit{Number of Households - High Income$_{z,y}$ (- Low Income$_{z,y}$)} is the number of individual tax returns (in logs) in zip-code \textit{z} in year \textit{y} with a household income equal or above (below) USD 50,000. \textit{Age$_{z,y}$} is the average age (in logs) of individuals with credit report in zip-code \textit{z} in year \textit{y}, compiled from Equifax Consumer Credit Panel. \textit{Child Credit$_{z,y}$} is the share of tax returns on income that claim a child credit in zip-code \textit{z} in year \textit{y}, obtained from the \textit{IRS}. *, **, and *** denote statistical significance at the 10\%, 5\%, and 1\% levels, respectively. See Table \ref{tab: variable definitions} in the Appendix for detailed variable definitions.\\
			\setlength{\tabcolsep}{16pt}
			\begin{tabular}{lcccccccc}
		\toprule
		\toprule
		\multicolumn{1}{c}{} & \multicolumn{2}{c}{Number of Households -} & \multicolumn{2}{c}{Number of Households -} & \multicolumn{2}{c}{Age$_{z,y}$} & \multicolumn{2}{c}{Child Credit$_{z,y}$} \\
		\multicolumn{1}{c}{} & \multicolumn{2}{c}{High Income$_{z,y}$} & \multicolumn{2}{c}{Low Income$_{z,y}$} & \multicolumn{2}{c}{} & \multicolumn{2}{c}{} \\
		\multicolumn{1}{c}{} & (1)   & (2)   & (3)   & (4)   & (5)   & (6)   & (7)   & (8) \\
		\midrule
		\multicolumn{1}{c}{} & \multicolumn{1}{c}{} & \multicolumn{1}{c}{} & \multicolumn{1}{c}{} & \multicolumn{1}{c}{} & \multicolumn{1}{c}{} & \multicolumn{1}{c}{} & \multicolumn{1}{c}{} & \multicolumn{1}{c}{} \\
		PreEntry 2-Years & 0.003 & 0.003 & -0.002 & 0.008 & -0.002 & 0.000 & 0.001 & -0.000 \\
		\multicolumn{1}{r}{} & (0.004) & (0.004) & (0.003) & (0.005) & (0.002) & (0.001) & (0.001) & (0.000) \\
		PreEntry 1-Years & 0.008 & 0.003 & -0.002 & 0.011 & -0.003 & 0.001 & 0.000 & -0.001 \\
		\multicolumn{1}{r}{} & (0.005) & (0.005) & (0.004) & (0.008) & (0.002) & (0.002) & (0.001) & (0.001) \\
		PostEntry 0-Years & 0.009*** & -0.002 & -0.010** & 0.009 & -0.002*** & 0.000 & -0.001 & -0.000 \\
		\multicolumn{1}{r}{} & (0.003) & (0.002) & (0.005) & (0.008) & (0.001) & (0.000) & (0.001) & (0.000) \\
		PostEntry 1-Years & 0.023*** & -0.003 & -0.015*** & 0.015 & -0.004*** & -0.001 & -0.003*** & 0.000 \\
		\multicolumn{1}{r}{} & (0.004) & (0.003) & (0.006) & (0.013) & (0.001) & (0.001) & (0.001) & (0.000) \\
		PostEntry 2-Years & 0.036*** & -0.002 & -0.014** & 0.020* & -0.008*** & -0.001 & -0.005*** & 0.001 \\
		\multicolumn{1}{r}{} & (0.005) & (0.004) & (0.006) & (0.004) & (0.001) & (0.001) & (0.001) & (0.001) \\
		PostEntry 3-Years & 0.049*** & -0.007 & -0.010 & 0.026*** & -0.011*** & -0.002 & -0.007*** & 0.002 \\
		\multicolumn{1}{r}{} & (0.007) & (0.006) & (0.007) & (0.006) & (0.002) & (0.002) & (0.002) & (0.001) \\
		PostEntry 4-Years & 0.067*** & -0.008 & 0.011 & 0.014* & -0.015*** & -0.003 & -0.012*** & 0.001 \\
		\multicolumn{1}{r}{} & (0.010) & (0.008) & (0.011) & (0.008) & (0.002) & (0.002) & (0.002) & (0.001) \\
		\multicolumn{1}{r}{} & \multicolumn{1}{c}{} & \multicolumn{1}{c}{} & \multicolumn{1}{c}{} & \multicolumn{1}{c}{} & \multicolumn{1}{c}{} & \multicolumn{1}{c}{} & \multicolumn{1}{c}{} & \multicolumn{1}{c}{} \\
		Observations & 23,356 & 21,559 & 23,356 & 21,559 & 12,940 & 10,424 & 20,148 & 18,989 \\
		Zip-code FE & Yes   & Yes   & Yes   & Yes   & Yes   & Yes   & Yes   & Yes \\
		State-Year FE & Yes   & Yes   & Yes   & Yes   & Yes   & Yes   & Yes   & Yes \\
		Sample of Zip-codes & Urban & Suburban & Urban & Suburban & Urban & Suburban & Urban & Suburban \\
            Avg. Dep. Variable & 8.2 & 7.8 & 8.6 & 8.3 & 4.3 & 4.5 & 0.13 & 0.17 \\
		\bottomrule
\end{tabular}%

			\label{tab: gentrification}
		\end{table}
	\end{landscape}
\end{scriptsize}
\clearpage

\begin{scriptsize}
	\begin{landscape}
		\begin{table}[ht!]
			\caption{\textbf{Entry of Rideshare Services, Mortgage Originations and Characteristics of Applicants}}
			\footnotesize 
			This table displays the relation between the entry of rideshare services with the number of mortgage originations and with  characteristics of mortgage applicants in urban and suburban zip-codes. \textit{Urban (Suburban)} zip-codes are in the top (bottom) quartile of a measure of perceived urbanicity compiled by the Census’ \textit{Urbanization Perceptions Small Area Index}. \textit{\#Mortgage Originations$_{z,y}$} is the number of mortgage originations (in logs) in zip-code \textit{z} in year \textit{y}. \textit{Applicant Income$_{i,y}$} is the gross annual income (in logs) of applicant \textit{i} in year \textit{y}. \textit{FHA-Insured$_{i,y}$} is an indicator that mortgage \textit{i} originated in year \textit{y} is insured by the Federal Housing Administration. Dependent variables obtained from HMDA confidential. *, **, and *** denote statistical significance at the 10\%, 5\%, and 1\% levels, respectively. See Table \ref{tab: variable definitions} in the Appendix for detailed variable definitions.\\
			\setlength{\tabcolsep}{26pt}
	\begin{tabular}{ccccccc}
		\toprule
		\toprule
		\multicolumn{1}{r}{} & \multicolumn{2}{c}{\#Mortgage Originations$_{z,y}$} & \multicolumn{2}{c}{Applicant Income$_{i,y}$} & \multicolumn{2}{c}{FHA-Insured$_{i,y}$} \\
		\multicolumn{1}{r}{} & \multicolumn{1}{c}{(1)} & \multicolumn{1}{c}{(2)} & \multicolumn{1}{c}{(3)} & \multicolumn{1}{c}{(4)} & \multicolumn{1}{c}{(5)} & \multicolumn{1}{c}{(6)} \\
		\midrule
		\multicolumn{1}{r}{} &       &       &       &       &       &  \\
		\multicolumn{1}{l}{PreEntry 2-Years} & \multicolumn{1}{c}{-0.003} & \multicolumn{1}{c}{-0.001} & \multicolumn{1}{c}{0.005} & \multicolumn{1}{c}{0.005} & \multicolumn{1}{c}{-0.002} & \multicolumn{1}{c}{-0.009} \\
		\multicolumn{1}{r}{} &                \multicolumn{1}{c}{(0.025)} & \multicolumn{1}{c}{(0.023)} & \multicolumn{1}{c}{(0.004)} & \multicolumn{1}{c}{(0.005)} & \multicolumn{1}{c}{(0.009)} & \multicolumn{1}{c}{(0.015)} \\
		\multicolumn{1}{l}{PreEntry 1-Years} & \multicolumn{1}{c}{0.023} & \multicolumn{1}{c}{-0.011} & \multicolumn{1}{c}{0.001} & \multicolumn{1}{c}{0.016**} & \multicolumn{1}{c}{-0.008} & \multicolumn{1}{c}{-0.025**} \\
		\multicolumn{1}{r}{} &                \multicolumn{1}{c}{(0.033)} & \multicolumn{1}{c}{(0.021)} & \multicolumn{1}{c}{(0.005)} & \multicolumn{1}{c}{(0.005)} & \multicolumn{1}{c}{(0.012)} & \multicolumn{1}{c}{(0.012)} \\
		\multicolumn{1}{l}{PostEntry 0-Years} & \multicolumn{1}{c}{0.017} & \multicolumn{1}{c}{0.012} & \multicolumn{1}{c}{0.020***} & \multicolumn{1}{c}{0.010**} & \multicolumn{1}{c}{-0.020***} & \multicolumn{1}{c}{-0.016} \\
		\multicolumn{1}{r}{} &                \multicolumn{1}{c}{(0.024)} & \multicolumn{1}{c}{(0.014)} & \multicolumn{1}{c}{(0.004)} & \multicolumn{1}{c}{(0.004)} & \multicolumn{1}{c}{(0.007)} & \multicolumn{1}{c}{(0.015)} \\
		\multicolumn{1}{l}{PostEntry 1-Years} & \multicolumn{1}{c}{0.014} & \multicolumn{1}{c}{0.006} & \multicolumn{1}{c}{0.041***} & \multicolumn{1}{c}{0.009} & \multicolumn{1}{c}{-0.034***} & \multicolumn{1}{c}{-0.012} \\
		\multicolumn{1}{r}{} &                \multicolumn{1}{c}{(0.022)} & \multicolumn{1}{c}{(0.012)} & \multicolumn{1}{c}{(0.006)} & \multicolumn{1}{c}{(0.006)} & \multicolumn{1}{c}{(0.007)} & \multicolumn{1}{c}{(0.016)} \\
		\multicolumn{1}{l}{PostEntry 2-Years} & \multicolumn{1}{c}{0.117***} & \multicolumn{1}{c}{0.033} & \multicolumn{1}{c}{0.069***} & \multicolumn{1}{c}{0.012} & \multicolumn{1}{c}{-0.075***} & \multicolumn{1}{c}{-0.024} \\
		\multicolumn{1}{r}{} &                \multicolumn{1}{c}{(0.020)} & \multicolumn{1}{c}{(0.034)} & \multicolumn{1}{c}{(0.009)} & \multicolumn{1}{c}{(0.008)} & \multicolumn{1}{c}{(0.016)} & \multicolumn{1}{c}{(0.018)} \\
		\multicolumn{1}{l}{PostEntry 3-Years} & \multicolumn{1}{c}{0.248***} & \multicolumn{1}{c}{0.010} & \multicolumn{1}{c}{0.094***} & \multicolumn{1}{c}{0.014} & \multicolumn{1}{c}{-0.108***} & \multicolumn{1}{c}{-0.029} \\
		\multicolumn{1}{r}{} &                \multicolumn{1}{c}{(0.039)} & \multicolumn{1}{c}{(0.031)} & \multicolumn{1}{c}{(0.012)} & \multicolumn{1}{c}{(0.011)} & \multicolumn{1}{c}{(0.013)} & \multicolumn{1}{c}{(0.019)} \\
		\multicolumn{1}{l}{PostEntry 4-Years} & \multicolumn{1}{c}{0.056*} & \multicolumn{1}{c}{0.031} & \multicolumn{1}{c}{0.114***} & \multicolumn{1}{c}{0.007} & \multicolumn{1}{c}{-0.085***} & \multicolumn{1}{c}{-0.031} \\
		\multicolumn{1}{r}{} &                   \multicolumn{1}{c}{(0.030)} & \multicolumn{1}{c}{(0.032)} & \multicolumn{1}{c}{(0.014)} & \multicolumn{1}{c}{(0.013)} & \multicolumn{1}{c}{(0.016)} & \multicolumn{1}{c}{(0.027)} \\
		\multicolumn{1}{r}{} &       &       &       &       &       &  \\
		\multicolumn{1}{l}{Observations} & \multicolumn{1}{c}{9,335} & \multicolumn{1}{c}{11,410} & \multicolumn{1}{c}{2,295,577} & \multicolumn{1}{c}{3,534,990} & \multicolumn{1}{c}{2,295,577} & \multicolumn{1}{c}{3,534,990} \\
		\multicolumn{1}{l}{Zip-code FE} & \multicolumn{1}{c}{Yes} & \multicolumn{1}{c}{Yes} & \multicolumn{1}{c}{Yes} & \multicolumn{1}{c}{Yes} & \multicolumn{1}{c}{Yes} & \multicolumn{1}{c}{Yes} \\
		\multicolumn{1}{l}{State-Year FE} & \multicolumn{1}{c}{Yes} & \multicolumn{1}{c}{Yes} & \multicolumn{1}{c}{Yes} & \multicolumn{1}{c}{Yes} & \multicolumn{1}{c}{Yes} & \multicolumn{1}{c}{Yes} \\
		\multicolumn{1}{l}{Zip-codes} & \multicolumn{1}{c}{Urban} & \multicolumn{1}{c}{Suburban} & \multicolumn{1}{c}{Urban} & \multicolumn{1}{c}{Suburban} & \multicolumn{1}{c}{Urban} & \multicolumn{1}{c}{Suburban} \\
		Avg. Dep. Variable & 4.7   & 5.3   & 4.6   & 4.5   & 0.21  & 0.18 \\
		\bottomrule
	\end{tabular}%

			\label{tab: mortgageoriginations}
		\end{table}
	\end{landscape}
\end{scriptsize}
\clearpage

\begin{scriptsize}
	\begin{landscape}
		\begin{table}[ht!]
			\caption{\textbf{Entry of Rideshare Services and Characteristics of Zip-Codes of Mortgage Originations}}
			\footnotesize 
			This table displays the results of a regression relating the entry of rideshare services with the characteristics of zip-codes of the properties for which the mortgages were issued in urban and suburban zip-codes. \textit{Urban (Suburban)} zip-codes are in the top (bottom) quartile of a measure of perceived urbanicity compiled by the Census’ \textit{Urbanization Perceptions Small Area Index}. \textit{Median House Price in Zip-code$_{i,y}$} is the median house price (in logs) in the zip-code of the property financed by mortgage \textit{i} originated in year \textit{y}. \textit{Average Poverty Rate in Zip-code$_{i,y}$} is the average poverty rate of the zip-code of the property financed by mortgage \textit{i} originated in year \textit{y}. \textit{Average Minority Share in Zip-code$_{i,y}$} is the average share of minority individuals of the zip-code of the property financed by mortgage \textit{i} originated in year \textit{y}. Mortgage information obtained from \textit{HMDA confidential}, while information on the dependent variables obtained from \textit{Zillow} and the \textit{Census Bureau}.*, **, and *** denote statistical significance at the 10\%, 5\%, and 1\% levels, respectively. See Table \ref{tab: variable definitions} in the Appendix for detailed variable definitions. \\
			\setlength{\tabcolsep}{18pt} 
			\begin{tabular}{lcccccc}
	\toprule
	\toprule
	\multicolumn{1}{r}{} & \multicolumn{2}{c}{Median House Price in Zip-code$_{i,y}$} & \multicolumn{2}{c}{Average Poverty Rate in Zip-code$_{i,y}$} & \multicolumn{2}{c}{Average Minority Share in Zip-code$_{i,y}$} \\
	\multicolumn{1}{r}{} & \multicolumn{1}{c}{(1)} & \multicolumn{1}{c}{(2)} & \multicolumn{1}{c}{(3)} & \multicolumn{1}{c}{(4)} & \multicolumn{1}{c}{(5)} & \multicolumn{1}{c}{(6)} \\
	\midrule
	\multicolumn{1}{r}{} &                &                &                &                &                &  \\
	PreEntry 2-Years & \multicolumn{1}{c}{0.001} & \multicolumn{1}{c}{-0.003} & \multicolumn{1}{c}{-0.001} & \multicolumn{1}{c}{0.001} & \multicolumn{1}{c}{0.001} & \multicolumn{1}{c}{-0.003**} \\
	\multicolumn{1}{r}{} & \multicolumn{1}{c}{(0.003)} & \multicolumn{1}{c}{(0.005)} & \multicolumn{1}{c}{(0.001)} & \multicolumn{1}{c}{(0.001)} & \multicolumn{1}{c}{(0.001)} & \multicolumn{1}{c}{(0.001)} \\
	PostEntry 1-Years & \multicolumn{1}{c}{0.004} & \multicolumn{1}{c}{-0.001} & \multicolumn{1}{c}{-0.002} & \multicolumn{1}{c}{0.000} & \multicolumn{1}{c}{0.000} & \multicolumn{1}{c}{-0.004***} \\
	\multicolumn{1}{r}{} & \multicolumn{1}{c}{(0.003)} & \multicolumn{1}{c}{(0.006)} & \multicolumn{1}{c}{(0.002)} & \multicolumn{1}{c}{(0.001)} & \multicolumn{1}{c}{(0.001)} & \multicolumn{1}{c}{(0.002)} \\
	PostEntry 0-Years & \multicolumn{1}{c}{-0.004} & \multicolumn{1}{c}{-0.003} & \multicolumn{1}{c}{0.000} & \multicolumn{1}{c}{-0.000} & \multicolumn{1}{c}{0.000} & \multicolumn{1}{c}{0.002} \\
	\multicolumn{1}{r}{} & \multicolumn{1}{c}{(0.003)} & \multicolumn{1}{c}{(0.004)} & \multicolumn{1}{c}{(0.001)} & \multicolumn{1}{c}{(0.001)} & \multicolumn{1}{c}{(0.001)} & \multicolumn{1}{c}{(0.001)} \\
	PostEntry 1-Years & \multicolumn{1}{c}{-0.009**} & \multicolumn{1}{c}{-0.006} & \multicolumn{1}{c}{0.000} & \multicolumn{1}{c}{-0.000} & \multicolumn{1}{c}{0.003**} & \multicolumn{1}{c}{0.000} \\
	\multicolumn{1}{r}{} & \multicolumn{1}{c}{(0.004)} & \multicolumn{1}{c}{(0.004)} & \multicolumn{1}{c}{(0.001)} & \multicolumn{1}{c}{(0.001)} & \multicolumn{1}{c}{(0.001)} & \multicolumn{1}{c}{(0.002)} \\
	PostEntry 2-Years & \multicolumn{1}{c}{-0.020***} & \multicolumn{1}{c}{-0.016***} & \multicolumn{1}{c}{0.003*} & \multicolumn{1}{c}{0.000} & \multicolumn{1}{c}{0.006***} & \multicolumn{1}{c}{0.002} \\
	\multicolumn{1}{r}{} & \multicolumn{1}{c}{(0.005)} & \multicolumn{1}{c}{(0.005)} & \multicolumn{1}{c}{(0.002)} & \multicolumn{1}{c}{(0.001)} & \multicolumn{1}{c}{(0.002)} & \multicolumn{1}{c}{(0.002)} \\
	PostEntry 3-Years & \multicolumn{1}{c}{-0.032***} & \multicolumn{1}{c}{-0.011***} & \multicolumn{1}{c}{0.004**} & \multicolumn{1}{c}{0.001} & \multicolumn{1}{c}{0.010***} & \multicolumn{1}{c}{0.003**} \\
	\multicolumn{1}{r}{} & \multicolumn{1}{c}{(0.006)} & \multicolumn{1}{c}{(0.004)} & \multicolumn{1}{c}{(0.002)} & \multicolumn{1}{c}{(0.000)} & \multicolumn{1}{c}{(0.002)} & \multicolumn{1}{c}{(0.001)} \\
	PostEntry 4-Years & \multicolumn{1}{c}{-0.050***} & \multicolumn{1}{c}{-0.019***} & \multicolumn{1}{c}{0.006***} & \multicolumn{1}{c}{0.001*} & \multicolumn{1}{c}{0.015***} & \multicolumn{1}{c}{0.004**} \\
	\multicolumn{1}{r}{} & \multicolumn{1}{c}{(0.008)} & \multicolumn{1}{c}{(0.006)} & \multicolumn{1}{c}{(0.002)} & \multicolumn{1}{c}{(0.001)} & \multicolumn{1}{c}{(0.003)} & \multicolumn{1}{c}{(0.002)} \\
	\multicolumn{1}{r}{} &                &                &                &                &                &  \\
	Observations   & 2,295,577        & 3,534,990        & 2,295,577        & 3,534,990        & 2,295,577        & 3,534,990 \\
	MSA FE         & \multicolumn{1}{c}{Yes} & \multicolumn{1}{c}{Yes} & \multicolumn{1}{c}{Yes} & \multicolumn{1}{c}{Yes} & \multicolumn{1}{c}{Yes} & \multicolumn{1}{c}{Yes} \\
	State-Year FE  & \multicolumn{1}{c}{Yes} & \multicolumn{1}{c}{Yes} & \multicolumn{1}{c}{Yes} & \multicolumn{1}{c}{Yes} & \multicolumn{1}{c}{Yes} & \multicolumn{1}{c}{Yes} \\
	Zip-codes       & \multicolumn{1}{c}{Urban} & \multicolumn{1}{c}{Suburban} & \multicolumn{1}{c}{Urban} & \multicolumn{1}{c}{Suburban} & \multicolumn{1}{c}{Urban} & \multicolumn{1}{c}{Suburban} \\
	Avg. Dep. Variable & 12.3           & 12.4           & 0.20            & 0.06           & 0.43          & 0.18 \\
	\bottomrule
\end{tabular}%
			\label{tab: zip code characteristics}
		\end{table}
	\end{landscape}
\end{scriptsize}
\clearpage

\begin{scriptsize}
	\begin{table}[ht!]
		\caption{\textbf{Entry of Rideshare Services and House and Rental Prices}}
		\footnotesize 
		This table displays the results of regressions relating the entry of rideshare services with housing and rental prices at the zip-code*year level, in urban and suburban zip-codes.\textit{ Urban (Suburban)} zip-codes are in the top (bottom) quartile of a measure of perceived urbanicity compiled by the Census’ \textit{Urbanization Perceptions Small Area Index}. \textit{House Price Index$_{z,y}$} is an index of house prices (in logs) at the zip-code \textit{z} in year \textit{y}, obtained from \textit{Corelogic}. \textit{Rental Prices$_{z,y}$} is the average monthly market rent (in logs) of zip-code \textit{z} in year \textit{y}, obtained from \textit{Zillow}. *, **, and *** denote statistical significance at the 10\%, 5\%, and 1\% levels, respectively. See Table \ref{tab: variable definitions} in the Appendix for detailed variable definitions. \\
		\setlength{\tabcolsep}{25pt}
		\begin{tabular}{lcccc}
	\toprule
	\toprule
	& \multicolumn{2}{c}{House Price Index$_{z,y}$} & \multicolumn{2}{c}{Rental Prices$_{z,y}$} \\
	& (1)            & (2)            & (3)            & (4) \\
	\midrule
	&                &                &                &  \\
	PreEntry 2-Years & -0.004         & 0.005          & 0.015          & -0.010 \\
	& (0.016)        & (0.012)        & (0.034)        & (0.029) \\
	PreEntry 1-Years & 0.005          & 0.013          & 0.033          & -0.005 \\
	& (0.021)        & (0.016)        & (0.042)        & (0.036) \\
	PostEntry 0-Years & 0.033***       & 0.014          & 0.025***       & 0.015*** \\
	& (0.008)        & (0.009)        & (0.005)        & (0.004) \\
	PostEntry 1-Years & 0.057***       & 0.019          & 0.042***       & 0.023*** \\
	& (0.011)        & (0.013)        & (0.007)        & (0.006) \\
	PostEntry 2-Years & 0.082***       & 0.017          & 0.061***       & 0.029*** \\
	& (0.016)        & (0.020)        & (0.012)        & (0.008) \\
	PostEntry 3-Years & 0.125***       & 0.042**        & 0.078***       & 0.024*** \\
	& (0.018)        & (0.022)        & (0.018)        & (0.009) \\
	PostEntry 4-Years & 0.147***       & 0.022          & 0.084**        & 0.013 \\
	& (0.029)        & (0.023)        & (0.035)        & (0.015) \\
	&                &                &                &  \\
	Observations   & 13,431          & 12,397          & 17,592          & 16,162 \\
	Zip-code FE     & Yes            & Yes            & Yes            & Yes \\
	State-Year FE  & No             & Yes            & No             & Yes \\
	Sample of Zip-codes & Urban          & Suburban       & Urban          & Suburban \\
	Avg. Dep. Variable & 5.2            & 5              & 7.3            & 7.5 \\
	\bottomrule
\end{tabular}%
		\label{tab: house prices}
	\end{table}
\end{scriptsize}
\clearpage 

\begin{scriptsize}
	\begin{table}[ht!]
		\caption{\textbf{Entry of Rideshare Services and Relocation Away from Zip-Code}}
		\footnotesize 
		This table displays the results of a regression relating the entry of rideshare services with the moving of individuals away from their current zip-code of residence. \textit{Urban (Suburban)} zip-codes are in the top (bottom) quartile of a measure of perceived urbanicity compiled by the Census’ \textit{Urbanization Perceptions Small Area Index}. \textit{Moving Out of Zip-Code$_{i,y}$} is an indicator of whether individual \textit{i} has a different residential zip-code in year \textit{y+1} relative to year \textit{y}. \textit{Homeowner (Non-Homeowner)} are individuals that have (not) had a mortgage or home-equity loan in their current zip-code. \textit{Controls} are a series of individual controls including age (in logs), FICO score, and an indicator of late repayment. *, **, and *** denote statistical significance at the 10\%, 5\%, and 1\% levels, respectively.  See Table \ref{tab: variable definitions} in the Appendix for detailed variable definitions.\\
		\setlength{\tabcolsep}{20.5pt}
		\begin{tabular}{lcccc}
	\toprule
	\toprule
	\multicolumn{1}{r}{} & \multicolumn{4}{c}{Moving Out of Zip-Code$_{i,y}$} \\
	\multicolumn{1}{r}{} & \multicolumn{1}{c}{(1)} & \multicolumn{1}{c}{(2)} & \multicolumn{1}{c}{(3)} & \multicolumn{1}{c}{(4)} \\
	\midrule
	\multicolumn{1}{r}{} &                &                &                &  \\
	PreEntry 2-Years & \multicolumn{1}{c}{-0.004} & \multicolumn{1}{c}{-0.005} & \multicolumn{1}{c}{0.001} & \multicolumn{1}{c}{0.002} \\
	\multicolumn{1}{r}{} & \multicolumn{1}{c}{(0.005)} & \multicolumn{1}{c}{(0.005)} & \multicolumn{1}{c}{(0.007)} & \multicolumn{1}{c}{(0.004)} \\
	PreEntry 1-Years & \multicolumn{1}{c}{0.000} & \multicolumn{1}{c}{0.001} & \multicolumn{1}{c}{0.005} & \multicolumn{1}{c}{0.001} \\
	\multicolumn{1}{r}{} & \multicolumn{1}{c}{(0.004)} & \multicolumn{1}{c}{(0.007)} & \multicolumn{1}{c}{(0.008)} & \multicolumn{1}{c}{(0.006)} \\
	PostEntry 0-Years & \multicolumn{1}{c}{0.004} & \multicolumn{1}{c}{0.005} & \multicolumn{1}{c}{-0.003} & \multicolumn{1}{c}{-0.003} \\
	\multicolumn{1}{r}{} & \multicolumn{1}{c}{(0.005)} & \multicolumn{1}{c}{(0.007)} & \multicolumn{1}{c}{(0.007)} & \multicolumn{1}{c}{(0.003)} \\
	PostEntry 1-Years & \multicolumn{1}{c}{0.004} & \multicolumn{1}{c}{0.003} & \multicolumn{1}{c}{0.000} & \multicolumn{1}{c}{-0.004} \\
	\multicolumn{1}{r}{} & \multicolumn{1}{c}{(0.004)} & \multicolumn{1}{c}{(0.005)} & \multicolumn{1}{c}{(0.006)} & \multicolumn{1}{c}{(0.007)} \\
	PostEntry 2-Years & \multicolumn{1}{c}{0.022***} & \multicolumn{1}{c}{0.027***} & \multicolumn{1}{c}{0.006} & \multicolumn{1}{c}{0.002} \\
	\multicolumn{1}{r}{} & \multicolumn{1}{c}{(0.007)} & \multicolumn{1}{c}{(0.004)} & \multicolumn{1}{c}{(0.012)} & \multicolumn{1}{c}{(0.006)} \\
	PostEntry 3-Years & \multicolumn{1}{c}{0.012*} & \multicolumn{1}{c}{0.014**} & \multicolumn{1}{c}{0.006} & \multicolumn{1}{c}{0.001} \\
	\multicolumn{1}{r}{} & \multicolumn{1}{c}{(0.007)} & \multicolumn{1}{c}{(0.007)} & \multicolumn{1}{c}{(0.015)} & \multicolumn{1}{c}{(0.008)} \\
	PostEntry 4-Years & \multicolumn{1}{c}{0.028***} & \multicolumn{1}{c}{0.031***} & \multicolumn{1}{c}{0.021*} & \multicolumn{1}{c}{0.005} \\
	\multicolumn{1}{r}{} & \multicolumn{1}{c}{(0.006)} & \multicolumn{1}{c}{(0.006)} & \multicolumn{1}{c}{(0.007)} & \multicolumn{1}{c}{(0.011)} \\
	\multicolumn{1}{r}{} &                &                &                &  \\
	Observations   & 668,649         & 513,131         & 155,506         & 537,769 \\
	Controls       & \multicolumn{1}{c}{Yes} & \multicolumn{1}{c}{Yes} & \multicolumn{1}{c}{Yes} & \multicolumn{1}{c}{Yes} \\
	Zip-code FE     & \multicolumn{1}{c}{Yes} & \multicolumn{1}{c}{Yes} & \multicolumn{1}{c}{Yes} & \multicolumn{1}{c}{Yes} \\
	State-Year FE  & \multicolumn{1}{c}{Yes} & \multicolumn{1}{c}{Yes} & \multicolumn{1}{c}{Yes} & \multicolumn{1}{c}{Yes} \\
	Zip-codes       & \multicolumn{1}{c}{Urban} & \multicolumn{1}{c}{Urban} & \multicolumn{1}{c}{Urban} & \multicolumn{1}{c}{Suburban} \\
	Sample         & All            & Non-Homeowner  & Homeowner      & All \\
	Avg. Dep. Variable & 0.11           & 0.12           & 0.07           & 0.08 \\
	\bottomrule
\end{tabular}%
		\label{tab: relocation}
	\end{table}
\end{scriptsize}
\clearpage

\begin{scriptsize}
	\begin{table}[ht!]
		\caption{\textbf{Entry of Rideshare Services and Change in Credit Quality}}
		\footnotesize 
		This table displays the results of a regression relating the entry of rideshare services with changes in the financial health of incumbent residents, proxied by an indicator of credit delinquency. \textit{Urban (Suburban)} zip-codes are in the top (bottom) quartile of a measure of perceived urbanicity compiled by the Census’ \textit{Urbanization Perceptions Small Area Index}. \textit{Late Repayment$_{i,y+2}$} is an indicator of whether individual \textit{i} is late on his credit payments in year \textit{y+2}. \textit{Controls} are a series of individual controls including age (in logs), fico score, and an indicator of current late repayment. \textit{Homeowner(Non-Homeowner)} are individuals that have (not) had a mortgage in their current zip-code. *, **, and *** denote statistical significance at the 10\%, 5\%, and 1\% levels, respectively. See Table \ref{tab: variable definitions} in the Appendix for detailed variable definitions.\\
		\setlength{\tabcolsep}{20.5pt}
		\begin{tabular}{lcccc}
	\toprule
	\toprule
	\multicolumn{1}{r}{} & \multicolumn{4}{c}{Late Repayment$_{i,y+2}$} \\
	\multicolumn{1}{r}{} & \multicolumn{1}{c}{(1)} & \multicolumn{1}{c}{(2)} & \multicolumn{1}{c}{(3)} & \multicolumn{1}{c}{(4)} \\
	\midrule
	\multicolumn{1}{r}{} &                &                &                &  \\
	PreEntry 2-Years & \multicolumn{1}{c}{-0.001} & \multicolumn{1}{c}{0.001} & \multicolumn{1}{c}{-0.007} & \multicolumn{1}{c}{-0.000} \\
	\multicolumn{1}{r}{} & \multicolumn{1}{c}{(0.002)} & \multicolumn{1}{c}{(0.002)} & \multicolumn{1}{c}{(0.006)} & \multicolumn{1}{c}{(0.002)} \\
	PostEntry 1-Years & \multicolumn{1}{c}{-0.000} & \multicolumn{1}{c}{0.002} & \multicolumn{1}{c}{-0.010} & \multicolumn{1}{c}{-0.002} \\
	\multicolumn{1}{r}{} & \multicolumn{1}{c}{(0.003)} & \multicolumn{1}{c}{(0.003)} & \multicolumn{1}{c}{(0.008)} & \multicolumn{1}{c}{(0.003)} \\
	PostEntry 0-Years & \multicolumn{1}{c}{-0.000} & \multicolumn{1}{c}{-0.001} & \multicolumn{1}{c}{0.002} & \multicolumn{1}{c}{-0.002} \\
	\multicolumn{1}{r}{} & \multicolumn{1}{c}{(0.002)} & \multicolumn{1}{c}{(0.002)} & \multicolumn{1}{c}{(0.007)} & \multicolumn{1}{c}{(0.002)} \\
	PostEntry 1-Years & \multicolumn{1}{c}{-0.003} & \multicolumn{1}{c}{-0.004} & \multicolumn{1}{c}{0.001} & \multicolumn{1}{c}{0.000} \\
	\multicolumn{1}{r}{} & \multicolumn{1}{c}{(0.002)} & \multicolumn{1}{c}{(0.003)} & \multicolumn{1}{c}{(0.007)} & \multicolumn{1}{c}{(0.002)} \\
	PostEntry 2-Years & \multicolumn{1}{c}{0.001} & \multicolumn{1}{c}{0.008**} & \multicolumn{1}{c}{-0.010} & \multicolumn{1}{c}{-0.004} \\
	\multicolumn{1}{r}{} & \multicolumn{1}{c}{(0.003)} & \multicolumn{1}{c}{(0.003)} & \multicolumn{1}{c}{(0.007)} & \multicolumn{1}{c}{(0.003)} \\
	PostEntry 3-Years & \multicolumn{1}{c}{0.006*} & \multicolumn{1}{c}{0.021***} & \multicolumn{1}{c}{-0.037***} & \multicolumn{1}{c}{-0.007} \\
	\multicolumn{1}{r}{} & \multicolumn{1}{c}{(0.003)} & \multicolumn{1}{c}{(0.005)} & \multicolumn{1}{c}{(0.008)} & \multicolumn{1}{c}{(0.004)} \\
	PostEntry 4-Years & \multicolumn{1}{c}{0.007*} & \multicolumn{1}{c}{0.015***} & \multicolumn{1}{c}{-0.023***} & \multicolumn{1}{c}{0.001} \\
	\multicolumn{1}{r}{} & \multicolumn{1}{c}{(0.004)} & \multicolumn{1}{c}{(0.004)} & \multicolumn{1}{c}{(0.009)} & \multicolumn{1}{c}{(0.006)} \\
	\multicolumn{1}{r}{} &                &                &                &  \\
	Observations   & 646,080         & 494,808         & 151,261         & 521,852 \\
	Controls       & \multicolumn{1}{c}{Yes} & \multicolumn{1}{c}{Yes} & \multicolumn{1}{c}{Yes} & \multicolumn{1}{c}{Yes} \\
	Zip-code FE     & \multicolumn{1}{c}{Yes} & \multicolumn{1}{c}{Yes} & \multicolumn{1}{c}{Yes} & \multicolumn{1}{c}{Yes} \\
	State-Year FE  & \multicolumn{1}{c}{Yes} & \multicolumn{1}{c}{Yes} & \multicolumn{1}{c}{Yes} & \multicolumn{1}{c}{Yes} \\
	Zip-codes       & \multicolumn{1}{c}{Urban} & \multicolumn{1}{c}{Urban} & \multicolumn{1}{c}{Urban} & \multicolumn{1}{c}{Suburban} \\
	Sample         & \multicolumn{1}{c}{All} & \multicolumn{1}{c}{Non-Homeowner} & \multicolumn{1}{c}{Homeowner} & \multicolumn{1}{c}{All} \\
	Avg. Dep. Variable & 0.023          & 0.021          & 0.034          & 0.024 \\
	\bottomrule
\end{tabular}%
		\label{tab: credit quality}
	\end{table}
\end{scriptsize}
\clearpage


\begin{scriptsize}
	\begin{landscape}
		\begin{figure}[!h]%
		\caption{\textbf{Entry of Rideshare Services and Income of Mortgage Applicants}} 
            \footnotesize{This figure displays the evolution of yearly coefficients of a regression relating the entry of rideshare services with the income of mortgage applicants in urban and suburban zip-codes, in the left and right panels respectively. \textit{Urban (Suburban)} zip-codes are in the top (bottom) quartile of a measure of perceived urbanicity compiled by the Census’ \textit{Urbanization Perceptions Small Area Index}. Regressions include zip-code and state*year fixed effects. Standard errors are clustered at the zip-code level. The blue vertical bars represent confidence intervals of the coefficients at the 95 percent significance level. See Table \ref{tab: variable definitions} in the Appendix for detailed variable definitions.} \
            \vspace{2pt} 
            \hrule height 1pt 
            \vspace{1.5pt} 
            \hrule height 0.5pt 
            \vspace{3.5pt} 
			\normalsize
			\begin{minipage}{0.65\textwidth}
				\centering
				\caption*{(A) Urban Zip-codes}
				\includegraphics[width=\linewidth]{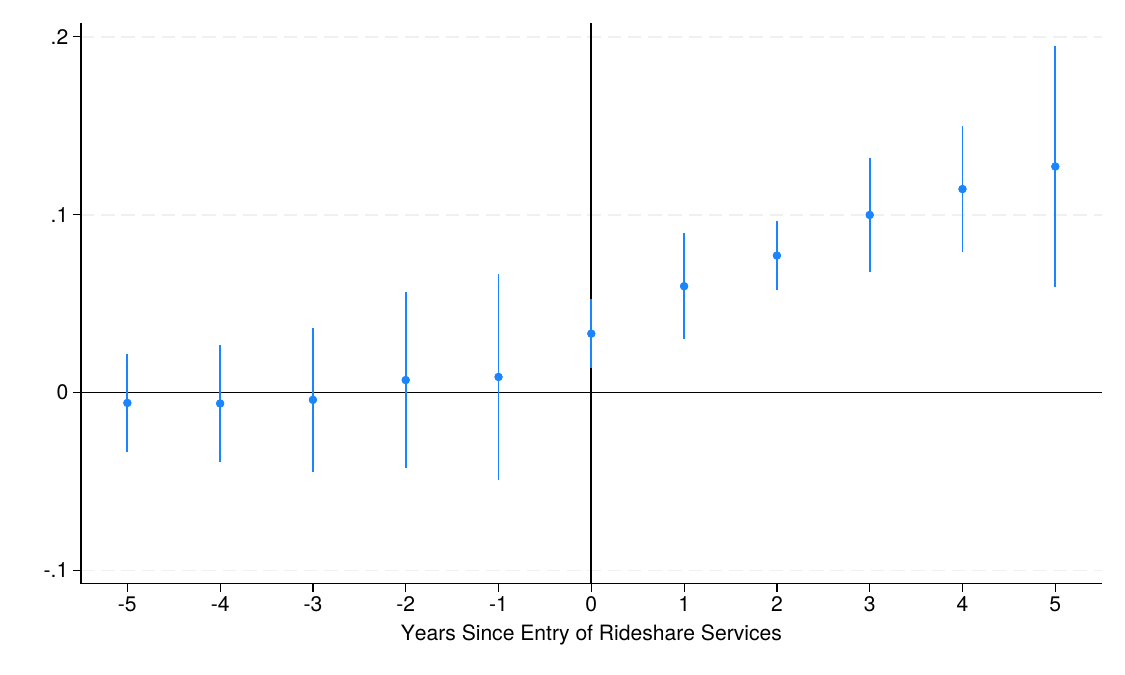}
			\end{minipage}\hfil
			\begin{minipage}{0.66\textwidth}
				\centering
				\caption*{(B) Suburban Zip-codes}
				\includegraphics[width=\linewidth]{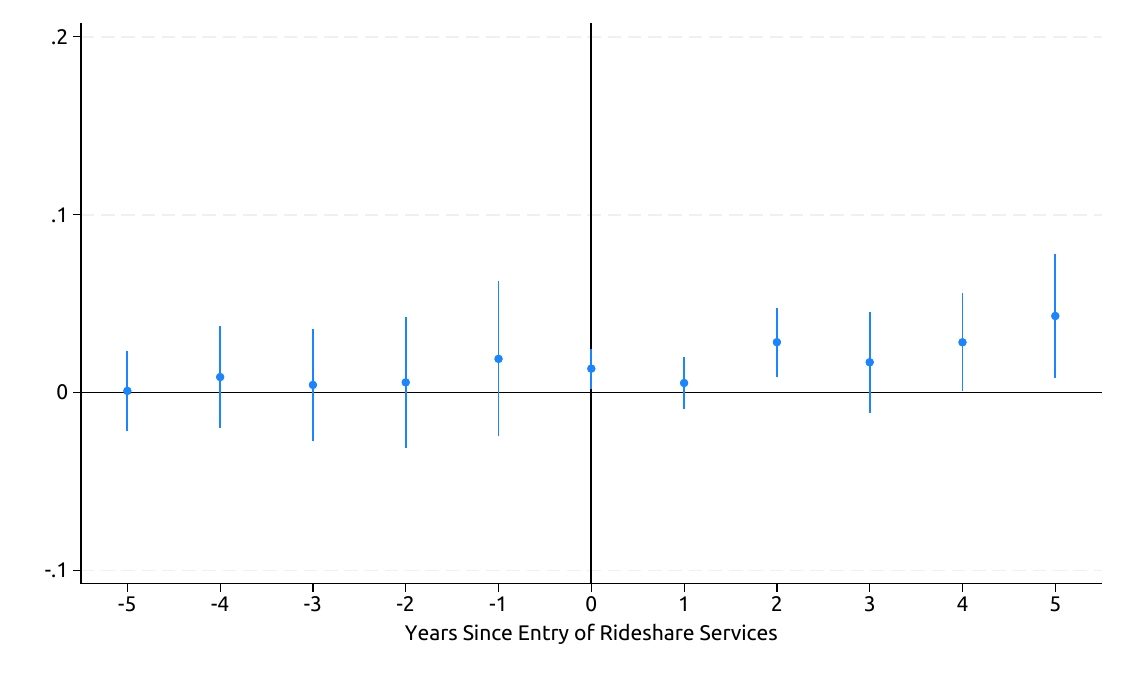}
			\end{minipage}
			\label{fig:incomeapplicants}
            \hrule height 0.5pt 
		\end{figure}
	\end{landscape}
\end{scriptsize}
\clearpage

\begin{scriptsize}
	\begin{landscape}
		\begin{figure}[!h]%
			\caption{\textbf{Entry of Rideshare Services and ex-ante Median Home Values of Zip-codes of Mortgage Originations}} 
			\footnotesize{This figure displays the evolution of yearly coefficients of a regression relating the entry of rideshare services with the median home values (in logs) of the zip-codes of the mortgage originations in urban and suburban zip-codes, in the left and right panels respectively. \textit{Urban (Suburban)} zip-codes are in the top (bottom) quartile of a measure of perceived urbanicity compiled by the Census’ \textit{Urbanization Perceptions Small Area Index}. Regressions include MSA and state*year fixed effects. Standard errors are clustered at the zip-code level. The blue vertical bars represent confidence intervals of the coefficients at the 95 percent significance level.  See Table \ref{tab: variable definitions} in the Appendix for detailed variable definitions.} \
            \vspace{2pt} 
            \hrule height 1pt 
            \vspace{1.5pt} 
            \hrule height 0.5pt 
            \vspace{3.5pt} 
			\normalsize
			\begin{minipage}{0.65\textwidth}
				\centering
				\caption*{(A) Urban Zip-codes}
				\includegraphics[width=\linewidth]{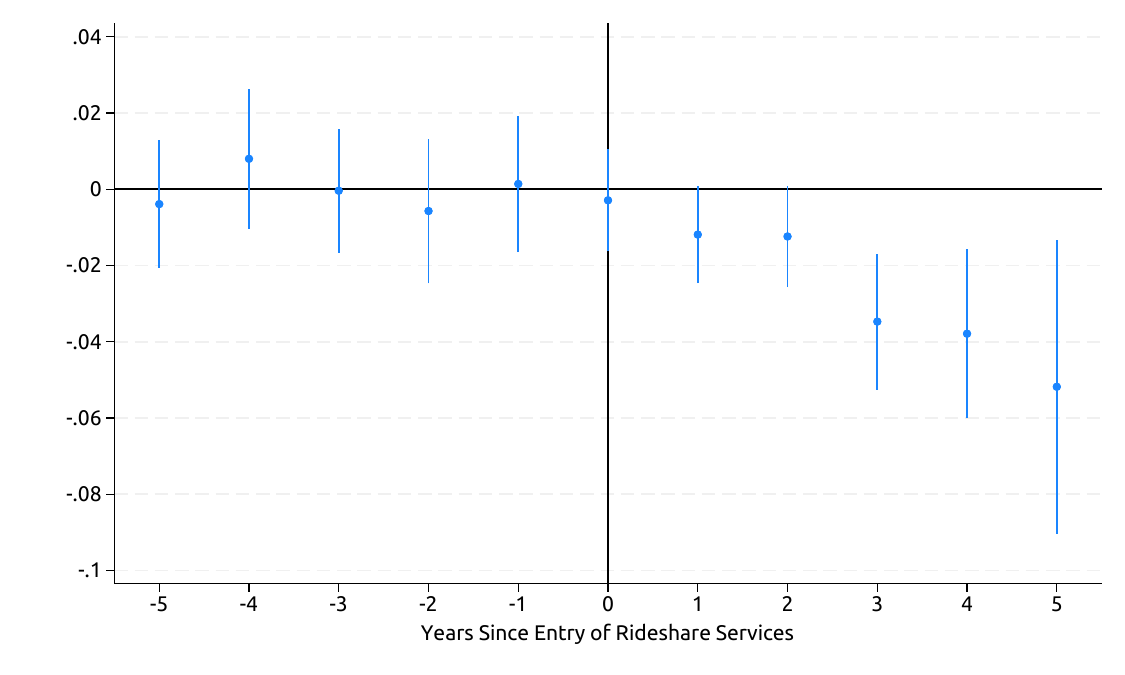}
			\end{minipage}\hfil
			\begin{minipage}{0.66\textwidth}
				\centering
				\caption*{(B) Suburban Zip-codes}
				\includegraphics[width=\linewidth]{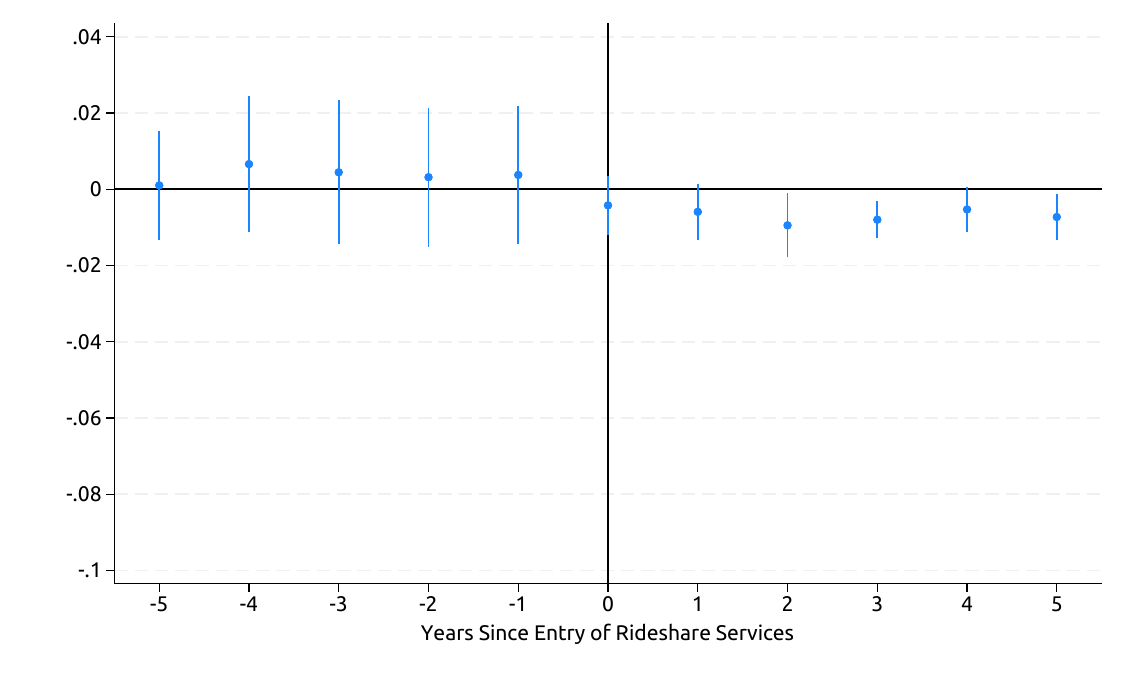}
			\end{minipage}
			\label{fig:homevalue_zip-codes}
                \hrule height 0.5pt 
		\end{figure}
	\end{landscape}
\end{scriptsize}
\clearpage

\begin{scriptsize}
	\begin{landscape}
		\begin{figure}[!h]%
			\caption{\textbf{Entry of Rideshare Services and House Prices}} 
			\footnotesize{This figure displays the evolution of yearly coefficients of a regression relating the entry of rideshare services with housing prices in urban and suburban zip-codes, in the left and right panels respectively. \textit{Urban (Suburban)} zip-codes are in the top (bottom) quartile of a measure of perceived urbanicity compiled by the Census’ \textit{Urbanization Perceptions Small Area Index}. Regressions include zip-code and state*year fixed effects. Standard errors are clustered at the zip-code level. The blue vertical bars represent confidence intervals of the coefficients at the 95 percent significance level. See Table \ref{tab: variable definitions} in the Appendix for detailed variable definitions.} \
          \vspace{2pt} 
            \hrule height 1pt 
            \vspace{1.5pt} 
            \hrule height 0.5pt 
            \vspace{3.5pt} 

			\normalsize
			\begin{minipage}{0.6\textwidth}
				\centering
				\caption*{(A) Urban Zip-codes}
				\includegraphics[width=\linewidth]{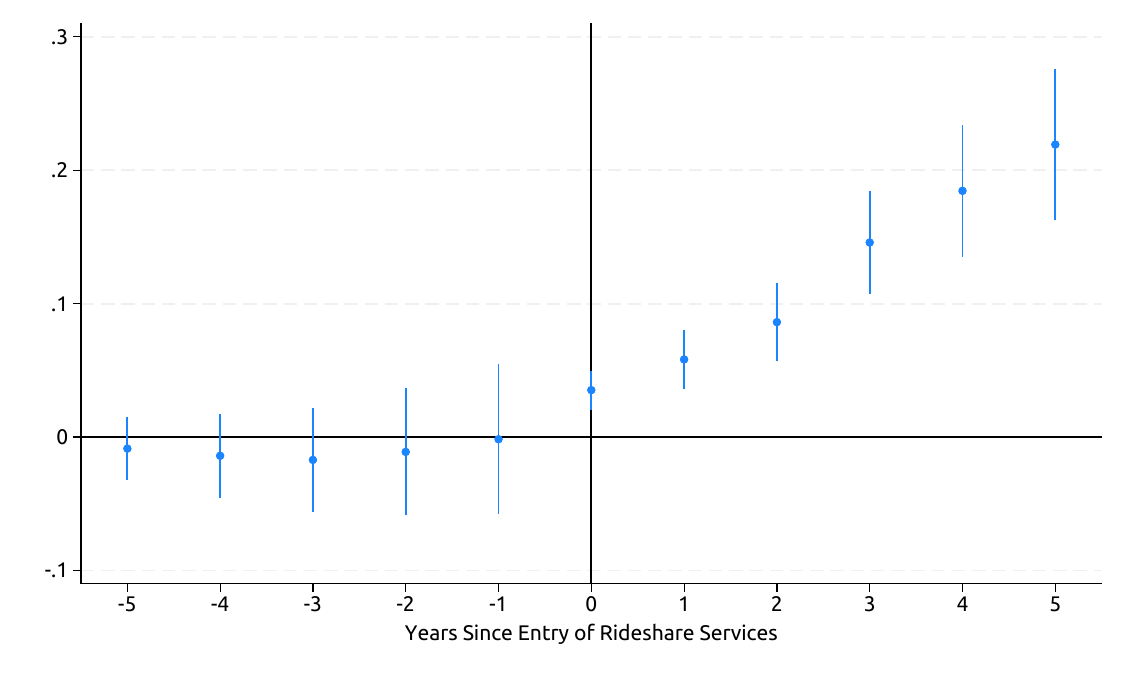}
			\end{minipage}\hfil
			\begin{minipage}{0.6\textwidth}
				\centering
				\caption*{(B) Suburban Zip-codes}
				\includegraphics[width=\linewidth]{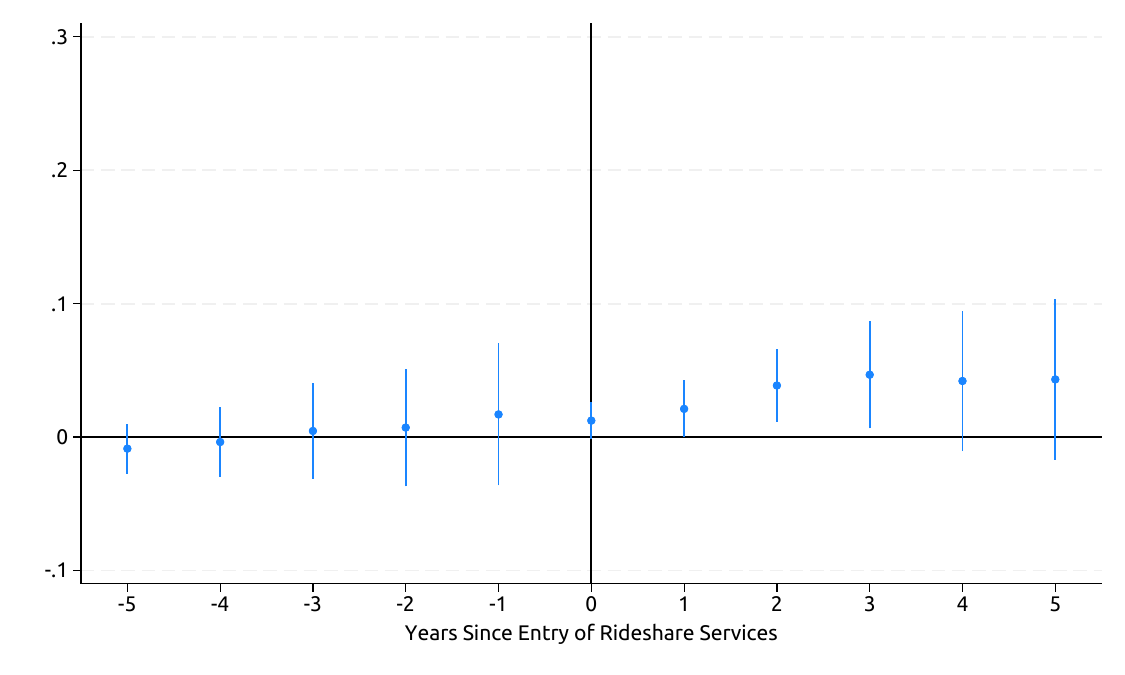}
			\end{minipage}
			\label{fig:houseprice}
        \hrule height 0.5pt 
        \end{figure}
	\end{landscape}
\end{scriptsize}
\clearpage

\begin{appendices}
\clearpage  	
\appendix

\renewcommand\thefigure{\thesection.\arabic{figure}}
\renewcommand\thetable{\thesection.\arabic{table}}
\setcounter{figure}{0}
\setcounter{table}{0}

\renewcommand{\thetable}{A\arabic{table}}
\renewcommand{\thefigure}{A\arabic{figure}}

\begin{center}
\section*{Appendix}
\label{appendix:analysis}
\end{center}




\begin{scriptsize}
        \singlespacing
        \begin{longtable}{lp{36em}}
		\caption[Variable Definitions]{\textbf{Variable Definitions}} \\
		\toprule
		\toprule
		Variable       & Definition \\
		\midrule
		\endfirsthead
		\caption[]{\textbf{Variable Definitions} (Continued)} \\
		\toprule
		\toprule
		Variable       & Definition \\
		\midrule
		\endhead
		\toprule
		\endfoot
		\bottomrule
		\endlastfoot
			Urban$_{z}$        & Measure of perceived urbanicity of zip-code \textit{ z} compiled by the\textit{ Census’ Urbanization Perceptions Small Area Index}.  \\
	MSA$_{m}$           & Geographical region \textit{m}, defined by the U.S. Office of Management and Budget, that represents a core urban area with a population of at least 50,000, along with adjacent communities that have a high degree of social and economic integration with the core, typically measure by commuting patterns. \\
	Entry$_{m,y} $      & Indicator of whether rideshare services start operations in an MSA \textit{m} in year \textit{y}. Data provided by rideshare companies \textit{Uber} and \textit{Lyft}. \\
	Number of Drivers$_{m,y}$ & Number of cab drivers (NAICS 4853) in MSA \textit{m} in year \textit{y} as reported by the \textit{Non-Employer Statistics} (NES) from the \textit{Census Bureau} (in logs). Dataset pools drivers of rideshare services with traditional "yellow-cab" drivers. \\
	Revenue Drivers$_{m,y}$ & Value of revenue obtained by cabdrivers in MSA \textit{m} in year \textit{y}, as reported by the \textit{NES} (in logs). \\
	Share of Number of Drivers$_{m,y}$ & Fraction of cabdrivers in MSA\textit{ m} over total workers in year \textit{y} as reported by the \textit{NES}. \\
	Share of Revenue of Drivers$_{m,y}$ & Fraction of revenue in MSA \textit{m} of cab-driver workers in year\textit{ y} as reported by the \textit{NES}. \\
	Trips Capita$_{z}$  & Number of rideshare services trips per capita in zip-code \textit{z} (in logs). Data—for Chicago, New York City, Massachusetts, and San Francisco—obtained from mandatory public reporting. \\
	Google Searches$_{z}$ & Frequency of per-capita searches of the terms "Uber", “Lyft” or "Rideshare services" in zip-code \textit{z} relative to total searches in the MSA. Data obtained from \textit{Google Trends} in the month following the entry of a rideshare services in the MSA. \\
	Number of Households$_{z,y}$ & Number of individual income tax returns (in logs) in zip-code \textit{z} and year \textit{y}. Data obtained from the \textit{SOI Tax Stats} from the \textit{IRS}.  \\
	Child Credit$_{z,y}$ & Share of individual income tax returns that claim a child credit in zip-code \textit{z} and year \textit{ y}. Data obtained from the \textit{SOI Tax Stats} from the \textit{IRS}.  \\
     Vacancy Rate$_{z,y}$ & Percentage of addresses vacant over total addresses in zip-code \textit{z} in year \textit{y}. Data obtained from \textit{HUD}. \\
    Number of Vacancies$_{z,y}$ & Number of addresses (in logs) in zip-code \textit{z} in year \textit{y} that \textit{USPS} delivery staff have identified as vacant for 90 days or longer. Data obtained from \textit{HUD}. \\
	Age$_{z,y}$         & Average age (in logs) of individuals with a credit report and social security residing in zip-code \textit{z} and year \textit{y}. Data obtained from the \textit{Equifax Consumer Credit Panel}. \\
	\#Mortgage Originations$_{z,y}$ & Number of mortgage originations (in logs) in zip-code \textit{z} in year \textit{y}. Data obtained from confidential \textit{HMDA}. \\
	Applicant Income$_{i,y}$ & Gross annual income (in logs) of applicant of mortgage \textit{i} in year \textit{y}. Data obtained from confidential \textit{HMDA}. \\
	FHA-Insured$_{i,y}$ & Indicator of whether the originated loan \textit{i} in year \textit{y} was insured by the Federal House Administration. Data obtained from confidential \textit{HMDA}. \\
	Mortgage Value$_{i,y}$ & Value (in logs) of mortgage \textit{i} originated in year \textit{y}. Data obtained from confidential \textit{HMDA}. \\
	House Price Index$_{z,y}$ & Home price index of zip-code\textit{ z} in year \textit{y} (in logs) calculated using weighted repeat sales. Data obtained from \textit{CoreLogic}. \\
	Rental Prices$_{z,y}$ & Average monthly market rent in zip-code\textit{ z} in year \textit{y} (in logs). Repeat-rent index weighted by the rental housing stock. Data obtained from \textit{Zillow}. \\
	Moving Out of Zip-code$_{i,y}$ & Indicator that individual \textit{i} will be living in a different zip-code in \textit{y+1} relative to the one living in \textit{y}. Data obtained from the \textit{Equifax Consumer Credit Panel}. \\
     Late Repayment$_{i,y}$ & Indicator that individual \textit{i} is late on a outstanding debt in year \textit{y}. Data obtained from the \textit{Equifax Consumer Credit Panel}. \\
	Median House Price in Zip-code$_{i,y}$ & Median house prices (in logs) in the zip-code of the property financed by mortgage \textit{i} originated in year \textit{y}. Data obtained from \textit{Zillow}\textit{ Home Value Index}. \\
	Average Poverty Rate in Zip-code$_{i,y}$ & Average poverty rate in the zip-code of the property financed by mortgage \textit{i} originated in year \textit{y}. Data obtained from \textit{Census Bureau}. \\
	Average Minority Share in Zip-code$_{i,y}$ & Fraction of non-white households in the zip-code of the property financed by mortgage \textit{ i} originated in year \textit{y}. Data obtained from \textit{Census Bureau}. \\
	Population$_{l,y}$  & Population in location \textit{l} in year \textit{y} (in logs). Data from \textit{Census Bureau}. \\
	Density$_{l,y}$     & Ratio of population to land area of location \textit{l} in year \textit{y}. Data from \textit{Census Bureau}. \\
	Car Ownership$_{z}$ & Percent of individuals living in zip-code \textit{z} who own a car. Data obtained from \textit{American Community Survey}. \\
	Commute Time$_{z }$ & Average commute time (in minutes) in zip-code \textit{z}. Data obtained from \textit{American Community Survey}. \\
	Parking Cost$_{z}$  & Average cost (in dollars) of monthly parking in zip-code \textit{z}. Data obtained from \textit{Spot Hero}. \\
	Walkscore$_{z}$     & Measure of how walkable a zip-code \textit{z} is on a scale from 0 (Car-Dependent) to 100 (Walker’s Paradise). Data obtained from \textit{Redfin}. \\
	Transitscore$_{z}$  & Measure of how well a location is served by public transit on a scale from 0 (“Minimal Transit”) to 100 (“Rider’s Paradise”). Data obtained from\textit{ Redfin}. \\
	Rideshare In State$_{m,y-1}$ & Indicator of whether rideshare services were operating in the state of MSA \textit{m} in year \textit{y-1}. Data obtained from \textit{Uber} and \textit{Lyft}. \\
	Income per Household$_{l,y}$ & Average income of households in location \textit{l} and year \textit{y} as declared in IRS return (in logs). \\
	Unemployment$_{m,y-1}$ & Average unemployment rate in MSA \textit{m} in year \textit{y-1}. Data obtained from \textit{IRS}. \
		\label{tab: variable definitions}
	\end{longtable}
\end{scriptsize}
\clearpage

\begin{scriptsize}
	\begin{table}[ht!]
		\caption{\textbf{Comparison of Urban and Suburban Zip-Codes}}
		\footnotesize 
		This table compares the average characteristics of \textit{urban} and \textit{suburban} zip-codes. \textit{Urban (Suburban)} zip-codes are in the top (bottom) quartile of a measure of perceived urbanicity compiled by the Census’ \textit{Urbanization Perceptions Small Area Index}. Standard errors are in parenthesis. See Table \ref{tab: variable definitions} in the Appendix for detailed variable definitions.\\
		\setlength{\tabcolsep}{52.5pt}
		\begin{tabular}{lcc}
	\toprule
	\toprule
	\multicolumn{1}{l}{} & \multicolumn{1}{c}{\textbf{Urban}} & \multicolumn{1}{c}{\textbf{Suburban}} \\
	\midrule
	\multicolumn{1}{r}{} &                &  \\
	Population$_{z}$(‘000) & \multicolumn{1}{c}{26.7} & \multicolumn{1}{c}{16.3} \\
	\multicolumn{1}{r}{} & \multicolumn{1}{c}{(19.0)} & \multicolumn{1}{c}{(12.2)} \\
	Density$_{z}$       & \multicolumn{1}{c}{3,906} & \multicolumn{1}{c}{427} \\
	\multicolumn{1}{r}{} & \multicolumn{1}{c}{(5,940)} & \multicolumn{1}{c}{(630)} \\
	Car Ownership$_{z}$ & \multicolumn{1}{c}{78.9} & \multicolumn{1}{c}{95.4} \\
	\multicolumn{1}{r}{} & \multicolumn{1}{c}{(13.9)} & \multicolumn{1}{c}{(3.4)} \\
	Commute Time$_{z}$  & \multicolumn{1}{c}{24.3} & \multicolumn{1}{c}{28.0} \\
	\multicolumn{1}{r}{} & \multicolumn{1}{c}{(8.1)} & \multicolumn{1}{c}{(7.3)} \\
	Parking Cost$_{z}$  & \multicolumn{1}{c}{180.5} & \multicolumn{1}{c}{124.4} \\
	\multicolumn{1}{r}{} & \multicolumn{1}{c}{(99.6)} & \multicolumn{1}{c}{(33.1)} \\
	Walkscore$_{z}$     & \multicolumn{1}{c}{62.4} & \multicolumn{1}{c}{18.9} \\
	\multicolumn{1}{r}{} & \multicolumn{1}{c}{(22.0)} & \multicolumn{1}{c}{(15.4)} \\
	Transitscore$_{z}$  & \multicolumn{1}{c}{36.6} & \multicolumn{1}{c}{16.1} \\
	\multicolumn{1}{r}{} & \multicolumn{1}{c}{(24.2)} & \multicolumn{1}{c}{(20.0)} \\
	Number of Zip-Codes & 1,952           & 1,833 \\
	Share of Population & 63.6           & 36.4 \\
	\bottomrule
\end{tabular}%
		\label{tab:urban_suburban}
	\end{table}
\end{scriptsize}
\clearpage

\begin{scriptsize}
	\begin{table}[ht!]
		\caption{\textbf{Impact of Rideshare Services on Supply of Drivers}}
		\footnotesize 
		This table displays the impact of entry of rideshare services on the supply and revenue of drivers at the MSA level. \textit{Number Drivers$_{m,y}$}, is the number of the drivers (in logs) in MSA \textit{m} in year \textit{y}. \textit{Revenue Drivers$_{m,y}$} is the value of total revenue (in logs) obtained from cab services in MSA \textit{m} in year \textit{y}. \textit{Share Number Drivers$_{m,y}$} is the share of drivers over total workers in MSA \textit{m} and year \textit{y}.\textit{ Share Revenue Drivers$_{m,y}$} is the share of revenue received by drivers over revenue of total workers in MSA \textit{m} and year \textit{y}. Data on all dependent variables from the \textit{Non-Employer Statistics} compiled by the \textit{Census Bureau}. *, **, and *** denote statistical significance at the 10\%, 5\%, and 1\% levels, respectively. See Table \ref{tab: variable definitions} in the Appendix for detailed variable definitions.\\
		\setlength{\tabcolsep}{19.5pt}
		\begin{tabular}{lcccc}
	\toprule
	\toprule
	& Number  & Revenue  & Share Number  & Share Revenue \\
	& Drivers$_{m,y}$ &  Drivers$_{m,y}$ & Drivers$_{m,y}$ & Drivers$_{m,y}$ \\
	& (1)            & (2)            & (3)            & (4) \\
	\midrule
	&                &                &                &  \\
	PreEntry 2-Years & -0.015         & 0.029          & -0.000         & 0.000 \\
	& (0.061)        & (0.086)        & (0.000)        & (0.000) \\
	PreEntry 1-Years & 0.040          & 0.103          & 0.000          & 0.000 \\
	& (0.069)        & (0.096)        & (0.000)        & (0.000) \\
	PostEntry 0-Years & 0.339***       & 0.172***       & 0.002***       & 0.000*** \\
	& (0.029)        & (0.038)        & (0.000)        & (0.000) \\
	PostEntry 1-Years & 0.634***       & 0.401***       & 0.005***       & 0.001*** \\
	& (0.038)        & (0.043)        & (0.000)        & (0.000) \\
	PostEntry 2-Years & 0.747***       & 0.530***       & 0.008***       & 0.002*** \\
	& (0.042)        & (0.047)        & (0.001)        & (0.000) \\
	PostEntry 3-Years & 0.754***       & 0.554***       & 0.011***       & 0.002*** \\
	& (0.044)        & (0.054)        & (0.001)        & (0.000) \\
	PostEntry 4-Years & 0.782***       & 0.656***       & 0.014***       & 0.004*** \\
	& (0.049)        & (0.057)        & (0.001)        & (0.000) \\
	&                &                &                &  \\
	Observations   & 5,072          & 5,072          & 5,072          & 5,072 \\
	MSA FE         & Yes            & Yes            & Yes            & Yes \\
	State-Year FE  & Yes            & Yes            & Yes            & Yes \\
 	Avg. Dep. Variable  & 4.4            & 7.1            & 0.004            & 0.011 \\
 	\bottomrule
\end{tabular}%
		\label{tab:driversupply}
	\end{table}
\end{scriptsize}
\clearpage

\begin{scriptsize}
	\begin{table}[ht!]
		\caption{\textbf{Characteristics of MSA Prior to Entry of Rideshare Services}}
		\footnotesize 
		This table relates the characteristics of an MSA in a given year with the entry of rideshare services in the MSA. The dependent variable \textit{Entry$_{m,y}$} is a yearly indicator of whether rideshare services entered MSA \textit{m} in year \textit{y}. Observations of an MSA are dropped for all years after entry. \textit{Population$_{m,y-1}$} is the population (in logs) living in MSA \textit{m} in year \textit{y-1}.\textit{Density$_{m,y-1}$} is measured as population of MSA \textit{m} in year \textit{y-1} over the area of the MSA.\textit{ Rideshare In-State$_{m,y-1}$} is an indicator of whether rideshare services have already entered the state of the MSA \textit{m} by year \textit{y-1}. Income per \textit{Household$_{m,y}$} are the average earnings (in logs) of households living in MSA \textit{m} in year \textit{y-1}. \textit{Unemployment$_{m,y-1}$ }is the unemployment rate of MSA \textit{m} in year \textit{y-1}. Data on population, income per household, and unemployment obtained from IRS’s individual tax returns. Data on density obtained from the Census Bureau. *, **, and *** denote statistical significance at the 10\%, 5\%, and 1\% levels, respectively. Standard errors clustered at the MSA level.  See Table \ref{tab: variable definitions} in the Appendix for detailed variable definitions.\\
		\setlength{\tabcolsep}{25pt}
		\begin{tabular}{lcccc}
	\toprule
	\toprule
	\multicolumn{1}{r}{} & \multicolumn{4}{c}{Entry$_{m,y}$} \\
	\multicolumn{1}{r}{} & (1)            & (2)            & (3)            & (4) \\
	\midrule
	Population$_{m,y-1}$  & \multicolumn{1}{c}{0.132**} & \multicolumn{1}{c}{0.132**} & \multicolumn{1}{c}{0.099**} & \multicolumn{1}{c}{0.093**} \\
	& \multicolumn{1}{c}{(0.055)} & \multicolumn{1}{c}{(0.054)} & \multicolumn{1}{c}{(0.037)} & \multicolumn{1}{c}{(0.035)} \\
	Rideshare In-State$_{m,y-1}$ &       & \multicolumn{1}{c}{0.105**} & \multicolumn{1}{c}{0.102**} & \multicolumn{1}{c}{0.101**} \\
	&       & \multicolumn{1}{c}{(0.033)} & \multicolumn{1}{c}{(0.033)} & \multicolumn{1}{c}{(0.033)} \\
	Density$_{m,y-1}$ &       &       & \multicolumn{1}{c}{0.132} & \multicolumn{1}{c}{0.135} \\
	&       &       & \multicolumn{1}{c}{(0.083)} & \multicolumn{1}{c}{(0.084)} \\
	Income per Household$_{m,y-1}$  &       &       &       & \multicolumn{1}{c}{0.066} \\
	&       &       &       & \multicolumn{1}{c}{(0.075)} \\
	Unemployment$_{m,y-1}$  &       &       &       & \multicolumn{1}{c}{-0.974} \\
	&       &       &       & \multicolumn{1}{c}{(0.679)} \\
	&       &       &       &  \\
	Observations   & \multicolumn{1}{c}{2,185} & \multicolumn{1}{c}{2,185} & \multicolumn{1}{c}{2,185} & \multicolumn{1}{c}{2,185} \\
	R-squared      & \multicolumn{1}{c}{0.291} & \multicolumn{1}{c}{0.298} & \multicolumn{1}{c}{0.301} & \multicolumn{1}{c}{0.301} \\
	Year FE        & Yes            & Yes            & Yes            & Yes \\
	\bottomrule
\end{tabular}%
		\label{tab:entry}
	\end{table}
\end{scriptsize}
\clearpage

\begin{scriptsize}
	\begin{table}[ht!]
		\caption{\textbf{Entry of Rideshare Services and Demographic Characteristics of Households – Pooled Zip-Codes}}
		\footnotesize 
		This table displays the results of a regression relating the entry of rideshare services with the number of high- and low-income households, average age and number of child-credit deductions in a zip-code. Regressions pool urban and suburban zip-codes together. \textit{Urban (Suburban)} zip-codes are in the top (bottom) quartile of a measure of perceived urbanicity compiled by the \textit{Census’ Urbanization Perceptions Small Area Index}. \textit{Urban} zip-codes are the treatment group for the entry of rideshare services in an MSA. \textit{Number of High-Income (Low-Income) Households$_{z,y}$} is the number of individual tax returns (in logs) in zip-code z in year y with a household income equal or above (below) USD 50,000. \textit{Age$_{z,y}$} is the average age (in logs) of individuals with credit report in zip-code \textit{z} in year \textit{y}, compiled from \textit{Equifax Consumer Credit Panel}. \textit{Child Credit$_{z,y}$} is the percentage of tax returns that claim a child credit in zip-code \textit{z} in year \textit{y}, obtained from the \textit{IRS}. *, **, and *** denote statistical significance at the 10\%, 5\%, and 1\% levels, respectively. See Table \ref{tab: variable definitions} in the Appendix for detailed variable definitions.\\
		\setlength{\tabcolsep}{11pt}
		\begin{tabular}{lcccc}
	\toprule
	\toprule
	& Number of High- & Number of Low-& Age$_{z,y}$         & Child Credit$_{z,y}$ \\
	&Income Households$_{z,y}$ & Income Households$_{z,y}$ &       & \\
	& (1)            & (2)            & (3)            & (4) \\
	\midrule
	&                &                &                &  \\
	PreEntry 2-Years & 0.012          & -0.016      & -0.001         & 0.000 \\
	& (0.009)        & (0.007)        & (0.015)        & (0.001) \\
	PreEntry 1-Years & 0.010          & -0.027      & -0.001         & 0.001 \\
	& (0.010)        & (0.007)        & (0.017)        & (0.001) \\
	PostEntry 0-Years & 0.023***       & -0.041***      & -0.006***      & 0.001 \\
	& (0.005)        & (0.006)        & (0.001)        & (0.001) \\
	PostEntry 1-Years & 0.033***       & -0.056***      & -0.007***      & -0.002* \\
	& (0.005)        & (0.007)        & (0.001)        & (0.001) \\
	PostEntry 2-Years & 0.040***       & -0.076***      & -0.010***      & -0.001 \\
	& (0.006)        & (0.008)        & (0.001)        & (0.001) \\
	PostEntry 3-Years & 0.060***       & -0.094***      & -0.011***      & -0.004*** \\
	& (0.007)        & (0.009)        & (0.001)        & (0.001) \\
	PostEntry 4-Years & 0.078***       & -0.115***      & -0.011***      & -0.008*** \\
	& (0.008)        & (0.011)        & (0.001)        & (0.002) \\
	&                &                &                &  \\
	Observations   & 52,198          & 52,198          & 36,384          & 43,769 \\
	Zip-code FE     & Yes            & Yes            & Yes            & Yes \\
	MSA-Year FE    & Yes            & Yes            & Yes            & Yes \\
	Average Dep. Variable & 8              & 8.5            & 4.3            & 0.15 \\
	\bottomrule
\end{tabular}%
		\label{tab:table_demography_pooled}
	\end{table}
\end{scriptsize}
\clearpage

\begin{scriptsize}
	\begin{table}[ht!]
		\caption{\textbf{Entry of Rideshare Services, Mortgage Originations and Characteristics of Applicants – Pooled Zip-Codes}}
		\footnotesize 
		This table displays the results of regressions relating the entry of rideshare services with the number and characteristics of mortgage applicants. Regressions pool urban and suburban zip-codes together. \textit{Urban (Suburban)} zip-codes are in the top (bottom) quartile of a measure of perceived urbanicity compiled by the Census’ \textit{Urbanization Perceptions Small Area Index}. \textit{Urban} zip-codes are the treatment group for the entry of rideshare services in an MSA. \#\textit{Mortgage Originations$_{z,y}$} is the number of mortgage originations (in logs) in zip-code \textit{z} in year \textit{y}. \textit{Applicant Income$_{i,y}$} is the gross annual income of applicant \textit{i} in year \textit{y} (in logs). \textit{FHA-Insured$_{i,y}$} is an indicator that mortgage \textit{i} originated in year \textit{y} is insured by the Federal Hosing Administration.  Dependent variables obtained from \textit{HMDA confidential}. *, **, and *** denote statistical significance at the 10\%, 5\%, and 1\% levels, respectively. See Table \ref{tab: variable definitions} in the Appendix for detailed variable definitions.\\
		\setlength{\tabcolsep}{17.0pt}
		\begin{tabular}{lccc}
		\toprule
		\toprule
		\multicolumn{1}{r}{} & \multicolumn{1}{c}{\#Mortgage Originations$_{z,y}$} & \multicolumn{1}{c}{Applicant Income$_{i,y}$} & \multicolumn{1}{c}{FHA-Insured$_{i,y}$} \\
		\multicolumn{1}{r}{} & \multicolumn{1}{c}{(1)} & \multicolumn{1}{c}{(2)} & \multicolumn{1}{c}{(3)} \\
		\midrule
		\multicolumn{1}{r}{} &       &       &  \\
		PreEntry 2-Years & \multicolumn{1}{c}{-0.047} & \multicolumn{1}{c}{0.004} & \multicolumn{1}{c}{0.005} \\
		\multicolumn{1}{r}{} & \multicolumn{1}{c}{(0.035)} & \multicolumn{1}{c}{(0.009)} & \multicolumn{1}{c}{(0.005)} \\
		PreEntry 1-Years & \multicolumn{1}{c}{-0.055} & \multicolumn{1}{c}{-0.001} & \multicolumn{1}{c}{0.003} \\
		\multicolumn{1}{r}{} & \multicolumn{1}{c}{(0.035)} & \multicolumn{1}{c}{(0.011)} & \multicolumn{1}{c}{(0.006)} \\
		PostEntry 0-Years & \multicolumn{1}{c}{0.005} & \multicolumn{1}{c}{0.002} & \multicolumn{1}{c}{-0.005} \\
		\multicolumn{1}{r}{} & \multicolumn{1}{c}{(0.012)} & \multicolumn{1}{c}{(0.006)} & \multicolumn{1}{c}{(0.004)} \\
		PostEntry 1-Years & \multicolumn{1}{c}{0.034**} & \multicolumn{1}{c}{0.011*} & \multicolumn{1}{c}{-0.014***} \\
		\multicolumn{1}{r}{} & \multicolumn{1}{c}{(0.016)} & \multicolumn{1}{c}{(0.006)} & \multicolumn{1}{c}{(0.005)} \\
		PostEntry 2-Years & \multicolumn{1}{c}{0.074***} & \multicolumn{1}{c}{0.022***} & \multicolumn{1}{c}{-0.018***} \\
		\multicolumn{1}{r}{} & \multicolumn{1}{c}{(0.022)} & \multicolumn{1}{c}{(0.006)} & \multicolumn{1}{c}{(0.005)} \\
		PostEntry 3-Years & \multicolumn{1}{c}{0.130***} & \multicolumn{1}{c}{0.036***} & \multicolumn{1}{c}{-0.029***} \\
		\multicolumn{1}{r}{} & \multicolumn{1}{c}{(0.027)} & \multicolumn{1}{c}{(0.009)} & \multicolumn{1}{c}{(0.005)} \\
		PostEntry 4-Years & \multicolumn{1}{c}{0.190***} & \multicolumn{1}{c}{0.040***} & \multicolumn{1}{c}{-0.036***} \\
		\multicolumn{1}{r}{} & \multicolumn{1}{c}{(0.032)} & \multicolumn{1}{c}{(0.013)} & \multicolumn{1}{c}{(0.005)} \\
		\multicolumn{1}{r}{} &       &       &  \\
		Observations & 32,921 & 7,804,222 & 7,804,222 \\
		Zip-code FE & \multicolumn{1}{c}{Yes} & \multicolumn{1}{c}{Yes} & \multicolumn{1}{c}{Yes} \\
		MSA-Year FE & \multicolumn{1}{c}{Yes} & \multicolumn{1}{c}{Yes} & \multicolumn{1}{c}{Yes} \\
		Average Dep. Variable & 5     & 4.5   & 0.19 \\
		
		\bottomrule
	\end{tabular}%

		\label{tab:table_mortgage_origination_pooled}
	\end{table}
\end{scriptsize}
\clearpage

\begin{scriptsize}
	\begin{table}[ht!]
		\caption[Entry of Rideshare Services and House Vacancies]{\textbf{Entry of Rideshare Services and Housing Vacancies}}
		\footnotesize 
		This table displays the results of a regression relating the entry of rideshare services with housing vacancies, in urban and suburban zip-codes. \textit{Urban (Suburban)} zip-codes are in the top (bottom) quartile of a measure of perceived urbanicity compiled by the Census’ \textit{Urbanization Perceptions Small Area Index}. \textit{Vacancy Rate$_{z,y}$} is the percentage of vacant addresses of a zip-code \textit{z} in year \textit{y}. \textit{Number of Vacancies$_{z,y}$} is the number (in logs) of vacant addresses of zip-code code \textit{z} in year \textit{y}, compiled by the USPS. *, **, and *** denote statistical significance at the 10\%, 5\%, and 1\% levels, respectively.  See Table \ref{tab: variable definitions} in the Appendix for detailed variable definitions.\\
		\setlength{\tabcolsep}{22pt}
		\begin{tabular}{lcccc}
	\toprule
	\toprule
	& \multicolumn{2}{c}{Vacancy Rate$_{z,y}$} & \multicolumn{2}{c}{Number of Vacancies$_{z,y}$} \\
	& (1)            & (2)            & (3)            & (4) \\
	\midrule
	&                &                &                &  \\
	PreEntry 2-Years & -0.047         & -0.068         & -0.002         & -0.003 \\
	& (0.126)        & (0.103)        & (0.020)        & (0.069) \\
	PreEntry 1-Years & -0.171         & -0.084         & -0.039         & -0.026 \\
	& (0.159)        & (0.118)        & (0.026)        & (0.086) \\
	PostEntry 0-Years & -0.180***      & 0.052          & -0.048***      & -0.015 \\
	& (0.046)        & (0.045)        & (0.014)        & (0.029) \\
	PostEntry 1-Years & -0.422***      & -0.068         & -0.145***      & -0.116*** \\
	& (0.085)        & (0.073)        & (0.025)        & (0.041) \\
	PostEntry 2-Years & -0.430***      & -0.144*        & -0.112***      & -0.143*** \\
	& (0.123)        & (0.080)        & (0.027)        & (0.054) \\
	PostEntry 3-Years & -0.632***      & -0.089         & -0.156***      & 0.002 \\
	& (0.142)        & (0.101)        & (0.028)        & (0.057) \\
	PostEntry 4-Years & -0.762***      & -0.131         & -0.193***      & -0.075 \\
	& (0.162)        & (0.118)        & (0.029)        & (0.086) \\
	&                &                &                &  \\
	Observations   & 11,515          & 10,093          & 11,515          & 10,093 \\
	Zip-code FE     & Yes            & Yes            & Yes            & Yes \\
	State-Year FE  & Yes            & Yes            & Yes            & Yes \\
	Sample of Zip-codes & Urban          & Suburban       & Urban          & Suburban \\
	Avg(Dependent Variable) & 4.3            & 1.4            & 9.2            & 8.7 \\
	\bottomrule
\end{tabular}%
		\label{tab:vacancies2}
	\end{table}
\end{scriptsize}
\clearpage

\begin{scriptsize}
	\begin{table}[ht!]
		\caption{\textbf{Entry of Rideshare Services and Housing Vacancies – Pooled Zip-Codes}}
		\footnotesize 
		This table displays the results of a regression relating the entry of rideshare services with housing vacancies, pooling urban and suburban zip-codes together. \textit{Urban (Suburban)} zip-codes are in the top (bottom) quartile of a measure of perceived urbanicity compiled by the \textit{Census’ Urbanization Perceptions Small Area Index}. \textit{Urban} zip-codes are the treatment group for the entry of rideshare services in an MSA. \textit{Vacancy Rate$_{z,y}$} is the percentage of vacant addresses of zip-code \textit{z} in year \textit{y}. \textit{Number of Vacancies$_{z,y}$} is the number (in logs) of vacant addresses of zip-code code \textit{z} in year \textit{y}, compiled by the USPS. *, **, and *** denote statistical significance at the 10\%, 5\%, and 1\% levels, respectively.  See Table \ref{tab: variable definitions} in the Appendix for detailed variable definitions.\\
		\setlength{\tabcolsep}{37pt}
		\begin{tabular}{lcc}
    \toprule
    \toprule
    & Vacancy Rate$_{z,y}$ & Number of Vacancies$_{z,y}$ \\
    & (1)            & (2)            \\
    \midrule
    &                &                 \\
    PreEntry 2-Years&-0.159&0.001 \\
    &(0.102)&(0.009) \\
    PreEntry 1-Years&-0.161&-0.003 \\
    &(0.113)&(0.009) \\
    PostEntry 0-Years&-0.303***&-0.011** \\
    &(0.057)&(0.005) \\
    PostEntry 1-Years&-0.823***&-0.029*** \\
    &(0.073)&(0.005) \\
    PostEntry 2-Years&-0.685***&-0.035*** \\
    &(0.083)&(0.007) \\
    PostEntry 3-Years&-0.626***&-0.041*** \\
    &(0.108)&(0.008) \\
    PostEntry 4-Years&-0.583***&-0.054*** \\
    &(0.142)&(0.009) \\
    && \\
    Observations&30,813&30,813 \\
    Zipcode FE&Yes&Yes \\
    MSA-Year FE&Yes&Yes \\
    Average Dep. Variable&2.9&8.9 \\
    \bottomrule
\end{tabular}%

		\label{tab:table_vacancies_pooled}
	\end{table}
\end{scriptsize}
\clearpage

\begin{scriptsize}
	\begin{landscape}
		\begin{table}[ht!]
			\caption{\textbf{Entry of Rideshare Services and Zip-codes of Mortgage Originations – Pooled Zip-Codes}}
			\footnotesize 
			This table displays the results of regressions relating the entry of rideshare services with the characteristics of zip-codes of the properties for which mortgages were originated. Regressions pool urban and suburban zip-codes together. \textit{Urban (Suburban)} zip-codes are in the top (bottom) quartile of a measure of perceived urbanicity compiled by the Census’ \textit{Urbanization Perceptions Small Area Index}. \textit{Urban} zip-codes are the treatment group for the entry of rideshare services in an MSA. \textit{Median House Price in Zip-code$_{i,y}$} is the median house price (in logs) in the zip-code of the property financed by mortgage \textit{i} originated in year \textit{y}. \textit{Average Poverty Rate in Zip-code$_{i,y}$} is the average poverty rate in the zip-code of the property financed by mortgage \textit{i} originated in year \textit{y}. \textit{Average Minority Share in Zip-code$_{i,y}$} is the average share of minority individuals of the zip-code of the property financed by mortgage \textit{i} originated in year \textit{y}. Mortgage information obtained from \textit{HMDA confidential}, while information on the dependent variables obtained from the \textit{Census Bureau}. *, **, and *** denote statistical significance at the 10\%, 5\%, and 1\% levels, respectively. See Table \ref{tab: variable definitions} in the Appendix for detailed variable definitions.\\
			\setlength{\tabcolsep}{18pt}
			\begin{tabular}{lccc}
	\toprule
	\toprule
	\multicolumn{1}{r}{} & \multicolumn{1}{c}{Median House Price in Zip-code$_{i,y}$} & \multicolumn{1}{c}{Average Poverty Rate in Zip-code$_{i,y}$} & \multicolumn{1}{c}{Average Minority Share in Zip-code$_{i,y}$} \\
	\multicolumn{1}{r}{} & \multicolumn{1}{c}{(1)} & \multicolumn{1}{c}{(2)} & \multicolumn{1}{c}{(3)} \\
	\midrule
	\multicolumn{1}{r}{} &                &                &  \\
	PreEntry 2-Years & \multicolumn{1}{c}{0.007} & \multicolumn{1}{c}{-0.001} & \multicolumn{1}{c}{-0.002} \\
	\multicolumn{1}{r}{} & \multicolumn{1}{c}{(0.005)} & \multicolumn{1}{c}{(0.001)} & \multicolumn{1}{c}{(0.002)} \\
	PostEntry 1-Years & \multicolumn{1}{c}{0.001} & \multicolumn{1}{c}{-0.001} & \multicolumn{1}{c}{-0.001} \\
	\multicolumn{1}{r}{} & \multicolumn{1}{c}{(0.005)} & \multicolumn{1}{c}{(0.001)} & \multicolumn{1}{c}{(0.002)} \\
	PostEntry 0-Years & \multicolumn{1}{c}{-0.007*} & \multicolumn{1}{c}{-0.001} & \multicolumn{1}{c}{0.000} \\
	\multicolumn{1}{r}{} & \multicolumn{1}{c}{(0.004)} & \multicolumn{1}{c}{(0.000)} & \multicolumn{1}{c}{(0.001)} \\
	PostEntry 1-Years & \multicolumn{1}{c}{-0.002} & \multicolumn{1}{c}{0.000} & \multicolumn{1}{c}{0.003**} \\
	\multicolumn{1}{r}{} & \multicolumn{1}{c}{(0.006)} & \multicolumn{1}{c}{(0.001)} & \multicolumn{1}{c}{(0.001)} \\
	PostEntry 2-Years & \multicolumn{1}{c}{-0.010} & \multicolumn{1}{c}{0.002**} & \multicolumn{1}{c}{0.008***} \\
	\multicolumn{1}{r}{} & \multicolumn{1}{c}{(0.008)} & \multicolumn{1}{c}{(0.001)} & \multicolumn{1}{c}{(0.002)} \\
	PostEntry 3-Years & \multicolumn{1}{c}{-0.018**} & \multicolumn{1}{c}{0.003***} & \multicolumn{1}{c}{0.011***} \\
	\multicolumn{1}{r}{} & \multicolumn{1}{c}{(0.009)} & \multicolumn{1}{c}{(0.001)} & \multicolumn{1}{c}{(0.002)} \\
	PostEntry 4-Years & \multicolumn{1}{c}{-0.032***} & \multicolumn{1}{c}{0.006***} & \multicolumn{1}{c}{0.015***} \\
	\multicolumn{1}{r}{} & \multicolumn{1}{c}{(0.009)} & \multicolumn{1}{c}{(0.001)} & \multicolumn{1}{c}{(0.003)} \\
	\multicolumn{1}{r}{} &                &                &  \\
	Observations   & 7,804,222      & 7,804,222      & 7,804,222 \\
	MSA-Urban FE   & \multicolumn{1}{c}{Yes} & \multicolumn{1}{c}{Yes} & \multicolumn{1}{c}{Yes} \\
	MSA-Year FE    & \multicolumn{1}{c}{Yes} & \multicolumn{1}{c}{Yes} & \multicolumn{1}{c}{Yes} \\
	Avg. Dep. Variable & 12.4           & 0.09           & 0.27 \\
	\bottomrule
\end{tabular}%
			\label{tab:tableA7}
		\end{table}
	\end{landscape}
\end{scriptsize}
\clearpage

\begin{scriptsize}
	\begin{table}[ht!]
		\caption{\textbf{Entry of Rideshare Services, House Prices and Rental Prices – Pooled Zip-Codes}}
		\footnotesize 
		This table displays the results of a regression relating the entry of rideshare services with housing prices and rental prices, pooling urban and suburban zip-codes together. \textit{Urban (Suburban)} zip-codes are in the top (bottom) quartile of a measure of perceived urbanicity compiled by the \textit{Census’ Urbanization Perceptions Small Area Index}. \textit{Urban} zip-codes are the treatment group for the entry of rideshare services in an MSA. \textit{House Price Index$_{z,y}$} is an index of house prices (in logs) at the zip-code \textit{z} year \textit{y}, obtained from \textit{Corelogic}. \textit{Rental Prices$_{z,y}$} is the average market rental prices (in logs) of zip-code \textit{z} in year \textit{y}, obtained from \textit{Zillow}. *, **, and *** denote statistical significance at the 10\%, 5\%, and 1\% levels, respectively.  See Table \ref{tab: variable definitions} in the Appendix for detailed variable definitions.\\
		\setlength{\tabcolsep}{39pt}
		\begin{tabular}{lcc}
    \toprule
    \toprule
          & House Price Index$_{z,y}$ & Rental Prices$_{z,y}$ \\
          & (1)   & (2) \\
    \midrule
         &       &  \\
    PreEntry 2-Years & -0.020 & 0.001 \\
          & (0.021) & (0.004) \\
    PreEntry 1-Years & -0.014 & 0.007 \\
          & (0.021) & (0.006) \\
    PostEntry 0-Years & 0.008*** & 0.007* \\
          & (0.002) & (0.004) \\
    PostEntry 1-Years & 0.013*** & 0.014*** \\
          & (0.003) & (0.004) \\
    PostEntry 2-Years & 0.023*** & 0.022*** \\
          & (0.003) & (0.005) \\
    PostEntry 3-Years & 0.035*** & 0.024*** \\
          & (0.004) & (0.006) \\
    PostEntry 4-Years & 0.045*** & 0.029*** \\
          & (0.004) & (0.007) \\
          &       &  \\
    Observations & 31,210 & 33,254 \\
    Zipcode FE & Yes   & Yes \\
    MSA-Year FE & Yes   & Yes \\
    Average Dep. Variable & 5.0   & 7.4 \\
    \bottomrule
\end{tabular}%

		\label{tab:table_houseprices_pool}
	\end{table}
\end{scriptsize}
\clearpage

\begin{scriptsize}
	\begin{table}[ht!]
		\caption{\textbf{Entry of Rideshare Services and Changes in Housing Prices given Vacancy Rates}}
		\footnotesize 
	This table displays the results of a regression relating the entry of rideshare services on housing prices in urban areas given housing vacancy rate in the zip-code. Sample is limited to urban zip-codes (i.e., zip-codes in the top quartile of a measure of perceived urbanicity). Zip-codes are grouped into terciles according to their vacancy rate, measured as the share of vacant addresses of zip-code. *, **, and *** denote statistical significance at the 10\%, 5\%, and 1\% levels, respectively.  See Table \ref{tab: variable definitions} in the Appendix for detailed variable definitions.\\
		\setlength{\tabcolsep}{26.5pt}
		\begin{tabular}{lccc}
	\toprule
	\toprule
	& \multicolumn{3}{c}{House Price Index$_{z,y}$} \\
	& (1)            & (2)            & (3)          \\
	\midrule
	&                &                &                  \\
	PreEntry 2-Years & \multicolumn{1}{c}{-0.004} & \multicolumn{1}{c}{-0.003} & \multicolumn{1}{c}{-0.003} \\
	\multicolumn{1}{r}{} & \multicolumn{1}{c}{(0.009)} & \multicolumn{1}{c}{(0.009)} & \multicolumn{1}{c}{(0.007)} \\
	PreEntry 1-Years & \multicolumn{1}{c}{0.015} & \multicolumn{1}{c}{-0.002} & \multicolumn{1}{c}{0.000} \\
	\multicolumn{1}{r}{} & \multicolumn{1}{c}{(0.012)} & \multicolumn{1}{c}{(0.012)} & \multicolumn{1}{c}{(0.008)} \\
	PostEntry 0-Years & 0.031*** & 0.016*** & \multicolumn{1}{c}{0.012} \\
	\multicolumn{1}{r}{} & \multicolumn{1}{c}{(0.007)} & \multicolumn{1}{c}{(0.005)} & \multicolumn{1}{c}{(0.007)} \\
	PostEntry 1-Years & 0.048*** & 0.025*** & \multicolumn{1}{c}{0.012} \\
	\multicolumn{1}{r}{} & \multicolumn{1}{c}{(0.014)} & \multicolumn{1}{c}{(0.008)} & \multicolumn{1}{c}{(0.010)} \\
	PostEntry 2-Years & 0.060*** & 0.042*** & 0.020* \\
	\multicolumn{1}{r}{} & \multicolumn{1}{c}{(0.017)} & \multicolumn{1}{c}{(0.010)} & \multicolumn{1}{c}{(0.011)} \\
	PostEntry 3-Years & 0.075*** & 0.061*** & 0.031* \\
	\multicolumn{1}{r}{} & \multicolumn{1}{c}{(0.02)} & \multicolumn{1}{c}{(0.013)} & \multicolumn{1}{c}{(0.016)} \\
	PostEntry 4-Years & 0.102*** & 0.070*** & 0.036* \\
	\multicolumn{1}{r}{} & \multicolumn{1}{c}{(0.022)} & \multicolumn{1}{c}{(0.017)} & \multicolumn{1}{c}{(0.019)} \\
	\multicolumn{1}{r}{} & \multicolumn{1}{c}{} & \multicolumn{1}{c}{} & \multicolumn{1}{c}{} \\
	&&&\\
	Observations & \multicolumn{1}{c}{9,984} & \multicolumn{1}{c}{9,548} & \multicolumn{1}{c}{10,444} \\
	Zip-code FE     & Yes            & Yes            & Yes            \\
	State-Year FE  & Yes            & Yes            & Yes           \\
	Sample of Zip-codes & Low Vacancy          & Medium Vacancy       & High Vacancy        \\
	\bottomrule
\end{tabular}%
		\label{tab:vacancies1}
	\end{table}
\end{scriptsize}
\clearpage

\begin{scriptsize}
	\begin{table}[ht!]
		\caption{\textbf{Entry of Rideshare Services and Housing Prices given Car Ownership Costs}}
		\footnotesize 
		This table displays the results of regressions relating the entry of rideshare services with housing and rental prices given average car-ownership costs in the MSA. Sample is restricted to urban zip-codes. \textit{Urban zip-codes} are the zip-codes in the top quartile of a measure of perceived urbanicity. An MSA has high (low) \textit{car-ownership} \textit{costs} if its average parking rates and insurance premium is above (below) median. Information of parking and car-insurance rates obtained from \textit{SpotHero} and \textit{CarInsurance} respectively. \textit{House Price Index$_{z,y}$} is an index of house prices at the zip-code \textit{z} year \textit{y}, obtained from \textit{Corelogic}. \textit{Rental Prices$_{z,y}$} is an index of average market rental prices of zip-code \textit{z} in year \textit{y}, obtained from \textit{Zillow}. *, **, and *** denote statistical significance at the 10\%, 5\%, and 1\% levels, respectively. See Table \ref{tab: variable definitions} in the Appendix for detailed variable definitions.\\
		\setlength{\tabcolsep}{26pt}
		\begin{tabular}{lcccc}
	\toprule
	\toprule
	& \multicolumn{2}{c}{House Price Index$_{z,y}$} & \multicolumn{2}{c}{Rental Prices$_{z,y}$} \\
	& (1)            & (2)            & (3)            & (4) \\
	\midrule
	&                &                &                &  \\
	PreEntry 2-Years & -0.004         & -0.011*        & -0.013         & -0.061*** \\
	& (0.010)        & (0.006)        & (0.011)        & (0.011) \\
	PreEntry 1-Years & 0.002          & -0.004         & -0.019         & -0.061*** \\
	& (0.010)        & (0.007)        & (0.012)        & (0.012) \\
	PostEntry 0-Years & 0.021***       & 0.009**        & 0.003          & 0.004 \\
	& (0.003)        & (0.004)        & (0.007)        & (0.004) \\
	PostEntry 1-Years & 0.036***       & 0.013***       & 0.016*         & 0.011** \\
	& (0.005)        & (0.004)        & (0.009)        & (0.005) \\
	PostEntry 2-Years & 0.049***       & 0.025***       & 0.027***       & 0.020*** \\
	& (0.007)        & (0.005)        & (0.009)        & (0.006) \\
	PostEntry 3-Years & 0.067***       & 0.040***       & 0.038***       & 0.024*** \\
	& (0.009)        & (0.005)        & (0.009)        & (0.007) \\
	PostEntry 4-Years & 0.085***       & 0.048***       & 0.056***       & 0.029*** \\
	& (0.011)        & (0.006)        & (0.012)        & (0.008) \\
	&                &                &                &  \\
	Observations   & 15,760         & 16,105         & 18,018         & 21,111 \\
	Zip-code FE     & Yes            & Yes            & Yes            & Yes \\
	State-Year FE  & No             & Yes            & No             & Yes \\
	Car Ownership Costs & High           & Low            & High           & Low \\
	\bottomrule
\end{tabular}%
		\label{tab:carownershipcost}
	\end{table}
\end{scriptsize}
\clearpage

\begin{scriptsize}
	\begin{table}[ht!]
		\caption{\textbf{Entry of Rideshare Services and Relocation Away From Zip-Code – Pooled Zip-codes}}
		\footnotesize 
		This table displays the results of regressions relating the entry of rideshare services with the moving of individuals away from their current zip-code of residence, pooling urban and suburban zip-codes together. \textit{Urban (Suburban)} zip-codes are in the top (bottom) quartile of a measure of perceived urbanicity compiled by the Census’ \textit{Urbanization Perceptions Small Area Index}. \textit{Urban} zip-codes are the treatment group for the entry of rideshare services in an MSA. \textit{Moving Out Zip-Code$_{i,y+1}$} is an indicator of whether individual \textit{i} has a different residential zip-code in year \textit{y+1} relative to year \textit{y}. \textit{Controls} are a series of individual variables including age (in logs), outstanding debt (in logs), fico score (in logs), and an indicator of current late repayment. \textit{Homeowner (Non-Homeowner)} are individuals that have (not) had a mortgage or home-equity loan in their current zip-code. *, **, and *** denote statistical significance at the 10\%, 5\%, and 1\% levels, respectively.  See Table \ref{tab: variable definitions} in the Appendix for detailed variable definitions.\\
		\setlength{\tabcolsep}{30pt}
		\begin{tabular}{lccc}
	\toprule
	\toprule
	\multicolumn{1}{r}{} & \multicolumn{3}{c}{Moving Out of Zip-Code$_{i,y+1}$} \\
	\multicolumn{1}{r}{} & \multicolumn{1}{c}{(1)} & \multicolumn{1}{c}{(2)} & \multicolumn{1}{c}{(3)} \\
	\midrule
	\multicolumn{1}{r}{} &                &                &  \\
	PreEntry 2-Years & \multicolumn{1}{c}{-0.004*} & \multicolumn{1}{c}{-0.004} & \multicolumn{1}{c}{-0.002} \\
	\multicolumn{1}{r}{} & \multicolumn{1}{c}{(0.002)} & \multicolumn{1}{c}{(0.003)} & \multicolumn{1}{c}{(0.003)} \\
	PostEntry 1-Years & \multicolumn{1}{c}{-0.004*} & \multicolumn{1}{c}{-0.005*} & \multicolumn{1}{c}{-0.002} \\
	\multicolumn{1}{r}{} & \multicolumn{1}{c}{(0.002)} & \multicolumn{1}{c}{(0.003)} & \multicolumn{1}{c}{(0.003)} \\
	PostEntry 0-Years & \multicolumn{1}{c}{0.003**} & \multicolumn{1}{c}{0.004**} & \multicolumn{1}{c}{-0.000} \\
	\multicolumn{1}{r}{} & \multicolumn{1}{c}{(0.001)} & \multicolumn{1}{c}{(0.002)} & \multicolumn{1}{c}{(0.003)} \\
	PostEntry 1-Years & \multicolumn{1}{c}{0.002} & \multicolumn{1}{c}{0.002} & \multicolumn{1}{c}{0.005*} \\
	\multicolumn{1}{r}{} & \multicolumn{1}{c}{(0.002)} & \multicolumn{1}{c}{(0.002)} & \multicolumn{1}{c}{(0.003)} \\
	PostEntry 2-Years & \multicolumn{1}{c}{0.002} & \multicolumn{1}{c}{0.002} & \multicolumn{1}{c}{0.001} \\
	\multicolumn{1}{r}{} & \multicolumn{1}{c}{(0.002)} & \multicolumn{1}{c}{(0.002)} & \multicolumn{1}{c}{(0.003)} \\
	PostEntry 3-Years & \multicolumn{1}{c}{0.005***} & \multicolumn{1}{c}{0.004*} & \multicolumn{1}{c}{0.009***} \\
	\multicolumn{1}{r}{} & \multicolumn{1}{c}{(0.002)} & \multicolumn{1}{c}{(0.002)} & \multicolumn{1}{c}{(0.003)} \\
	PostEntry 4-Years & \multicolumn{1}{c}{0.007***} & \multicolumn{1}{c}{0.006**} & \multicolumn{1}{c}{0.007***} \\
	\multicolumn{1}{r}{} & \multicolumn{1}{c}{(0.002)} & \multicolumn{1}{c}{(0.003)} & \multicolumn{1}{c}{(0.003)} \\
	\multicolumn{1}{r}{} &                &                &  \\
	Observations   & 2,412,835        & 1,411,011        & 1,001,804 \\
	Controls       & \multicolumn{1}{c}{Yes} & \multicolumn{1}{c}{Yes} & \multicolumn{1}{c}{Yes} \\
	Zip-code FE     & \multicolumn{1}{c}{Yes} & \multicolumn{1}{c}{Yes} & \multicolumn{1}{c}{Yes} \\
	MSA-Year FE    & \multicolumn{1}{c}{Yes} & \multicolumn{1}{c}{Yes} & \multicolumn{1}{c}{Yes} \\
	Sample         & \multicolumn{1}{c}{All} & \multicolumn{1}{c}{Non-Homeowner} & \multicolumn{1}{c}{Homeowner} \\
	Avg. Dep. Variable & 0.1            & 0.12           & 0.06 \\
	\bottomrule
\end{tabular}%
		\label{tab:tableA12}
	\end{table}
\end{scriptsize}
\clearpage

\begin{scriptsize}
	\begin{table}[ht!]
		\caption{\textbf{Entry of Rideshare Services and Late Credit – Pooled Zip-Codes}}
		\footnotesize 
		This table displays the results of regressions relating the entry of rideshare services with the short-term credit quality of individuals, pooling urban and suburban zip-codes together. \textit{Urban (Suburban)} zip-codes are in the top (bottom) quartile of a measure of perceived urbanicity compiled by the Census’ \textit{Urbanization Perceptions Small Area Index}. \textit{Urban} zip-codes are the treatment group for the entry of rideshare services in an MSA. \textit{Late Repayment$_{i,y+2}$} is an indicator of whether individual \textit{i} is late on his/her payments in year \textit{y+2}. \textit{Controls} are a series of individual variables including age (in logs), outstanding debt (in logs), fico score (in logs), and an indicator of current late repayment. \textit{Homeowner (Non-Homeowner)} are individuals that have (not) had a mortgage or home-equity loan in their current zip-code. *, **, and *** denote statistical significance at the 10\%, 5\%, and 1\% levels, respectively.  See Table \ref{tab: variable definitions} in the Appendix for detailed variable definitions.\\
		\setlength{\tabcolsep}{30pt}
		\begin{tabular}{lccc}
	\toprule
	\toprule
	\multicolumn{1}{r}{} & \multicolumn{3}{c}{Late Repayment$_{i,y+2}$} \\
	\multicolumn{1}{r}{} & \multicolumn{1}{c}{(1)} & \multicolumn{1}{c}{(2)} & \multicolumn{1}{c}{(3)} \\
	\midrule
	\multicolumn{1}{r}{} &                &                &  \\
	PreEntry 2-Years & \multicolumn{1}{c}{-0.000} & \multicolumn{1}{c}{0.001} & \multicolumn{1}{c}{-0.006**} \\
	\multicolumn{1}{r}{} & \multicolumn{1}{c}{(0.001)} & \multicolumn{1}{c}{(0.002)} & \multicolumn{1}{c}{(0.003)} \\
	PostEntry 1-Years & \multicolumn{1}{c}{0.001} & \multicolumn{1}{c}{0.001} & \multicolumn{1}{c}{-0.007**} \\
	\multicolumn{1}{r}{} & \multicolumn{1}{c}{(0.002)} & \multicolumn{1}{c}{(0.002)} & \multicolumn{1}{c}{(0.003)} \\
	PostEntry 0-Years & \multicolumn{1}{c}{0.002***} & \multicolumn{1}{c}{0.002*} & \multicolumn{1}{c}{0.000} \\
	\multicolumn{1}{r}{} & \multicolumn{1}{c}{(0.001)} & \multicolumn{1}{c}{(0.001)} & \multicolumn{1}{c}{(0.001)} \\
	PostEntry 1-Years & \multicolumn{1}{c}{0.003***} & \multicolumn{1}{c}{0.003***} & \multicolumn{1}{c}{-0.001} \\
	\multicolumn{1}{r}{} & \multicolumn{1}{c}{(0.001)} & \multicolumn{1}{c}{(0.001)} & \multicolumn{1}{c}{(0.001)} \\
	PostEntry 2-Years & \multicolumn{1}{c}{0.004***} & \multicolumn{1}{c}{0.005***} & \multicolumn{1}{c}{0.001} \\
	\multicolumn{1}{r}{} & \multicolumn{1}{c}{(0.001)} & \multicolumn{1}{c}{(0.001)} & \multicolumn{1}{c}{(0.001)} \\
	PostEntry 3-Years & \multicolumn{1}{c}{0.004***} & \multicolumn{1}{c}{0.004***} & \multicolumn{1}{c}{0.001} \\
	\multicolumn{1}{r}{} & \multicolumn{1}{c}{(0.001)} & \multicolumn{1}{c}{(0.001)} & \multicolumn{1}{c}{(0.002)} \\
	PostEntry 4-Years & \multicolumn{1}{c}{0.005***} & \multicolumn{1}{c}{0.004***} & \multicolumn{1}{c}{-0.001} \\
	\multicolumn{1}{r}{} & \multicolumn{1}{c}{(0.001)} & \multicolumn{1}{c}{(0.001)} & \multicolumn{1}{c}{(0.002)} \\
	\multicolumn{1}{r}{} &                &                &  \\
	Observations   & 2,175,947      & 1,279,590      & 896,338 \\
	Controls       & \multicolumn{1}{c}{Yes} & \multicolumn{1}{c}{Yes} & \multicolumn{1}{c}{Yes} \\
	Zip-code FE     & \multicolumn{1}{c}{Yes} & \multicolumn{1}{c}{Yes} & \multicolumn{1}{c}{Yes} \\
	MSA-Year FE    & \multicolumn{1}{c}{Yes} & \multicolumn{1}{c}{Yes} & \multicolumn{1}{c}{Yes} \\
	Sample         & \multicolumn{1}{c}{All} & \multicolumn{1}{c}{Non-Homeowner} & \multicolumn{1}{c}{Homeowner} \\
	Avg. Dep. Variable & 0.023          & 0.021          & 0.027 \\
	\bottomrule
\end{tabular}%
		\label{tab:tableA13}
	\end{table}
\end{scriptsize}
\clearpage


\begin{scriptsize}
	\begin{landscape}
		\begin{figure}[!h]%
			\caption{\textbf{Entry of Rideshare Services and Evolution of the Share of Drivers and Revenues of "Taxi and Limousine Services"}}
			\footnotesize{\raggedright This figure, displays the relation between the entry of rideshare services and the share of drivers  (left panel) and the share of revenues (right panel) of the sector \textit{Taxi and Limousine Service} in an MSA. The lines represent the sample average conditional on year to/from entry of rideshare services in the MSA. Blue vertical bars represent 95 percent confidence intervals.}
        \vspace{2pt} 
        \hrule height 0.8pt 
        \vspace{1.5pt} 
        \hrule height 0.5pt 
			\normalsize
			\begin{minipage}{0.6\textwidth}
				\centering
				\includegraphics[width=\linewidth]{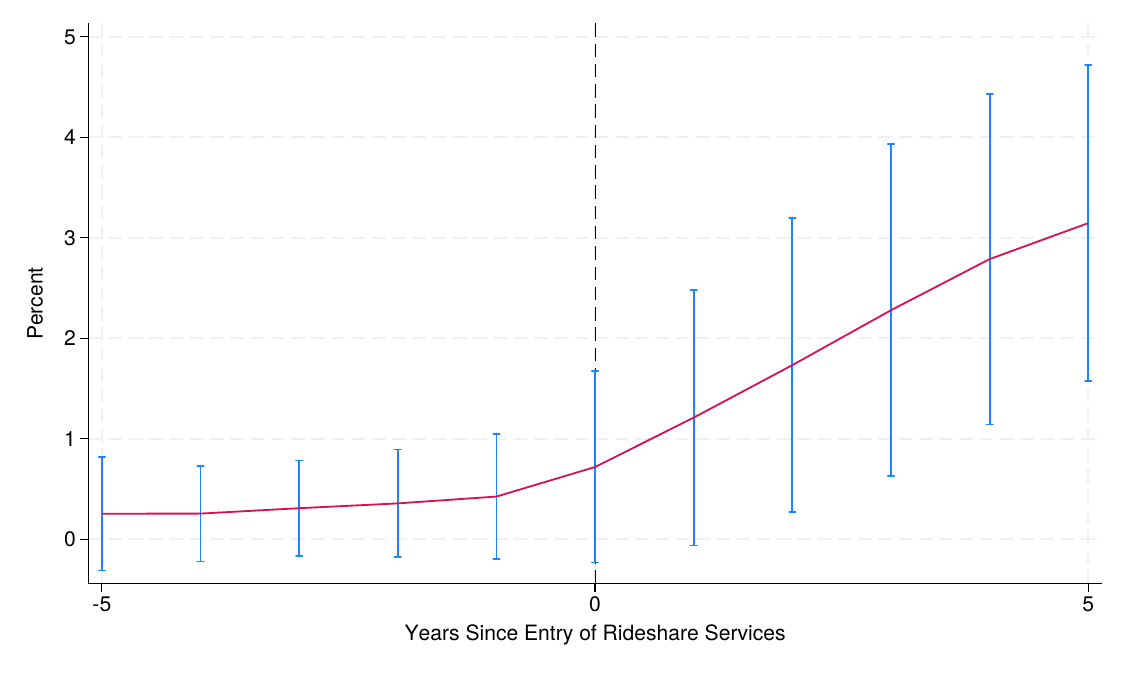}
			\end{minipage}\hfil
			\begin{minipage}{0.66\textwidth}
				\centering
				\includegraphics[width=\linewidth]{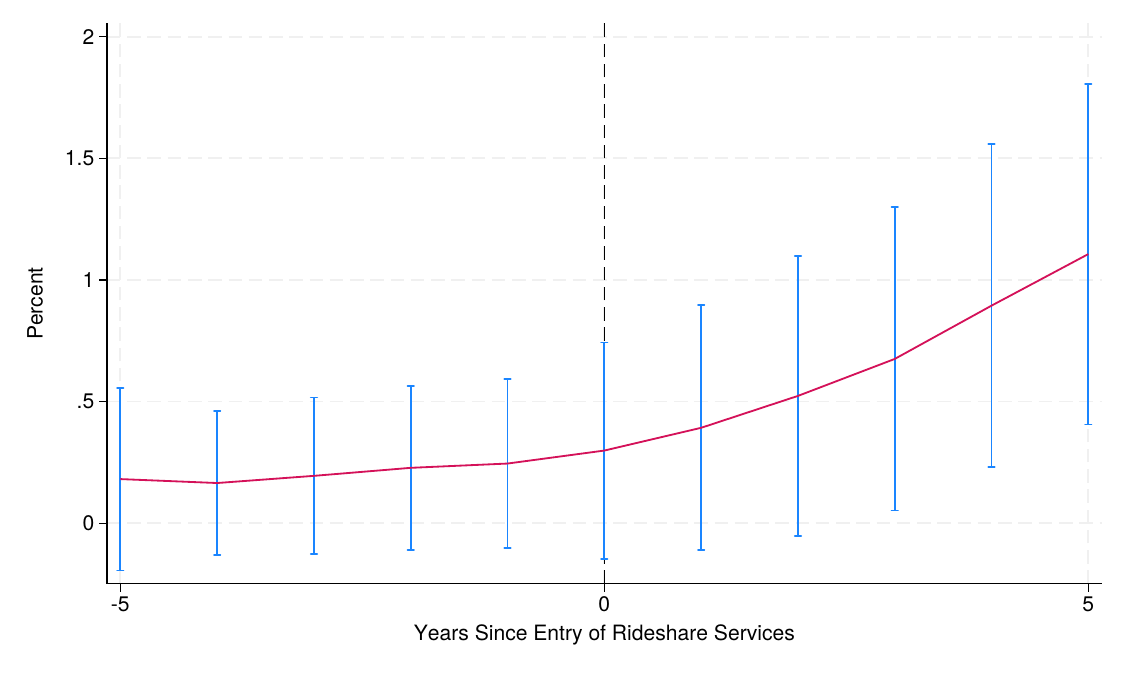}
			\end{minipage}
			\label{fig:supply_cabdrivers}
            \hrule height 0.5pt 
		\end{figure}
	\end{landscape}
\end{scriptsize}
\clearpage

\begin{scriptsize}
	\begin{landscape}
		\begin{figure}[!h]%
			\caption{\textbf{Entry of Rideshare Services and Number of Mortgage Originations}}
			\footnotesize{\raggedright This figure displays the evolution of yearly coefficients of a regression relating the entry of rideshare services and the number of mortgage originations in urban and suburban zip-codes, in the left and right panels respectively. \textit{Urban (Suburban)} zip-codes are in the top (bottom) quartile of a measure of perceived urbanicity compiled by the Census’ \textit{Urbanization Perceptions Small Area Index}. Regressions include zip-code and state*year fixed effects. Standard errors are clustered at the zip-code level. The blue vertical bars represent confidence intervals of the coefficients at the 95 percent significance level. See Table \ref{tab: variable definitions} in the Appendix for detailed variable definitions.}
        \vspace{2pt} 
        \hrule height 0.8pt 
        \vspace{1.5pt} 
        \hrule height 0.5pt 
			\normalsize
			\begin{minipage}{0.65\textwidth}
				\centering
				\caption*{(A) Urban Zip-codes}
				\includegraphics[width=\linewidth]{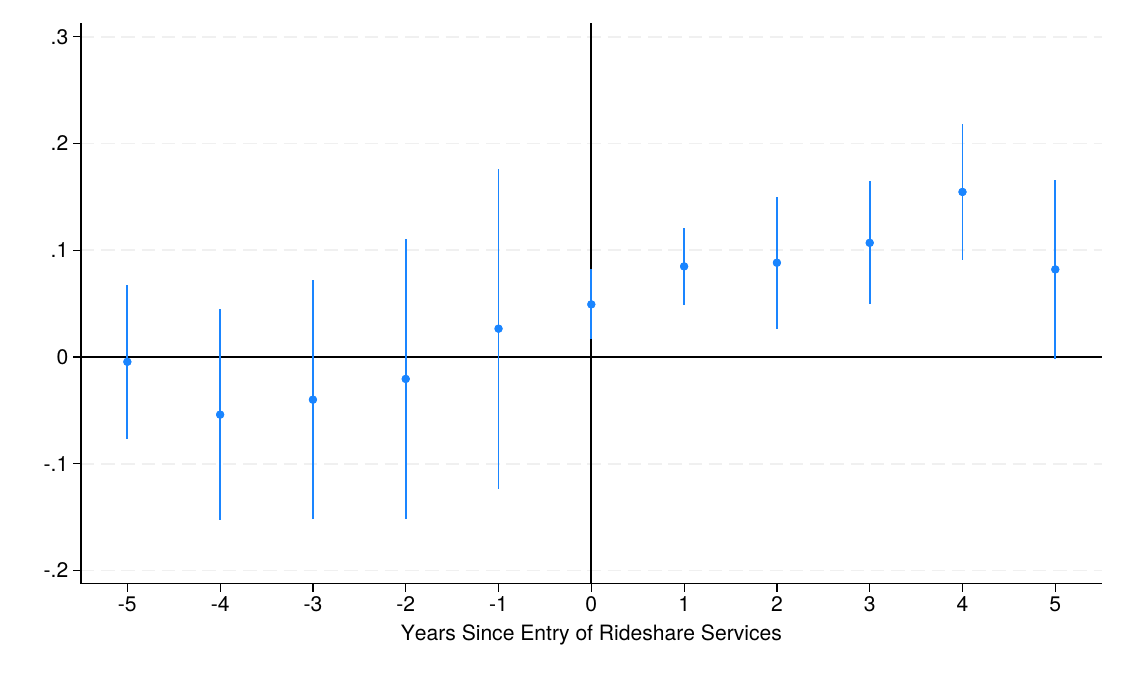}
			\end{minipage}\hfil
			\begin{minipage}{0.66\textwidth}
				\centering
				\caption*{(B) Suburban Zip-codes}
				\includegraphics[width=\linewidth]{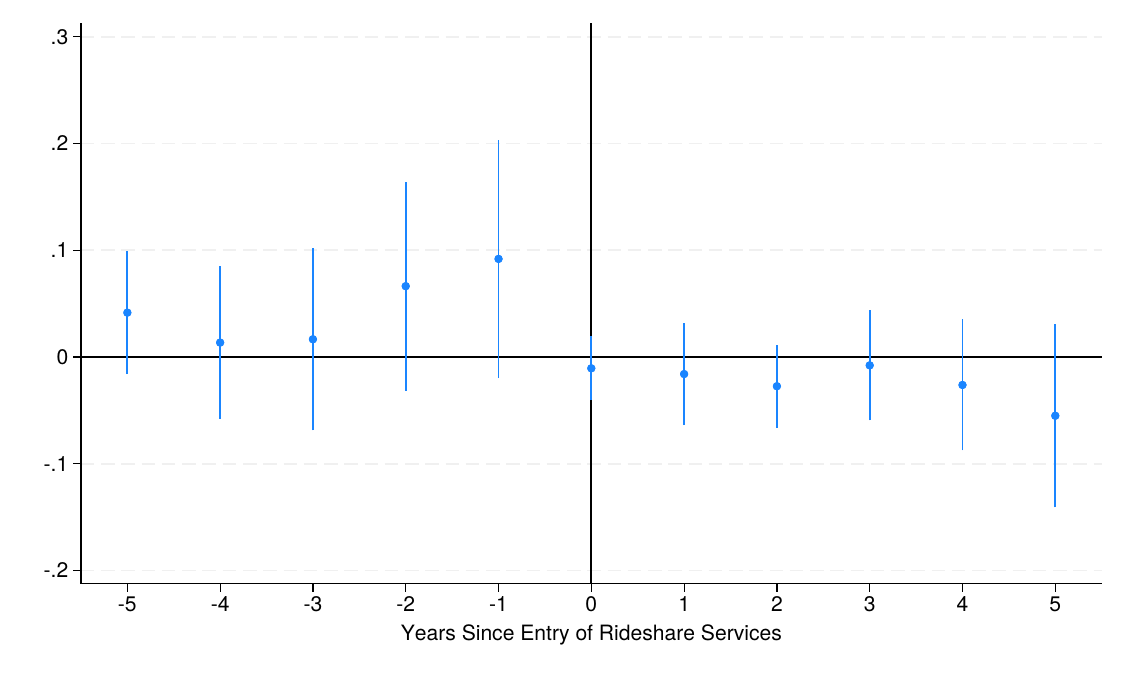}
			\end{minipage}
			\label{fig:mortgageorigination}
            \hrule height 0.5pt 
		\end{figure}
	\end{landscape}
\end{scriptsize}
\clearpage

\begin{scriptsize}
	\begin{landscape}
		\begin{figure}[!h]%
			\caption{\textbf{Entry of Rideshare Services and Vacancy Rates}}
			\footnotesize{\raggedright This figure displays the evolution of yearly coefficients of a regression relating the entry of rideshare services with housing-vacancy rates in urban and suburban zip-codes, in the left and right panels respectively. \textit{Urban (Suburban)} zip-codes are in the top (bottom) quartile of a measure of perceived urbanicity compiled by the Census’ \textit{Urbanization Perceptions Small Area Index.} Regressions include zip-code and state*year fixed effects. Standard errors are clustered at the zip-code level. The blue vertical bars represent confidence intervals of the coefficients at the 95 percent significance level. See Table \ref{tab: variable definitions} in the Appendix for detailed variable definitions.}
        \vspace{2pt} 
        \hrule height 0.8pt 
        \vspace{1.5pt} 
        \hrule height 0.5pt 
			\normalsize
			\begin{minipage}{0.66\textwidth}
				\centering
				\caption*{(A) Urban Zip-codes}
				\includegraphics[width=\linewidth]{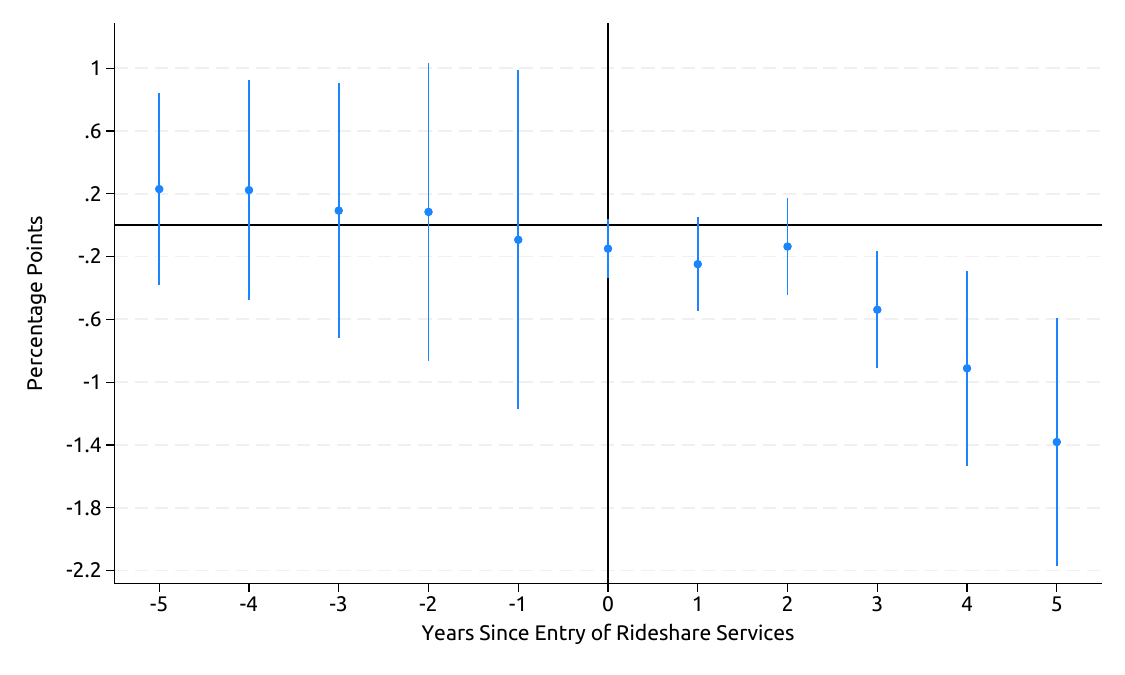}
			\end{minipage}\hfil
			\begin{minipage}{0.65\textwidth}
				\centering
				\caption*{(B) Suburban Zip-codes}
				\includegraphics[width=\linewidth]{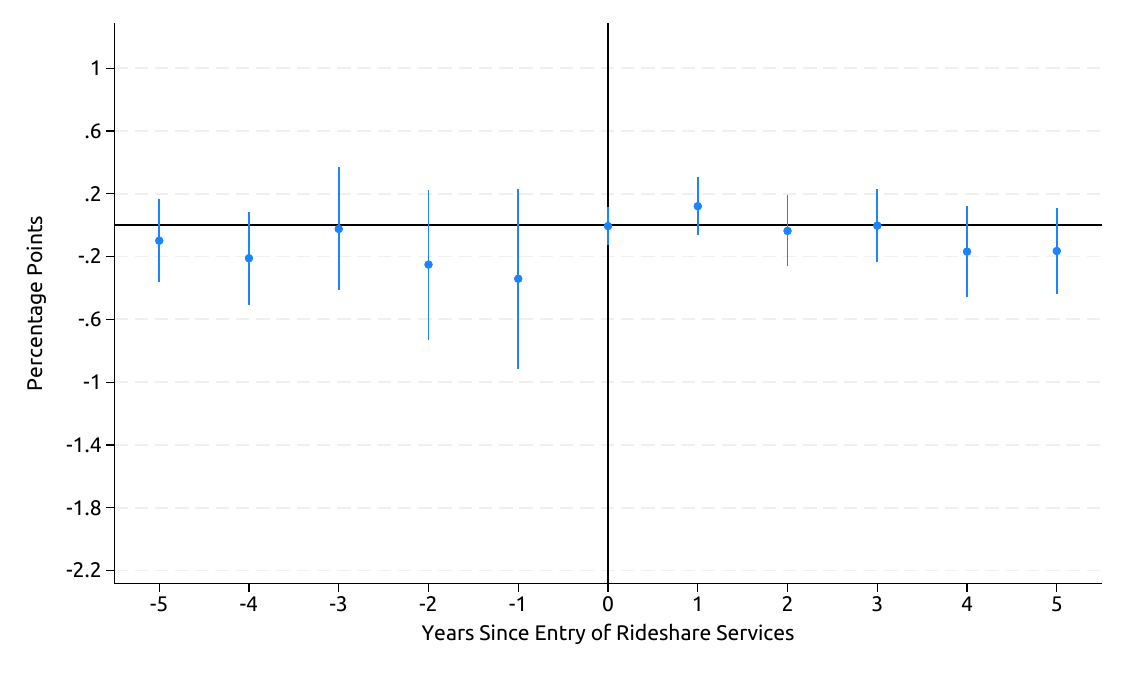}
			\end{minipage}
			\label{fig:vacancies}
            \hrule height 0.5pt 
		\end{figure}
	\end{landscape}
\end{scriptsize}
\clearpage

\begin{scriptsize}
	\begin{landscape}
		\begin{figure}[!h]%
			\caption{\textbf{Entry of Rideshare Services and Poverty Rate of Zip-codes of Mortgage Originations}}
			\footnotesize{\raggedright This figure displays the evolution of yearly coefficients of a regression relating the entry of rideshare services with the poverty rate of the zip-codes of the mortgage originations in urban and suburban zip-codes, in the left and right panels respectively. \textit{Urban (Suburban)} zip-codes are in the top (bottom) quartile of a measure of perceived urbanicity compiled by the Census’ \textit{Urbanization Perceptions Small Area Index}. Regressions include MSA and state*year fixed effects. Standard errors are clustered at the zip-code level. The blue vertical bars represent confidence intervals of the coefficients at the 95 percent significance level. See Table \ref{tab: variable definitions} in the Appendix for detailed variable definitions.}
        \vspace{2pt} 
        \hrule height 0.8pt 
        \vspace{1.5pt} 
        \hrule height 0.5pt 
			\normalsize
			\begin{minipage}{0.65\textwidth}
				\centering
				\caption*{(A) Urban Zip-codes}
				\includegraphics[width=\linewidth]{figures/fig\_A3\_1.png}
			\end{minipage}\hfil
			\begin{minipage}{0.66\textwidth}
				\centering
				\caption*{(B) Suburban Zip-codes}
				\includegraphics[width=\linewidth]{figures/fig\_A3\_2.png}
			\end{minipage}
			\label{fig:A3}
            \hrule height 0.5pt 
		\end{figure}
	\end{landscape}
\end{scriptsize}
\clearpage

\begin{scriptsize}
	\begin{landscape}
		\begin{figure}[!h]%
			\caption{\textbf{Exit of Rideshare Services and Rental Prices - The Case of Austin}}
			\footnotesize{\raggedright This figure displays the evolution of monthly coefficients of a regression relating the exit of rideshare services from Austin and the average rental prices in urban and suburban zip-codes, in the left and right panels respectively. \textit{Urban (Suburban)} zip-codes are in the top (bottom) quartile of a measure of perceived urbanicity compiled by the Census’ \textit{Urbanization Perceptions Small Area Index}. Regressions include zip-code and state*month fixed effects. Standard errors are clustered at the zip-code level. The vertical dashed black line to the left represents the month in which Austin legislature passed the ordinance regulating rideshare companies. The vertical dashed black line to the right represents the month in which \textit{Uber} and \textit{Lyft} stopped operations in Austin. The blue vertical bars represent confidence intervals of the coefficients at the 95 percent significance level. See Table \ref{tab: variable definitions} in the Appendix for detailed variable definitions.} 
        \vspace{2pt} 
        \hrule height 0.8pt 
        \vspace{1.5pt} 
        \hrule height 0.5pt 
			\normalsize
			\begin{minipage}{0.65\textwidth}
				\centering
				\caption*{(A) Urban Zip-Codes}
				\includegraphics[width=\linewidth]{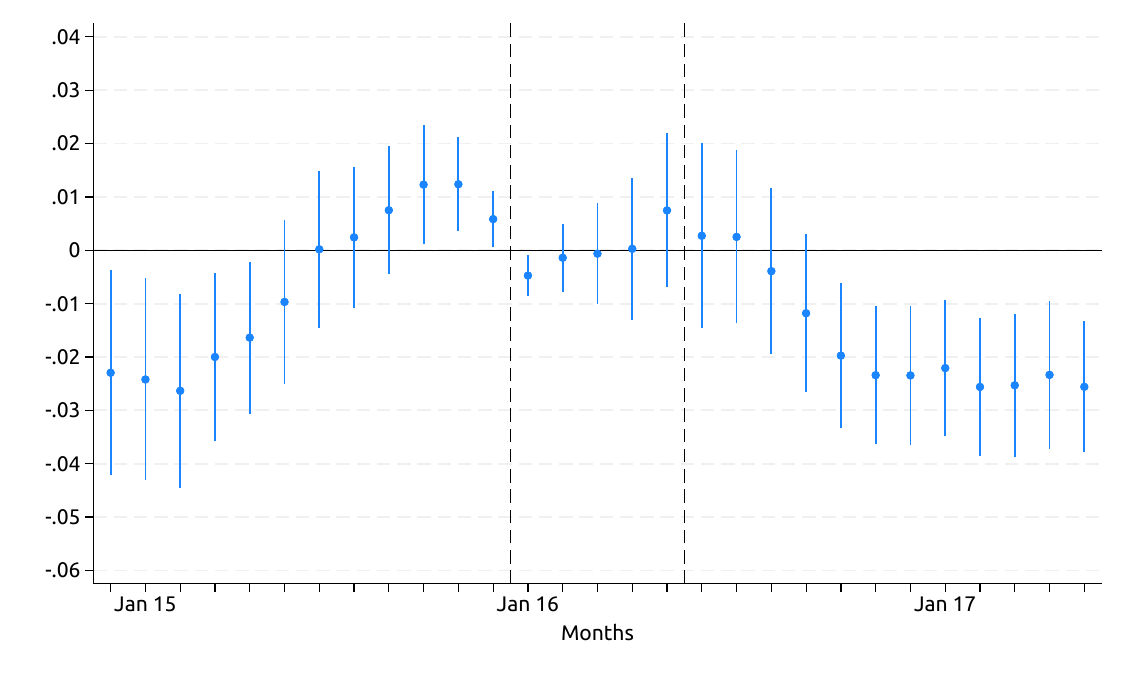}
			\end{minipage}\hfil
			\begin{minipage}{0.66\textwidth}
				\centering
				\caption*{(B) Suburban Zip-Codes}
				\includegraphics[width=\linewidth]{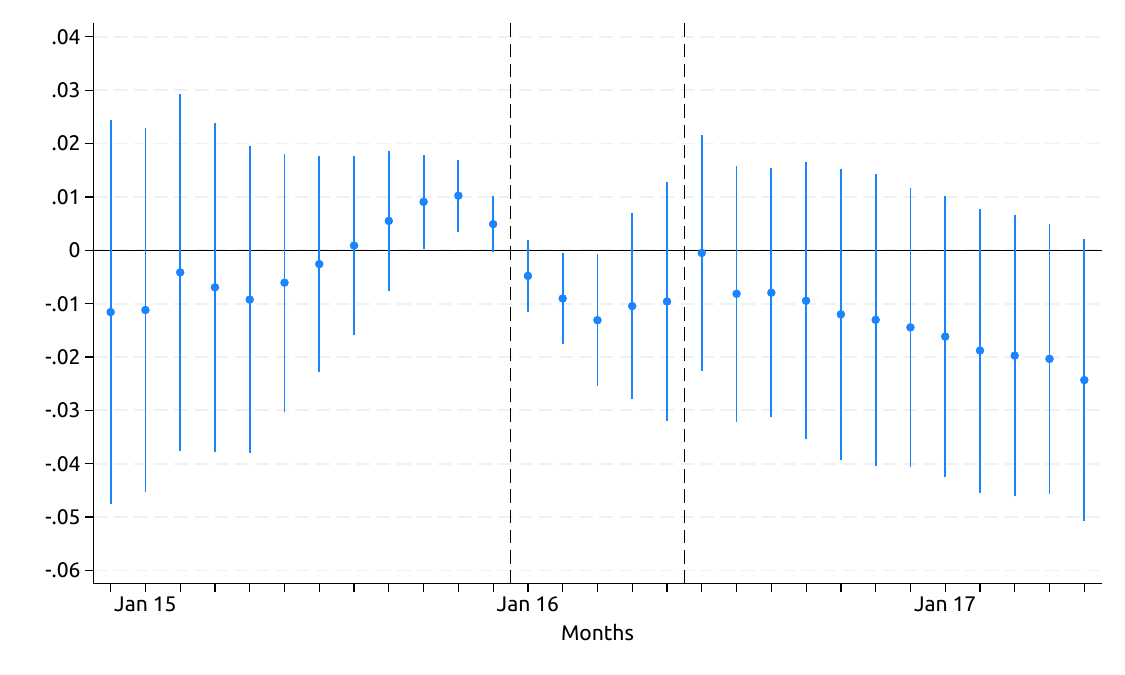}
			\end{minipage}
			\label{fig:exit}
            \hrule height 0.5pt 
		\end{figure}
	\end{landscape}
\end{scriptsize}
\clearpage


\renewcommand\thefigure{\thesection.\arabic{figure}}
\renewcommand\thetable{\thesection.\arabic{table}}
\setcounter{figure}{0}
\setcounter{table}{0}

\renewcommand{\thetable}{IA\arabic{table}}
\renewcommand{\thefigure}{IA\arabic{figure}}

\begin{center}
	\section*{Internet Appendix Table}
	\label{internetappendix:analysis}
\end{center}



\begin{scriptsize}
	\begin{table}[ht!]
		\caption[Characteristics of Homeowners vs non-Homeowners]{\textbf{Characteristics of Homeowners vs non-Homeowners}}
		\footnotesize 
		This table compares the average characteristics of homeowners and non-homeowners in our sample. Standard errors are in parenthesis. \textit{Homeowners (Non-Homeowners)}, are those individuals that have had (have not had) either a mortgage or an home-equity loan in current zip-code. See Table \ref{tab: variable definitions} in the Appendix for detailed variable definitions.\\
		\setlength{\tabcolsep}{41pt}
		\begin{tabular}{lcc}
	\toprule
	\toprule
	& \multicolumn{1}{c}{\textbf{Homeowner}} & \multicolumn{1}{c}{\textbf{Non-Homeowner}} \\
	\midrule
	Age$_i$ & 48.80  & 42.20  \\
	& (9.90) & (11.20) \\
	FICO$_i$ & 748   & 659 \\
	& (89)  & (161) \\
	Moving Out of Zip-Code$_i$& 0.07  & 0.12 \\
	& (0.26) & (0.33) \\
	Income$_z$ & 76,912 & 56,965 \\
	& (21,118) & (20,254) \\
	Poverty$_z$ & 0.11  & 0.19 \\
	& (0.09) & (0.10) \\
	\bottomrule
\end{tabular}%
		\label{tab:sumstathomeowner}
	\end{table}
\end{scriptsize}


\begin{scriptsize}
	\begin{landscape}
		\begin{figure}[!h]%
			\caption{\textbf{Urbanicity and Average Monthly Parking Costs in New York City}}
			\footnotesize{\raggedright This figure compares the urbanicity of zip-codes in New York City with average monthly parking costs. \textit{urbanicity} is measure of perceived urbanicity compiled by the Census’ \textit{Urbanization Perceptions Small Area Index}.} 
            \vspace{2pt} 
            \hrule height 0.8pt 
            \vspace{1.5pt} 
            \hrule height 0.5pt 
            \vspace{1.5pt} 
			\normalsize
				\centering
				\includegraphics[width=0.75\linewidth]{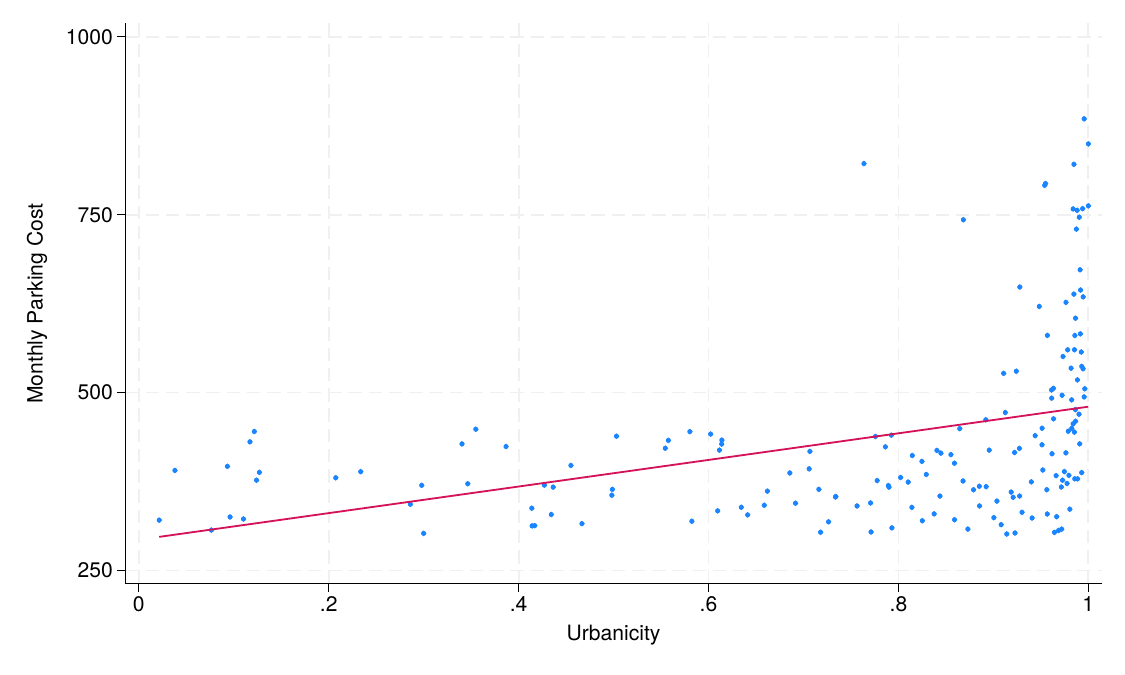}
			\label{fig: caronwershipcosts}
            \hrule height 0.5pt 
		\end{figure}
	\end{landscape}
\end{scriptsize}
\clearpage

\begin{scriptsize}
	\begin{landscape}
		\begin{figure}[!h]%
			\caption{\textbf{Entry of Rideshare Services and Value of Originated Mortgages}}
			\footnotesize{\raggedright This figure displays the evolution of yearly coefficients of a regression relating the entry of rideshare services with the average value of originated mortgages (in logs) in urban and suburban zip-codes, in the left and right panels respectively. \textit{Urban (Suburban)} zip-codes are in the top (bottom) quartile of a measure of perceived urbanicity compiled by the Census’ \textit{Urbanization Perceptions Small Area Index}. Regressions include zip-code and state*year fixed effects. Standard errors are clustered at the zip-code level. The blue vertical bars represent confidence intervals of the coefficients at the 95 percent significance level. See Table \ref{tab: variable definitions} in the Appendix for detailed variable definitions.} 
            \vspace{2pt} 
            \hrule height 0.8pt 
            \vspace{1.5pt} 
            \hrule height 0.5pt 
			\normalsize
			\begin{minipage}{0.6\textwidth}
				\centering
				\caption*{(A) Urban Zip-Codes}
				\includegraphics[width=\linewidth]{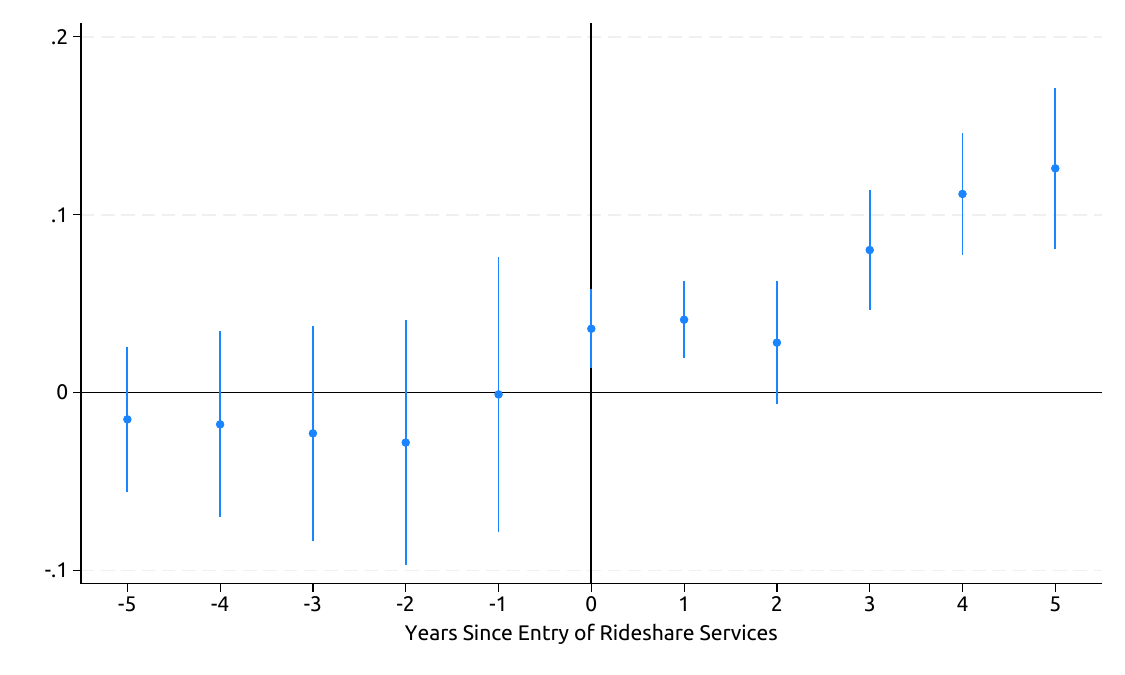}
			\end{minipage}\hfil
			\begin{minipage}{0.6\textwidth}
				\centering
				\caption*{(B) Suburban Zip-Codes}
				\includegraphics[width=\linewidth]{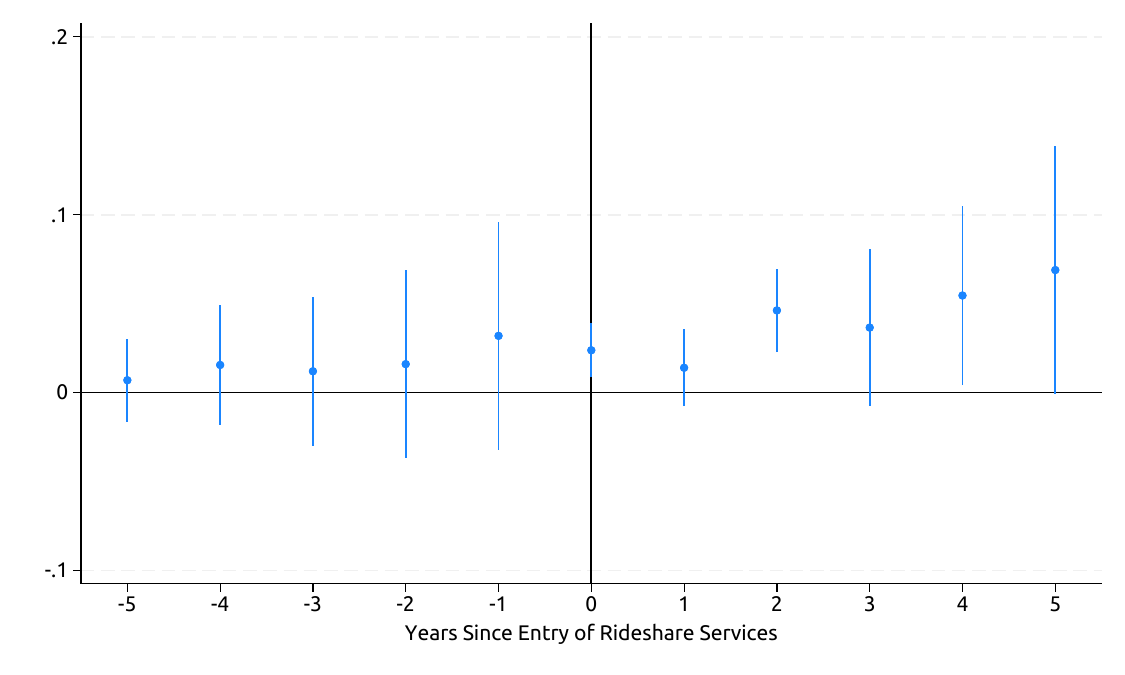}
			\end{minipage}
			\label{fig:mortgagevalue}
            \hrule height 0.5pt 
		\end{figure}
	\end{landscape}
\end{scriptsize}
\clearpage

\end{appendices}

\end{document}